\documentclass[aps,pra,twocolumn,superscriptaddress]{revtex4-1}

\usepackage[english]{babel}
\usepackage{amsmath}
\usepackage{amsthm}
\usepackage{amssymb}
\usepackage{mathtools}
\usepackage{ragged2e}
\usepackage{etoolbox}
\usepackage{booktabs}
\usepackage{color}
\usepackage{hyperref}
\usepackage{cases}
\usepackage{caption}
\usepackage{enumerate}
\usepackage{enumitem}
\usepackage{multirow}
\hypersetup{
  colorlinks   = true,  
  urlcolor     = blue,  
  linkcolor    = blue,  
  citecolor    = red    
}
\DeclarePairedDelimiter\ceil{\lceil}{\rceil}
\DeclarePairedDelimiter\floor{\lfloor}{\rfloor}

\usepackage[font=small,labelfont=bf,justification=justified,format=plain]{caption} 
\usepackage{natbib}
\usepackage{pgfplots}
\pgfplotsset{compat=1.17}
\usepackage{subcaption}
\usepackage{url}
\newcommand{\ket}[1]{ | #1 \rangle }
\newcommand{\bra}[1]{ \langle #1 | }

\newcommand\myeq{\mathrel{\stackrel{\makebox[0pt]{\mbox{\normalfont\scriptsize def}}}{=}}}
\DeclareMathOperator*{\argmin}{arg\,min}

\theoremstyle{definition}

\theoremstyle{remark}

\usepackage{scalerel,stackengine}
\stackMath
\newcommand\reallywidehat[1]{%
\savestack{\tmpbox}{\stretchto{%
  \scaleto{%
    \scalerel*[\widthof{\ensuremath{#1}}]{\kern-.6pt\bigwedge\kern-.6pt}%
    {\rule[-\textheight/2]{1ex}{\textheight}}
  }{\textheight}%
}{0.5ex}}%
\stackon[1pt]{#1}{\tmpbox}%
}
\usepackage{comment}
\usepackage{float}

\begin{document}
\definecolor{darkblue}{HTML}{5A7197}
\definecolor{darkorange}{HTML}{FFA500}
\definecolor{plotgreen}{HTML}{77AC30}
\definecolor{lightgreen}{HTML}{82BF85}
\definecolor{normalgreen}{HTML}{2BA02D}
\definecolor{darkgreen}{HTML}{09622A}

\title{Hybrid Quantum Singular Spectrum Decomposition for Time Series Analysis}

\author{J.J. \surname{Postema}}
\email{j.j.postema@tue.nl}
\affiliation{Department of Applied Physics and Science Education, Eindhoven University of Technology, P.~O.~Box 513, 5600 MB Eindhoven, The Netherlands}
\affiliation{Eindhoven Hendrik Casimir Institute, Eindhoven University of Technology, P.~O.~Box 513, 5600 MB Eindhoven, The Netherlands}

\author{P. \surname{Bonizzi}}
\affiliation{Department of Advanced Computing Sciences, Maastricht University, P.~O. ~Box 616, 6200 MD Maastricht, The Netherlands}

\author{G. \surname{Koekoek}}
\affiliation{Department of Gravitational Waves and Fundamental Physics, Maastricht University, P.~O. ~Box 616, 6200 MD Maastricht, The Netherlands}\affiliation{Nikhef, Science Park 105, 1098 XG Amsterdam, The Netherlands}

\author{R.L. \surname{Westra}}
\affiliation{Department of Advanced Computing Sciences, Maastricht University, P.~O. ~Box 616, 6200 MD Maastricht, The Netherlands}

\author{S.J.J.M.F. \surname{Kokkelmans}}
\affiliation{Department of Applied Physics and Science Education, Eindhoven University of Technology, P.~O.~Box 513, 5600 MB Eindhoven, The Netherlands}
\affiliation{Eindhoven Hendrik Casimir Institute, Eindhoven University of Technology, P.~O.~Box 513, 5600 MB Eindhoven, The Netherlands}

\date{\today}

\begin{abstract}
Classical data analysis requires computational efforts that become intractable in the age of Big Data. An essential task in time series analysis is the extraction of physically meaningful information from a noisy time series. One algorithm devised for this very purpose is singular spectrum decomposition (SSD), an adaptive method that allows for the extraction of narrow-banded components from non-stationary and non-linear time series. The main computational bottleneck of this algorithm is the singular value decomposition (SVD). Quantum computing could facilitate a speedup in this domain through superior scaling laws. We propose quantum SSD by assigning the SVD subroutine to a quantum computer. The viability for implementation and performance of this hybrid algorithm on a near term hybrid quantum computer is investigated. In this work we show that by employing randomised SVD, we can impose a qubit limit on one of the circuits to improve scalibility. Using this, we efficiently perform quantum SSD on simulations of local field potentials recorded in brain tissue, as well as GW150914, the first detected gravitational wave event.
\end{abstract}

\maketitle

\section{Introduction}
A recently proposed approach to classical time series analysis is \textit{singular spectrum decomposition} (SSD) \cite{ssd}. SSD aims to decompose non-linear and non-stationary time series into physically meaningful components. In stark contrast to a Fourier decomposition which employs a basis set of harmonic functions, SSD prepares an adaptive basis set of functions, whose elements depend uniquely on the time series. Applications include, but are not limited to, sleep apnea detection from an unprocessed ECG \cite{sleep} and the processing and analysis of brain waves \cite{macaque, macaque2}. In this work we will also cover application to gravitational wave analysis. The latter is interesting as real-time data analysis poses a challenge, and will be vital for next generations of gravitational wave detection technology. The main bottleneck for scaling this algorithm up for larger time series is the singular value decomposition (SVD) that is performed in each subroutine. For general $m\times n$ matrices, the classical SVD algorithm yields a computational complexity of $\mathcal{O}\left(mn\text{min}(m,n)\right)$ \cite{svdcomplexity}.\\

There exist clues that allude to quantum computing being able to provide a computational speedup for the computationally taxing SVD subroutine. Currently, quantum computing is in the \textit{Noisy Intermediate Scale Quantum} (NISQ) era \cite{nisq}. While some forms of error mitigation are proposed, noise still limits the amount of qubits and the fidelity of gate operations, such that their numbers are too little for full-fledged quantum computers running quantum error correction codes \cite{kandalarda}. Despite these shortcomings, hybrid quantum computers running variational quantum algorithms provide a useful platform for classically taxing calculations, one major example being the \textit{variational quantum eigensolver} (VQE) for quantum chemistry \cite{firstvqe?, vqe, molecularspectra}. Hybrid algorithms complement the strengths of both processors: classically efficient tasks are performed on a classical computer, while unitary calculations such as quantum state preparation are handed to a quantum computer. Because of the iterative nature of variational algorithms, noise can be mitigated at every step, increasing robustness of the algorithm.\\

The SVD subroutine can be performed on a quantum computer through the \textit{variational quantum singular value decomposition} (VQSVD) algorithm \cite{vqsvd}. Besides its application to SSD, SVD forms a compelling application of hybrid quantum computing in general, because of its versatile utility. In a similar fashion to VQE, one can train two parameterisable quantum circuits (PQCs) to learn the left and right singular vectors of a matrix, and sample the singular values through Hadamard tests. In this work we explore both a \textit{gate-based} paradigm \cite{vqe}, where the parameterisation is facilitated through rotational quantum gates, and a \textit{pulse-based} paradigm  \cite{robert}, where we optimise laser pulse profiles on a neutral atom quantum computing platform. The latter is derived from quantum optimal control (QOC) theory, and is thought to be more NISQ efficient.\\
\indent Our aim is to convert SSD into a variational hybrid quantum-classical algorithm called quantum singular spectrum decomposition (QSSD). Here, we pose the question whether quantum computing can provide an inherent computational speedup for SVD by surpassing the classical complexity. We employ randomised SVD to improve scalibility of the algorithm \cite{halko}. The classical theory behind SSD is introduced in Sec. \ref{sec:sectionII}. The implementation of QSSD on a hybrid quantum computer is elaborated on in Sec. \ref{sec:sectionIII}, using a gate-based approach, while in Sec. \ref{sec:sectionIV} we discuss a pulse-based approach. Simulation results for three different time series, among which the first observation of a gravitational wave, are presented in Sec. \ref{sec:sectionV}. We conclude with a summary and discussion about the viability of QSSD in Sec. \ref{sec:sectionVI}.

\section{Singular Spectrum Decomposition}\label{sec:sectionII}

The SSD algorithm is based on the singular spectrum analysis (SSA) method, whose basic ideas are presented in Appendix \ref{app:appendixPrel} \cite{ssd, ssa}. SSA is designed to explore the presence of specific patterns within a time series, while SSD is designed to provide a decomposition into its constituents components. Let $x(n)=\{x_1,x_2,\cdots,x_{N-1},x_N\}$ be an $N$-dimensional string of numbers, sampled at a uniform sampling frequency $F_S$ from some time series. We are interested in extracting (physically) meaningful components from the signal. By embedding the signal in a matrix, correlation functions between pairs of sample points can be extracted through a singular value decomposition of said matrix. Several types of components can be found, among which trends, oscillations and noise for instance. In SSD, such identification is completely automated by focusing on extracting narrow frequency band components.\\

The time series is first embedded in a trajectory matrix, however, in comparison to SSA, the number of columns is extended to $N$ and the time series is wrapped around, yielding:
\begin{equation}
X=\left[
\begin{array}{cccc|ccc}
x_1 & x_2 & \cdots & x_K & x_{K+1} & \cdots & x_N \\
x_2 & x_3 & \cdots & x_{K+1} & x_{K+2} & \cdots & x_1 \\
\vdots & \smash{\vdots} & \smash{\ddots} & \smash{\vdots} & \smash{\vdots} & \smash{\ddots} & \smash{\vdots} \\
x_M & x_{M+1} & \cdots & x_N & x_{1} & \cdots & x_{N+M-1} 
\end{array}
\right].
\end{equation}
It is to be understood that every subscript is $\text{mod } N$. The singular value decomposition (SVD) of such a matrix yields 
\begin{equation}
        X=\sum_{i} X_i\myeq\sum_{i} \sigma_i \boldsymbol{u}_i \boldsymbol{v}^\top_i, 
    \end{equation} 
where $\sigma_i$ are the singular values, and $\boldsymbol{u}_i$ and $\boldsymbol{v}_i$ the left and right singular vectors. As explained above, SSD aims to find physically meaningful components $g_j(n)$ through a data-driven and iterative approach. Starting from the original signal $x(n)$, at each iteration, a component $g^{(j)}(n)$ is extracted, such that a residue 
\begin{equation}
    v^{(j)}(n)=v^{(j-1)}(n)-g^{(j)}(n)
\end{equation}
remains, where $x(n)\myeq v^{(0)}(n)$.\\

\textit{For iteration $j=1$:}\\

The power spectral density (PSD) of the signal is calculated. The frequency $f_{\text{max}}$ associated with the dominant peak in the PSD is then estimated. If the ratio criterion $\frac{f_\text{max}}{F_S}<\delta$ is met, a trend is said to characterise the spectral contents of the residual. Ref. \cite{ssd} sets $\delta=10^{-3}$. In such a case, the embedding dimension is set to $M=\floor{\frac{N}{3}}$, as described in Ref. \cite{m=n3}. The first component $g^{(1)}(n)$ is then extracted from Hankelisation of the matrix $X_1=\sigma_1\boldsymbol{u}_1\boldsymbol{v}_1^\top$. Since most of the energy must be contained within the trend component, only 1 singular value is required for its reconstruction. If the ratio criterion is not met, the window length is defined by $M=\floor{1.2\frac{F_S}{f_\text{max}}}$ \cite{ssd}.\\

\textit{For iterations $j\geq 2$:}\\

For all iterations $j\geq 2$, the embedding dimension is always set equal to $M=\floor{1.2\frac{F_S}{f_\text{max}}}$. To extract meaningful components, only principal components with a dominant frequency in the band $[f_{\text{max}}-\delta f,f_{\text{max}}+\delta f]$ are selected. In order to estimate the value of the peak width $\delta f$, the Levenberg–Marquardt (LM) algorithm is used. First, the PSD of the residual is approximated with a sum of three Gaussian functions
\begin{equation}
    \gamma\left(f,\vec{A},\vec{\sigma}\right)=\sum_{i=1}^{3}A_i e^{-\frac{(f-\mu_i)^2}{2\sigma_i^2}},
\end{equation}
where $\vec{A}$ is the vector of amplitudes, $\vec{\sigma}$ is the vector of Gaussian peak widths and $\mu_i$ are the peak locations, given by
\begin{equation}
    \mu_1=f_\text{max},\quad \mu_2=f_2,\quad \mu_3=\frac{1}{2}(f_\text{max}+f_2)\myeq f_3,
\end{equation}
where $f_2$ is the frequency at which the second most dominant peak is situated. Starting the algorithm from the initial state vector components
\begin{equation}
    \vec{A}=\left(\frac{1}{2}\text{PSD}(f_\text{max}),\,\frac{1}{2}\text{PSD}(f_2),\,\frac{1}{4}\text{PSD}(f_3)\right),
\end{equation}
\begin{equation}
    \vec{\sigma}=\left(\frac{2}{3}\text{PSD}(f_\text{max}),\,\frac{2}{3}\text{PSD}(f_2),\,4|f_\text{max}-f_2|\right),
\end{equation}
the algorithm will eventually converge towards optimal $\vec{A}^\star, \vec{\sigma}^\star$. The peak width is then given by $\delta f=2.5\sigma_1^\star$. Finally, the component $g^{(j)}(n)$ is rescaled with a factor $a$, so that $\tilde{g}^{(j)}(n)=ag^{(j)}(n)$. Such rescaling makes sure the variance of the residue is minimised. Solving for
\begin{equation}\label{eq:rescale}
    a=\argmin_a ||v^{(j-1)}(n)-ag^{(j)}(n)||_2^2
\end{equation}
yields
$a=\frac{g^{(j),\top} v^{(j-1)}}{g^{(j),\top} g^{(j)}}$ \cite{ssd}. The LM algorithm stops when the energy of the remainder has been reduced to a predefined threshold (default at 1\%), ensuring proper convergence of SSD. This corresponds to the criterion
\begin{equation}
    \frac{||v^{(j+1)}(n)||_2^2}{||x(n)||_2^2}<0.01,
\end{equation}
so that the reconstructed signal reads
\begin{equation}
    x(n) = \sum_{i=1}^m \tilde{g}^{(i)}(n)+\mathcal{O}(v_m(n)).
\end{equation}

\section{Hybrid gate-based Singular value decomposition}\label{sec:sectionIII}

\subsection{\label{sec:OutlineAlgorithm}SVD quantum algorithm}

The singular value decomposition (SVD) of a general $m\times n$ matrix $X\in\mathbb{C}^{m\times n}$ is given by
\begin{equation}\label{eq:SVD}
    X=U\Sigma V^\dagger,
\end{equation}
where $U\in\mathbb{C}^{m\times m}$ and $V\in\mathbb{C}^{n\times n}$ are unitary square matrices, and $\Sigma=\text{diag}(\sigma_1,\sigma_2,\cdots,\sigma_r)\in\mathbb{R}^{m\times n}$ is a diagonal matrix with all $r=\text{rank}(X)$ singular values ordered from highest to lowest on the diagonal. Through a suitable loss function, Wang et al. \cite{vqsvd} have shown that this decomposition can be implemented on a quantum computer by training two quantum circuits to learn the unitary matrices $U(\boldsymbol{\alpha})$ and $V(\boldsymbol{\beta})$, parameterised by two sets of trainable parameters $\boldsymbol{\alpha}$ and $\boldsymbol{\beta}$. For real matrices, $U$ and $V$ will be orthogonal, which can be efficiently prepared on a quantum computer since $\text{SO}(N)\subset\text{SU}(N)$. We focus on this particular problem since data matrices are real. Through inversion of eq. (\ref{eq:SVD}), one finds that the singular values can be extracted from
\begin{equation}\label{eq:umv}
    \sigma_i=\bra{u_i}X\ket{v_i},
\end{equation}
where $\bra{u_i}$ is the Hermitian conjugate of the $i$-th column of $U$, and $\ket{v_i}$ is the $i$-th column of $V$. The singular values and their associated left and right orthonormal singular vectors allow for a reconstruction of the original matrix $X$ through
\begin{equation}\label{eq:m=suv}
    X=\sum_{i=1}^r \sigma_i \ket{u_i}\bra{v_i}.
\end{equation}
It is desirable to keep only the first few $T$ eigentriples such that they approximate the matrix $X$ to precision $\epsilon$. Suppose that $X^{(T)}=\sum_{i=1}^T \sigma_i \ket{u_i}\bra{v_i}$, then we find through the Cauchy-Schwarz inequality that the Frobenius norm of the difference is bounded by the squared sum of neglected singular values according to
\begin{equation}
    ||X-X^{(T)}||_F^2\leq\sum_{i=T+1}^r \sigma_i^2.
\end{equation}
More particularly, the Eckart-Young theorem states that this approximation minimises this error norm, and is therefore the optimal matrix approximation \cite{eckartyoung}.

\subsection{Trajectory matrix implementation}

In order to prepare inner products of the form (\ref{eq:umv}) on a quantum computer, we must perform a unitary decomposition, coined a linear combination of unitaries (LCU), of the trajectory matrix:
\begin{equation}\label{eq:lcu}
    X=\sum_{i=1}^{K} x_i e_i
\end{equation}
for $K\in\mathbb{N}$ unitaries $e_i$. Because of the non-square nature of $X$, we delay the discussion on the implementation of this decomposition to Sec. \ref{sec:hadamard}. This set of unitaries is closely related to Pauli operator strings of the form $\sigma_{i_1}^{(i)}\otimes\sigma_{i_2}^{(i)}\otimes\cdots$, but different choices are possible. Throughout this paper we adopt the notation
$i_j\in\{0,1,2,3\}=\{0,x,y,z\}$ and
\begin{equation}\label{eq:pauli}
    \sigma^0=\begin{pmatrix}1 &0\\0&1\end{pmatrix},\sigma^x=\begin{pmatrix}0 &1\\1&0\end{pmatrix},\sigma^y=\begin{pmatrix}0 &-i\\i&0\end{pmatrix},\sigma^z=\begin{pmatrix}1 &0\\0&-1\end{pmatrix}.
\end{equation}
Since $m$-qubit states are elements of a $2^m$-dimensional Hilbert space $\mathfrak{H}_m = \text{span}\{\ket{0},\ket{1}\}^{\otimes m}\simeq \mathbb{C}^{2^m}$, the dimensions of the trajectory matrix must be enforced to be equal to a power of 2. The optimal dimensions of the trajectory matrix $X$, as dictated by SSD, do not necessarily satisfy these constraints, so that we slightly alter the trajectory matrix dimensions for a VQSVD implementation. First, the optimal embedding dimension is rounded to the nearest power of two. Then, the number of columns is increased to the first power of two through \textit{periodic continuation}, where the signal is artificially extended through a wrap-around. Defining
\begin{equation}
    q_m=\argmin_{a\in\mathbb{N}}|M-2^a|,
\end{equation}
the qubit numbers of the $U(\boldsymbol{\alpha})$ and $V(\boldsymbol{\beta})$ circuits are given by $q_m$ and $q_n=\ceil{\log_2(N)}$ respectively. The trajectory matrix then takes the form

\onecolumngrid

\begin{figure}[H]
\centering{
\captionsetup{justification=justified}
\includegraphics[width=0.9\textwidth]{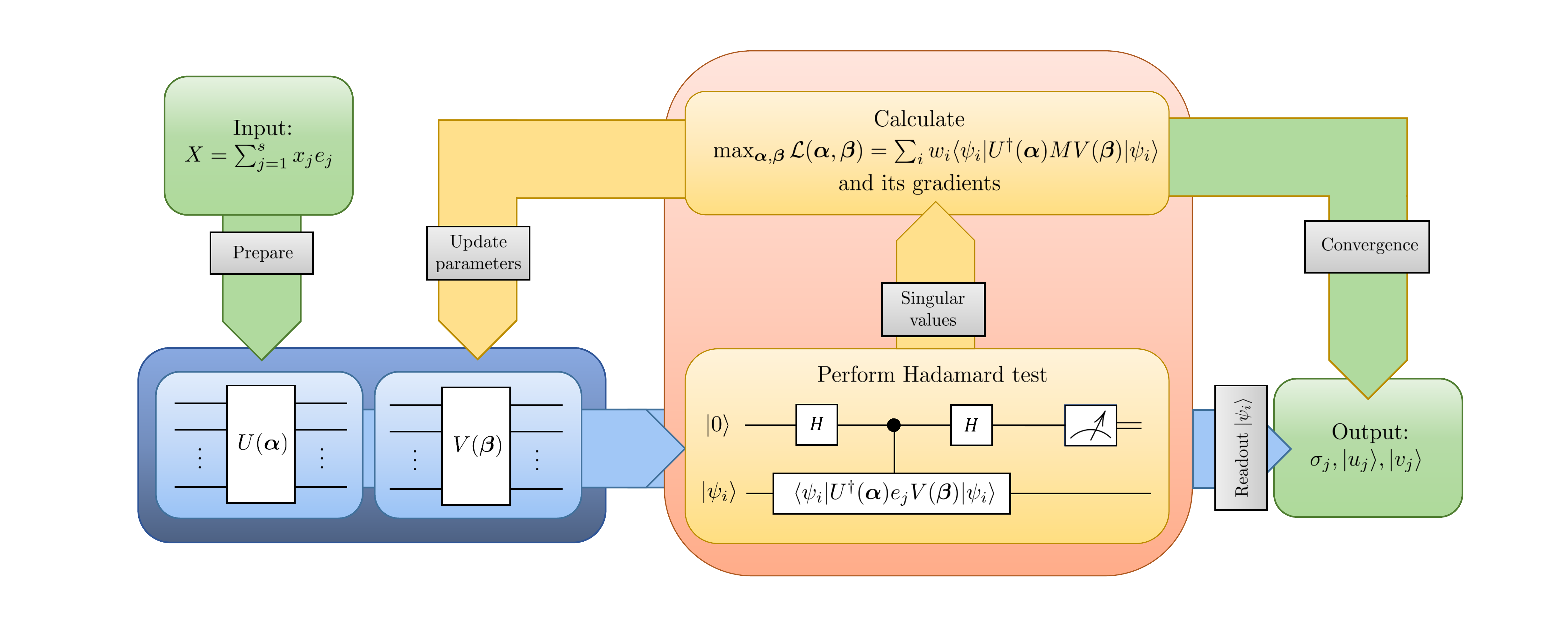}
\caption{Flowchart representation of the hybrid VQSVD algorithm. As an input, we choose the trajectory matrix $X$, decomposed into a non-square basis provided in Sec. \ref{sec:hadamard}. Two parameterisable quantum circuits $U(\boldsymbol{\alpha})$ and $V(\boldsymbol{\beta})$ are initiated, and subsequently trained to learn the set of left and right singular vectors (\textbf{\textcolor{darkblue}{blue}} block). To achieve this, a hybrid loop is initiated (\textbf{\textcolor{darkorange}{orange}} block): on a quantum processing unit, Hadamard tests approximate the singular values, while a classical loss function optimisation routine updates the variational parameters. Once the optimisation procedure has hit a threshold convergence condition, all relevant singular vectors are read out by sending in computational basis states through $U(\boldsymbol{\alpha})$ and $V(\boldsymbol{\beta})$ to extract all columns, and all estimated singular values $\sigma_j$ are given by the final set of Hadamard tests. In Sec. \ref{sec:sectionIV}, we employ a pulse-based approach to approximate the left and right singular matrices (\textbf{\textcolor{darkblue}{blue}} block).}
\label{fig:circuit}
}
\end{figure}

\twocolumngrid

\begin{widetext}
\begin{equation}
X=\left[
\begin{array}{ccccccc}
x_1 & x_2 & \cdots & x_N & x_1 & \cdots & x_{2^{qn}-N} \\
x_2 & x_3 & \cdots & x_{1} & x_2 & \cdots & x_{2^{qn}-N+1} \\
\vdots & \smash{\vdots} & \smash{\ddots} & \smash{\vdots} & \smash{\vdots} & \smash{\ddots} & \smash{\vdots} \\
x_{2^{q_m}} & x_{2^{q_m}+1} & \cdots & x_{N+2^{q_m}-1} & x_{N+2^{q_m}} & \cdots & x_{2^{q_n}+2^{q_m}-N-1}\\
\end{array}
\right].
\end{equation}
\end{widetext}

Again, it is to be understood that every index is mod $N$. While not obeying optimal SSD dimensions, this definition makes the most use of the space constrained by the dimensionality restrictions. Additionally, this definition reduces the number of unitaries $U_i$ required for the LCU in eq. (\ref{eq:lcu}), compared to embedding the trajectory matrix in a large null matrix.

\subsection{Randomised QSSD}

To obtain a more economic SVD routine, we approximate the SVD of the $m\times n$ trajectory matrix $X$ in a low-rank fashion, such that we maintain only the first $k$ singular values:
\begin{equation}
    X=U_k\Sigma_k V_k^\dagger,
\end{equation}
where $\Sigma_k$ is the singular value matrix, with $\sigma_p=0$ for all $p>k$. The computational complexity yields $\mathcal{O}(mnk)$, which is a 'lower' complexity than the conventional $\mathcal{O}(mn\text{min}(m,n))$ for $k\ll n$. By performing this low-rank approximation, we hope to improve the scaling laws of the QSSD algorithm and mitigate the two major problems which persist within the theory laid out by Ref. \cite{vqsvd}: the \textit{input problem} and the \textit{output problem}. The former refers to the fact that the decomposition of a matrix in a Pauli basis requires us to have a basis set that grows exponentially in the number of qubits. The latter refers to the read-out of the quantum circuits to obtain the matrices $U$ and $V$, which is also an operation whose computation time is exponential in the number of qubits.\\

First, we aim to find an optimal basis $\boldsymbol{Q}$ with as few columns as possible, such that \cite{rsvd}
\begin{equation}
    ||X-\boldsymbol{Q}\boldsymbol{Q}^\top X||\to\min_{\text{rank}(x)=k}||X-x||.
\end{equation}
This optimal basis can be approximated through a randomised matrix decomposition. Multiplication of the trajectory matrix by a random matrix
\begin{equation}
    \Omega=\begin{pmatrix}
    \vdots & \vdots &  & \vdots & \vdots\\
    \omega_1 & \omega_2 & \cdots & \omega_{k-1} & \omega_k\\
    \vdots & \vdots &  & \vdots & \vdots
    \end{pmatrix},
\end{equation}
where the vectors $\omega_i$ are drawn from the standard distribution $\mathcal{N}(\mu=0,\sigma=1)$, yields $Y = X^\top\Omega$. This allows for a random sampling of $\text{range}(X^\top)$. Performing a QR-decomposition on $Y$, and trimming the $Q$-matrix to a size $Q_{\text{max}(m,n)\times k}$, gives the optimal basis $\boldsymbol{Q}$. This matrix encompasses a basis transformation that lifts $X$ to a lower-dimensional space where the SVD is approximated up to the $k$-th value. Constructing this transformation gives $B = \boldsymbol{Q}^\top X$, whose SVD yields
\begin{equation}
    B = U_k \Sigma_k V_k^\top.
\end{equation}

We can establish a natural qubit cut-off number that defines a low-dimensional space onto which our trajectory matrix is projected. Because we focus on narrow frequency bands in the PSD, we can predict that we do not need plenty of singular values to reconstruct a signal. Every dominant frequency component yields 2 singular values, and simulations have shown that to obtain a faithful reconstruction of the corresponding signal, we need at most a number of singular values close to $\sim 10-12$. Therefore, we establish a dimensional cut-off $k \leq k_\Lambda = 16$ with associated qubit cut-off $q = 4$. Henceforth, we shall denote the reduced trajectory matrix as $X$.\\

\subsection{\label{sec:Circuit}The parameterisable quantum circuits}
We prepare two quantum circuits that learn to approximate the matrices of left ($U(\boldsymbol{\alpha})$) and right ($V(\boldsymbol{\beta})$) singular vectors of the randomised trajectory matrix $X$ simultaneously. Let $N$ and $\mathcal{U}(\boldsymbol{\theta})$ represent the qubit number and the unitary circuit representation of one circuit, agnostic to which one we refer to. For both circuits, we initialise the qubits in the vacuum state: $\ket{\psi_0}=\ket{\text{vac}}\myeq \ket{0}^{\otimes N}$, and employ the hardware efficient ansatz:
\begin{equation}
    \mathcal{U}(\boldsymbol{\theta})=\prod_{d=0}^{D-1} \left(\mathcal{U}_{\text{ent}}\times\bigotimes_{n=1}^N R_Y(\theta_{n+Nd})\right),
\end{equation}
where $D$ represents the depth of the circuit, $\mathcal{U}_{\text{ent}}\in\text{SU}(2^N)$ is a general entanglement operator over at most $N$ qubits, and $R_Y(\theta)$ is a rotational operator generated by the $\sigma^y$ Pauli operator:
\begin{equation}
    R_Y(\theta)=\exp{\left(-i\frac{\theta}{2}\sigma^y\right)}.
\end{equation}
We opt for this ansatz rather than a set of Euler rotations around the Bloch sphere, since this approach generates real amplitudes which are sufficient for the construction of orthonormal matrices, while simultaneously requiring a lower-dimensional parameter space. We choose a CNOT chain as our entanglement operation:

\begin{equation}
    \mathcal{U}_{\text{ent}} = \prod_{i=1}^{N-1}\text{CNOT}_{i,i+1}.
\end{equation}
\vspace{0.01cm}

\subsection{\label{sec:LossFunction}The loss function}

Knowing the columns of $U$ and $V$, the maximum singular value is retrieved from
\begin{equation}
    \sigma_1=\max_{\ket{u}\in\mathcal{S}_m,\ket{v}\in\mathcal{S}_n}\bra{u}X\ket{v},
\end{equation}
where $\mathcal{S}_i,\; i\in\{m,n\}$ is the set of normalised vectors of length $m$ and $n$ respectively. Other singular values are retrieved through a similar fashion, where the vectors $\ket{u}$ and $\ket{v}$ are restricted to be orthogonal to previous ones. This yields
\begin{equation}
    \sigma_k=\max_{\ket{u}\in\mathcal{S}_m,\ket{v}\in\mathcal{S}_n}\bra{u}X\ket{v}, \quad u\perp\mathfrak{o}^{(u)}_{k-1}, \quad v\perp\mathfrak{o}^{(v)}_{k-1},
\end{equation}
where 
\begin{equation}
    \mathfrak{o}^{(u)}_{k-1}=\text{span}\left(\ket{u_1},\cdots,\ket{u_{k-1}}\right).
\end{equation}
Through the Ky Fan theorem \cite{KyFan}, we arrive at a suitable loss function for VQSVD, given by a linear combination of the measured singular values:
\begin{equation}\label{eq:loss}
    \mathcal{L}(\boldsymbol{\alpha},\boldsymbol{\beta})=\sum_{j=1}^T w_j\cdot\mathfrak{Re}\bra{\psi_j^{(U)}}U^\dagger(\boldsymbol{\alpha})XV(\boldsymbol{\beta})\ket{\psi_j^{(V)}},
\end{equation}
where $w_1>w_2>\cdots>w_T>0$ form a set of ordered real-valued weights and $\{\ket{\psi_j^{(U)}}\}_{j=1}^T$ and $\{\ket{\psi_j^{(V)}}\}_{j=1}^T$ are sets of orthonormal vectors that select the right columns and vectors from $U$ and $V$, respectively. Here, Wang et al.'s linear parameterisation $w_j=T+1-j$ is adopted \cite{vqsvd}. Naturally, optimisation aims to find optimal $\boldsymbol{\alpha}^\star,\boldsymbol{\beta}^\star$ such that
\begin{equation}
    \boldsymbol{\alpha}^\star,\boldsymbol{\beta}^\star=\argmin_{\boldsymbol{\alpha},\boldsymbol{\beta}} \mathcal{L}(\boldsymbol{\alpha},\boldsymbol{\beta}).
\end{equation}
The terms $\mathfrak{Re}\bigg\{\bra{\psi_j^{(U)}}U^\dagger(\boldsymbol{\alpha})XV(\boldsymbol{\beta})\ket{\psi_j^{(V)}}\bigg\}$ can be measured on a quantum computer through a Hadamard test, if the matrix $X$ is decomposed into a basis of unitary operators $\{e_i\}_i$ such that $X=\sum_i x_i e_i$.

\subsection{Pseudo-unitary Hadamard Tests}\label{sec:hadamard}

Quantum computing usually deals with square matrices, since they are natural dimensions of operators in a Hilbert space. Decomposition of the trajectory matrix requires us to find a decomposition in a non-square basis, however. We can do so, if we relax the condition of unitarity of $U$ to
\begin{equation}
   U^+=U^\dagger\quad\text{where}\quad U^+ = (U^\dagger U)^{-1}U^\dagger,
\end{equation}

\onecolumngrid

\begin{figure}[H]
\captionsetup{justification=justified}
\captionsetup[subfigure]{labelformat=empty}
\centering
\subfloat[\textbf{(a)}]{\includegraphics[scale=0.5]{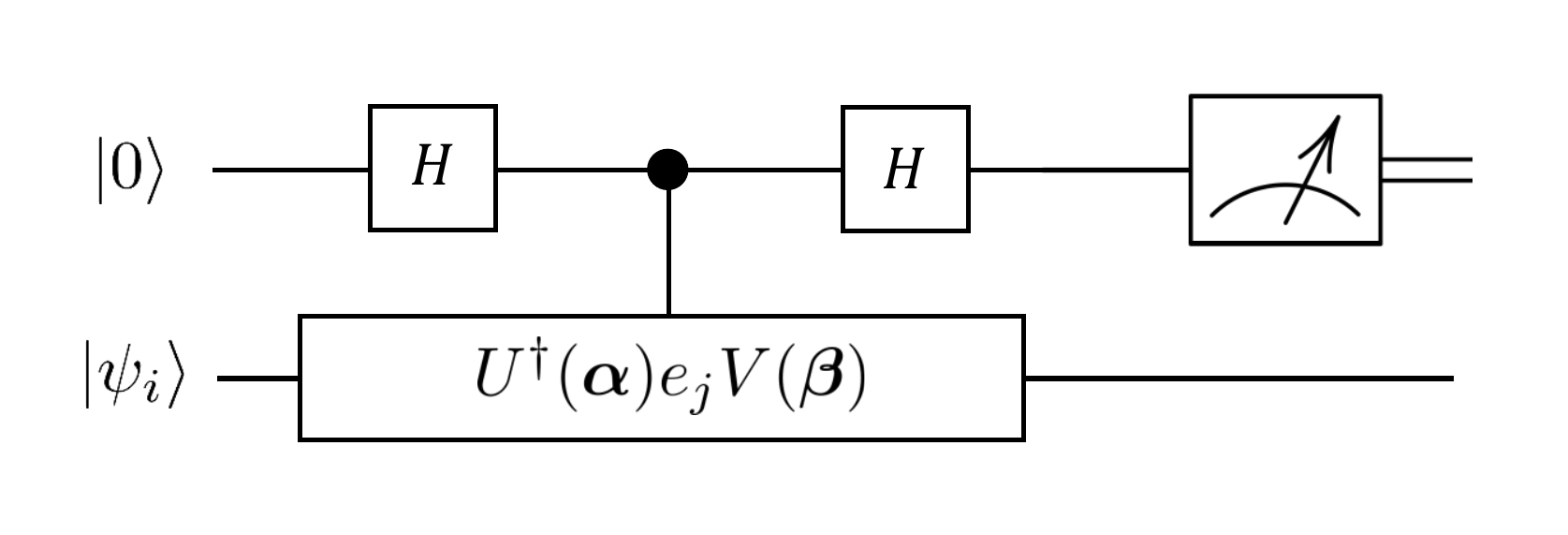}}
\subfloat[\textbf{(b)}]{\includegraphics[scale=0.45]{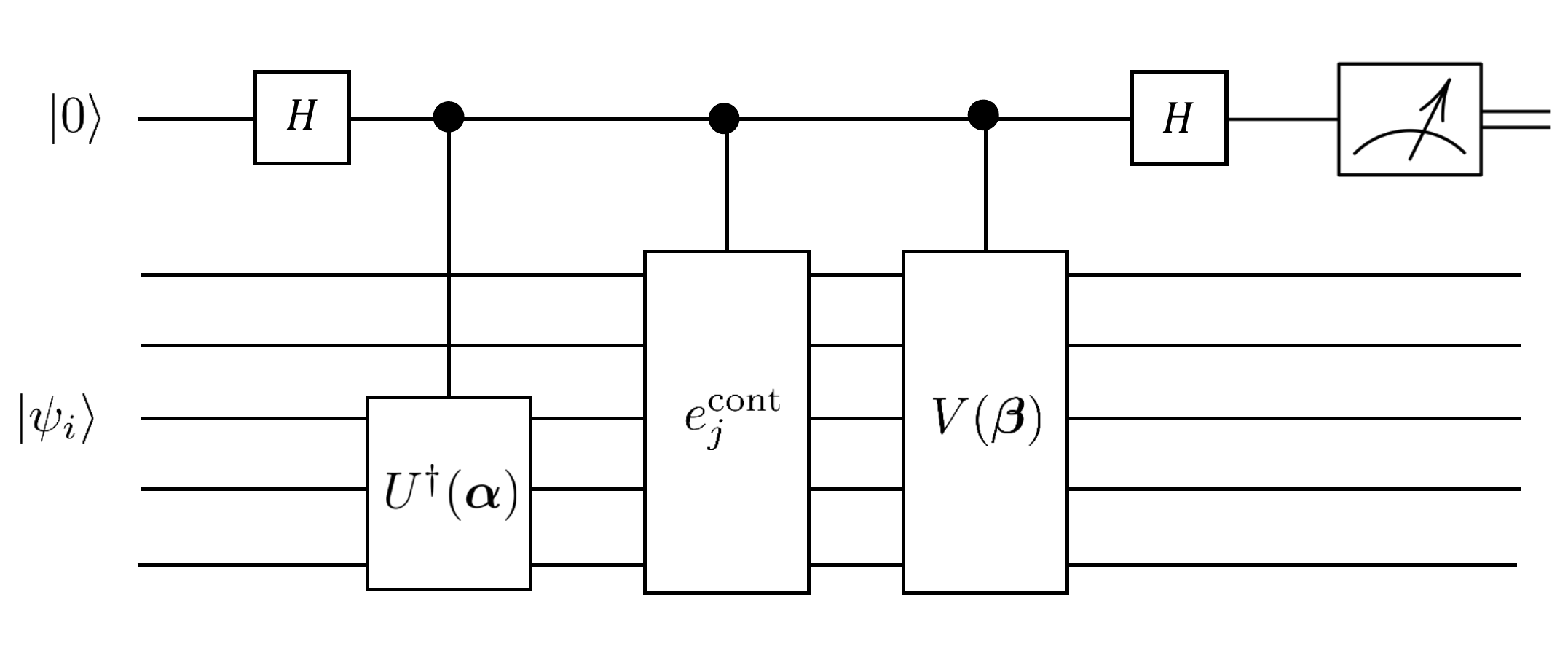}}
\caption{The circuit to perform a Hadamard test as given in Ref. \cite{vqsvd} (\textbf{(a)}) versus the Hadamard test circuit we employed for non-square matrix decompositions (\textbf{(b)}). $H$ represents the Hadamard gate, and the number of qubits in the bottom register equals $q_n$. $e_j^\text{cont}$ is the pseudo-unitary continuation of the basis element $e_j$, as given in Appendix \ref{app:pseudounitary}.}
\label{fig:pseudohadamard}
\end{figure}

\twocolumngrid

\noindent so that we find a non-square \textit{pseudo-unitary} basis decomposition. Because we adapted new optimal dimensions for the trajectory matrix to conform to quantum computing, we can construct a pseudo-Pauli basis:
\begin{equation}
    \{e_i\}=\{(P_i | O | \cdots | O),(O | P_i | \cdots | O),\cdots,(O | O | \cdots | P_i )\},
\end{equation}
where $P_i$ is the set of all possible Pauli operators of string length $q_m$, and $O$ is the $2^{q_m}\times 2^{q_m}$ null matrix. It is readily shown that every $e_i$ is pseudo-unitary and that together, they form a complete basis for any $2^{q_m}\times 2^{q_n}$ trajectory matrix, so that we can decompose $X$ according to
\begin{equation}\label{eq:pseudodecomp}
    X = \sum_i x_i e_i,
\end{equation}
where the expansion coefficients read
\begin{equation}
    x_i = \frac{1}{2^{q_m}}\text{Tr}(X^\dagger e_i).
\end{equation}
Implementation of these basis elements in a Hadamard test still requires square matrices. We employ \textit{pseudo-unitary continuation} to easily implement these elements by extending them to square matrices that have an easy tensor decomposition in a square Pauli basis, a protocol which is provided in Appendix \ref{app:pseudounitary}. Because the pseudo-Hadamard tests employ matrices $U$ and $V$ of different size, operating on different numbers of qubits, we must alter the controlled circuit subroutine of the Hadamard test as well. This is displayed in Fig. \ref{fig:pseudohadamard}. $U$ must be altered through
\begin{equation}
    \tilde{U}=\mathbb{I}_{q_n-q_m \text{qubits}}\otimes U,
\end{equation}
because it only acts on the last $q_m$ qubits.

\subsection{Analytical Gradient Optimiser}
For the classical optimisation routine, we employ an optimiser based on analytical gradients with adaptive Armijo conditions. Such optimisers circumvent finite difference inaccuracies by directly sampling the loss function at a point in parameter space through a parameter shift \cite{schuld}. The gradient of the loss function $\nabla\mathcal{L}$ is given by its entries:
\begin{equation}
    \frac{\partial\mathcal{L}}{\partial \alpha_\mu}=\frac{1}{2}\mathcal{L}\left(\boldsymbol{\alpha}+\pi\boldsymbol{e}_\mu,\boldsymbol{\beta}\right),
\end{equation}
\begin{equation}
    \frac{\partial\mathcal{L}}{\partial \beta_\nu}=\frac{1}{2}\mathcal{L}\left(\boldsymbol{\alpha},\boldsymbol{\beta}-\pi\boldsymbol{e}_\nu\right),
\end{equation}
where $\boldsymbol{e}_\mu,\boldsymbol{e}_\nu$ are unit vectors along the $\alpha_\mu/\beta_\nu$ direction in parameter space. A brief proof is provided in Appendix \ref{app:appgradient}. At every iteration $k$, we update the variational parameters along its gradient. The step size $\eta\in(0,1)$ is determined at every iteration and adapts to the loss function evaluation as well as past iterations. Generally, the parameter update reads
\begin{equation}
\theta^{(k+1)}=\theta^{(k)}+\eta\nabla\mathcal{L}(\theta^{(k)}).    
\end{equation}
At every iteration, we check for the first Wolfe condition, given by
\begin{equation}
    \mathcal{L}(\theta^{(k+1)})\geq \mathcal{L}(\theta^{(k)})+\mu\eta||\nabla\mathcal{L}(\theta^{(k)})||^2
\end{equation}
for some control parameter $\mu\in(0,1)$. If the condition is satisfied, we keep the step size equal to $\eta$, otherwise the step size is halved and the condition is checked again. Additionally, we introduce a tracker mechanism $\tau$ that can rescale the step size dependent on past steps. Initially set to 0, we update the tracker through $\tau \to \tau + 1$ if the Wolfe condition is met, otherwise, we set $\tau \to \tau - 1$. If the tracker hits a threshold $\pm\tau^\star$, it is reset to 0, and the step size is doubled or halved. Throughout our work, we adopt an initial step size of $\eta_0 = 10^{-6}$, a control parameter $\mu = 10^{-4}$ and a threshold of $\tau^\star = 3$.

\subsection{Orthonormal Matrix Reconstruction}
In order to reconstruct Hankel matrices from $\sum_i \sigma_i \ket{u_i} \bra{v_i}$, we must perform qubit state tomography over both circuits. Each column of the circuit matrix representation can be read out by sending the orthonormal set $\{\ket{\psi_i}\}_i$ through the circuits, and sampling all Pauli moments. Such approach is known to require an exponential number of samplings in the number of qubits. Matrix elements of the form (\ref{eq:umv}) are unprotected against global phase invariance, though. In contrast, the expectation value of an Hermitian operator $H$
\begin{equation}
    \bra{\psi}H\ket{\psi},
\end{equation}
are invariant under $\ket{\psi}\to e^{i\varphi}\ket{\psi}$ for arbitrary phases $\varphi$. The columns of $U$ and $V$ allow relative phase differences of $k\cdot\pi$ for $k\in\mathbb{Z}$, so that VQSVD finds local maxima where the singular values can be measured to be
\begin{equation}
    \sigma_i = \pm \sigma_i^{(\text{true})}.
\end{equation}
If the $i$-th singular value is measured to be negative, the sign needs to flipped, and the $i$-th column of $U$ gains an additional minus sign as well. 

\section{Pulse-based implementation for QSSD}\label{sec:sectionIV}
Besides the gate paradigm, variational quantum optimal control (VQOC) \cite{robert} offers an alternative approach to searching through a Hilbert space $\mathfrak{H}$ (see Ref. \cite{koch} for a review on QOC in general). Rather than optimising parameterised gates, we directly optimise the physical controls through which quantum operations are implemented. It is believed that for highly entangled systems, VQOC is able to outperform VQE in terms of accuracy, when resources are comparible. This is a result of the higher expressibility of optimal control, which one can prove by showing that a set of randomly initialised pulses produces unitary operations that resemble the Haar distribution more closely than a set of randomly initialised gates in the hardware efficient ansatz \cite{expressibilityhaar}.\\

In this paper, we consider application to a neutral atom quantum computer that is realised on a lattice of trapped Rydberg atoms, whose electronic state manifold offers an embedding of qubit states $\ket{0}$ and $\ket{1}$. Consider 2 linear arrays of respectively $q_m$ and $q_n$ atoms, which are treated as 2-level systems where the $\ket{0}$-state is mapped to a hyperfine electronic ground state, and the $\ket{1}$-state is mapped to a Rydberg state. Qubit states are addressed by monochromatic laser pulses, which we aim to optimise in order to find an optimal pulse that approximates the target state through propagators $U(t)\in\mathfrak{L}(\mathfrak{H}_{q_m})$ and $V(t)\in\mathfrak{L}(\mathfrak{H}_{q_n})$. For a total evolution time $T$ and $t\in(0,T)$, the circuit unitaries are subject to the propagator Schrödinger equation
\begin{equation}\label{eq:prop1}
    i\partial_t U(t) - H_U(t) U(t) = 0\quad \text{with} \quad U(0)=\mathbb{I},
\end{equation}
\begin{equation}\label{eq:prop2}
    i\partial_t V(t) - H_V(t) V(t) = 0\quad \text{with} \quad V(0)=\mathbb{I}.
\end{equation}
Here, $H_i$, $i\in\{U,V\}$ are the total Hamiltonians that drive the evolution of the qubit states. On a neutral atom quantum computing platform, the native qubit Hamiltonian reads
\begin{equation}
    H(t) = H_d + H_c[Z(t)],
\end{equation}
where $H_d$ is the drift Hamiltonian, which represents 'always-on' interactions in the system, and $H_c[Z(t)]$ is the control Hamiltonian, parameterised by the controls $Z(t)$ we impose on our system. For a linear array of $N$ qubits and $L\in\mathbb{N}$ controls, we find the van der Waals interaction
\begin{equation}
    H_d = \sum_{i<j}^N \frac{C_6}{R^6|i-j|^6}\ket{11}\bra{11}_{ij}
\end{equation}
and the control Hamiltonian
\begin{equation}
    H_c[Z(t)] = \sum_{l=1}^L \left(Q_l Z_l(t) + Q_l^\dagger \overline{Z_l(t)}\right).
\end{equation}
Here, $C_6$ characterises the Rydberg-Rydberg interaction strength, $R$ is the interatomic distance between neighbouring qubits, and $Q_l$ are control operators associated with our control parameters. In this work, we control the Rabi frequency and laser phase $|\Omega(t)|e^{i\varphi(t)}$, and the detuning $\Delta$ of the $\ket{1}$-state, so that their controls are given by
\begin{equation}
    Q_\Omega = \ket{0}\bra{1} + \text{h.c.},\quad Q_\Delta = \ket{1}\bra{1}.
\end{equation}
These controls parameterise both circuits by modeling pulses as a set of $T/\delta t \in \mathbb{N}$ time-discretised control pulses with duration $\delta t$. The physical implementation is a result from smoothing out the piecewise constant equidistant pulses at a minor loss of fidelity.
We aim to optimise a new loss function 
\begin{multline}
    \mathcal{L}_{\text{QOC}}=\mathfrak{Re}\bigg\{\text{Tr}\left[U^
    \dagger(T) M V(T) \rho \right]\bigg\} \\- \frac{\lambda}{2}\sum_{l=1}^L \int_0^T \left(|Z_{U,l}(t)|^2+|Z_{V,l}(t)|^2\right) dt,
\end{multline}
by adjusting (\ref{eq:loss}) to penalise strong controls for $\lambda > 0$. We enforce these constraints by adding time-dependent Lagrange multiplier terms to $\mathcal{L}_{\text{QOC}}$:
\begin{equation}
    J_3(U,Z_U,\eta_U)=\langle i\partial_t U(t) - H_U(t) U(t), \eta_U (t)\rangle_{\Omega_{q_m}},
\end{equation}
\begin{equation}
    J_4(V,Z_V,\eta_V)=\langle i\partial_t V(t) - H_V(t) V(t), \eta_V (t)\rangle_{\Omega_{q_n}},
\end{equation}
where the multipliers are called \textit{adjoints}, the inner product is given by
\begin{equation}
    \langle A, B \rangle_{\Omega_N} = \mathfrak{Re}\bigg\{\int_0^T \text{Tr}\left[A^\dagger B\right] dt\bigg\},
\end{equation}
and $\Omega_N = L^2([0,T],\mathfrak{H}_N)$. To ensure a realistic implementation, we restrict our controls to the space $\mathcal{Z}_{\text{ad}}$ of admissible pulses
\begin{equation}
    \mathcal{Z}_{\text{ad}} = \big\{Z(t) \in L^2([0,T],\mathbb{C}^L)\quad \big|\quad ||Z||_{\infty} \leq Z_{\text{max}} \big\}.
\end{equation}
In Appendix \ref{app:qoc}, we prove that it is possible to optimise this Lagrangian under these constraints by sampling the necessary quantities on a quantum computer, where we have omitted the proof of existence and uniqueness of the solutions. Our update scheme at iteration $k$ becomes
\begin{multline}\label{eq:grad}
    Z_{U,l}^{k+1}(t) = Z_{U,l}^k(t) + \eta\big(\lambda Z_{U,l}^k(t)\\-\text{Tr}\left[Q_l^\dagger\left(\eta_U(t)U^\dagger(t)+U(t)\eta_U^\dagger(t)\right)\right]\big)
\end{multline}
for the $U$-controls, and idem dito for $V$. For optimisation, we adopt the same Armijo optimiser protocol where we check for the first Wolfe condition at every step. In Appendix $\ref{app:qoc}$ we also explicitly solve for the adjoints. In conclusion, the VQOC paradigm requires to solve the propagator Schrödinger equations (\ref{eq:prop1}), (\ref{eq:prop2}), the adjoint equations (\ref{eq:adj1}) and (\ref{eq:adj2}), and the analytical gradient (\ref{eq:grad}).

\section{\label{sec:sectionV}Analyses and Results}

To benchmark the performance of QSSD, several simulations have been run, using a classical emulator of a noise-free quantum circuit. For efficiency purposes we assume full access to all relevant matrix elements, without resorting to sampling amplitudes through measurements. We compare gate-based and pulse-based results to the retrieved classical SSD components for a variety of applications, though we stress that because of the inherent inefficiency of the algorithm, we have not attempted to make a fair comparison between the two implementations. In order to draw a fair contrast, either implementation needs to be run on \textit{equivalent evolution circuits}, where both the gate-based and pulse-based have equivalent resources. This alludes to a similar circuit evolution time $T$, equivalent controls such as Bloch sphere rotational control only, and the number of quantum evaluations $\#$QE$_{\text{VQOC}}$ should scale linearly in the quantum resources compared to the number of quantum evaluations $\#$QE$_{\text{VQE}}$. \\

In this work, we adopted the hardware efficient ansatz for the gate-based circuits, with depth $D = 10$ and an iteration number it = 1000, and we assumed full control for the pulse-based circuits, in natural units where $\hbar = 1$ and $C_6 = 1$, for a total evolution time $T = 10$, a step size discretisation $T/\delta t = 100$ and an iteration number it = 150. We aim to show that with general resources, one can approximate the function space of the SVD reasonably well on a quantum computer, without employing equivalent evolution circuits. Note that the VQOC parameters are not fine-tuned for any implementation in particular. For a more realistic implementation on a neutral atom quantum computer, the parameters should lie in experimental bounds, and the space $\mathcal{Z}_{\text{ad}}$ should be adjusted accordingly.\\

We first apply QSSD to a simple toy example to showcase that properties of the SSD algorithm, such as pinpointing non-stationarity of a signal, carry over to the quantum algorithm. Then we apply QSSD to simulated cortical local field potentials (LFPs), electric potentials originating from the extracellular space in brain tissue, as well as GW150914 data, resulting from the first observation of a gravitational wave event. As a figure of merit we introduce the error measure
\begin{equation}\label{eq:error}
    \epsilon = \sqrt{\frac{1}{|\Lambda_\omega|}\int_{\Lambda_\omega}|p(\omega)-p_{\text{SSD}}(\omega)|^2d\omega},
\end{equation}
where $p(\omega)$ is the PSD of the simulated individual band components, and where $p_{\text{SSD}}(\omega)$ is the PSD of the SSD-recovered component. $\Lambda_\omega$ is the relevant frequency band of the PSD over which we integrate, and $|\Lambda_\omega|$ is its measure. This error focuses more closely on the frequency content of the retrieved components than an $l_2$-norm would, and is not cumulative over time due to small displacements. \\

\vspace{-0.5cm}
\subsection{Non-stationary signal}

\begin{figure}[H]
\captionsetup{justification=justified}
\captionsetup[subfigure]{labelformat=empty}
\centering{
\subfloat[\textbf{(a)}]{\includegraphics[scale=0.21]{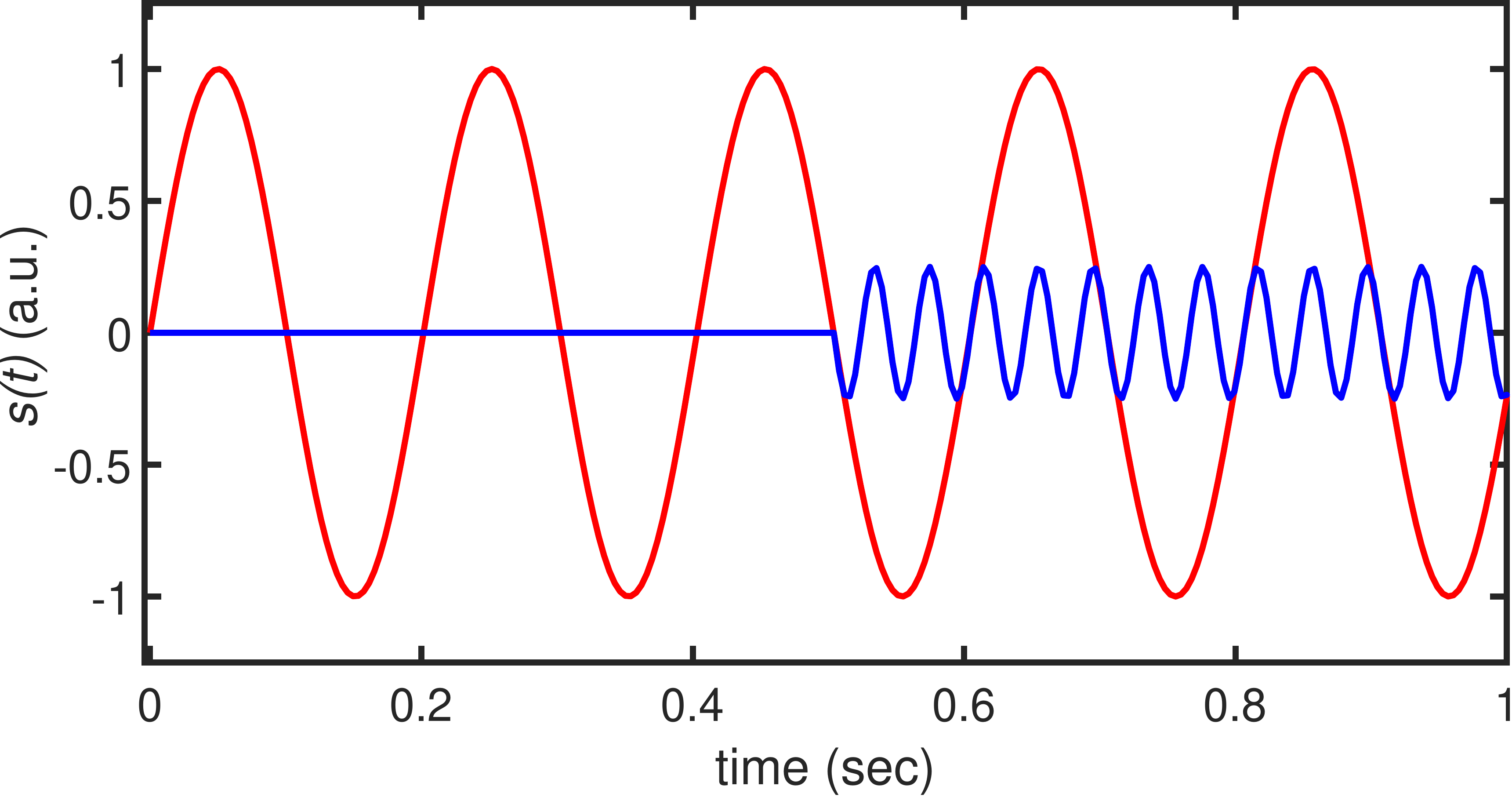}}\;\;\;
\subfloat[\textbf{(b)}]{\includegraphics[scale=0.21]{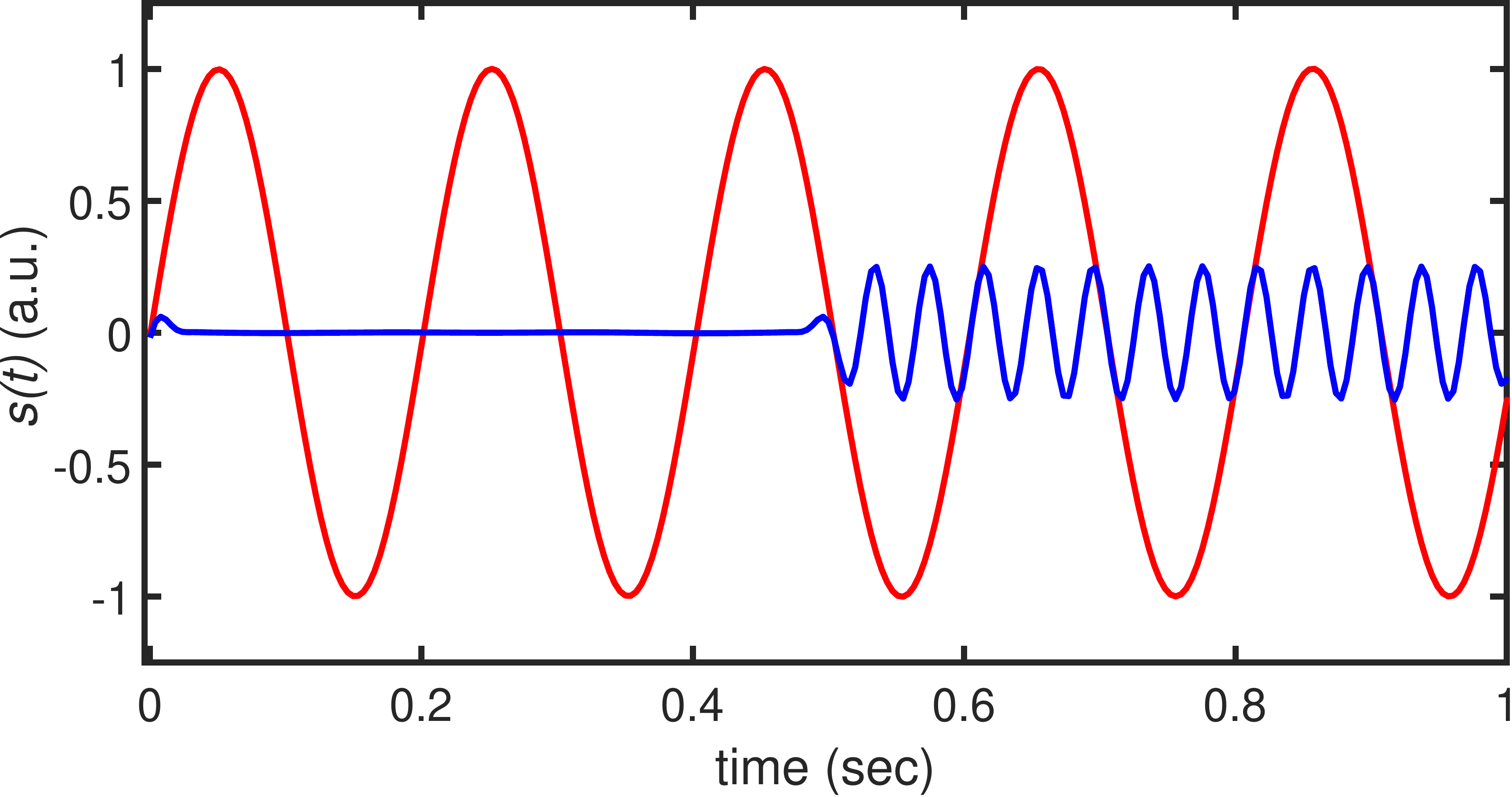}}

\subfloat[\textbf{(c)}]{\includegraphics[scale=0.21]{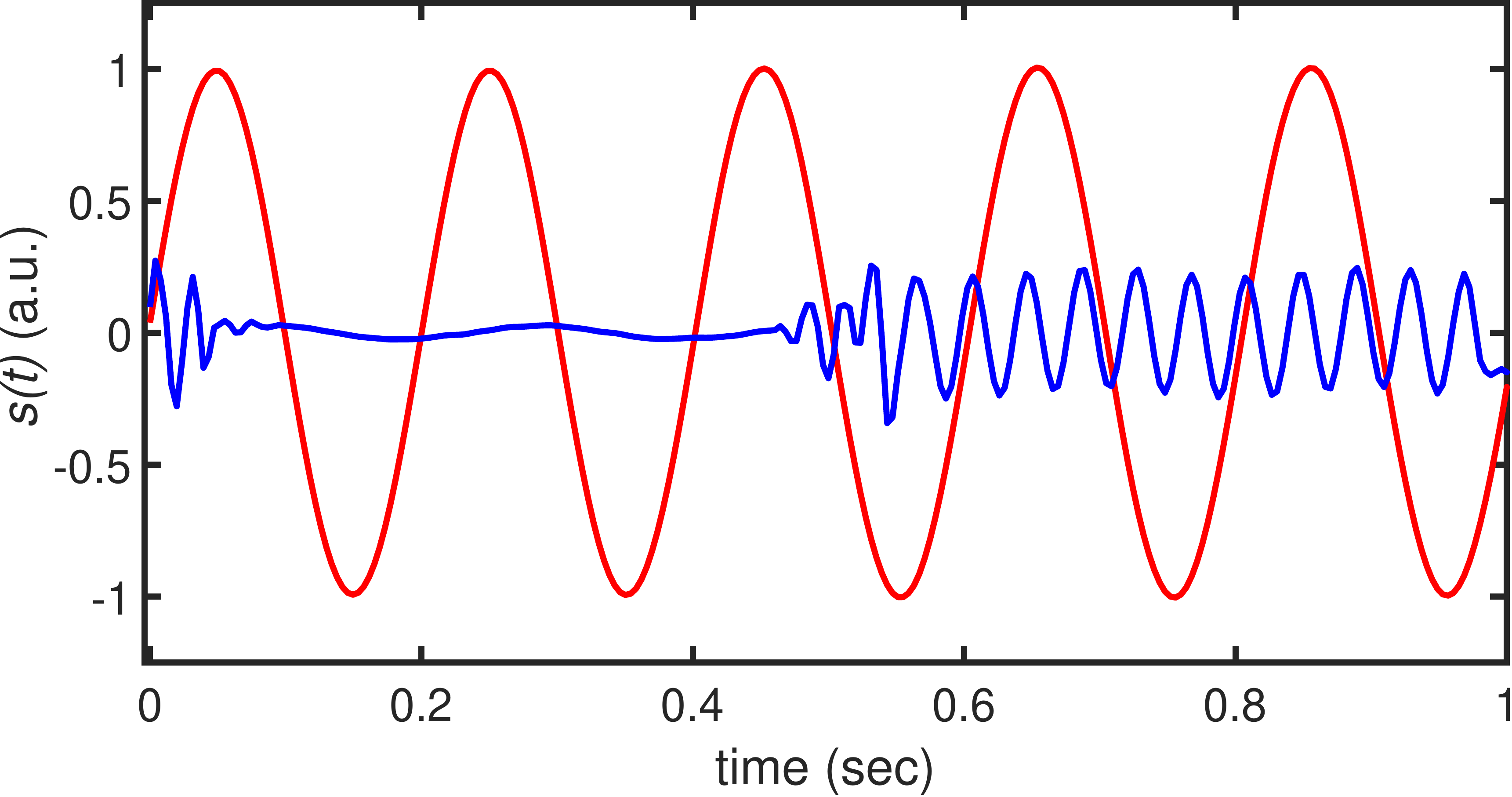}}\;\;\;
\subfloat[\textbf{(d)}]{\includegraphics[scale=0.21]{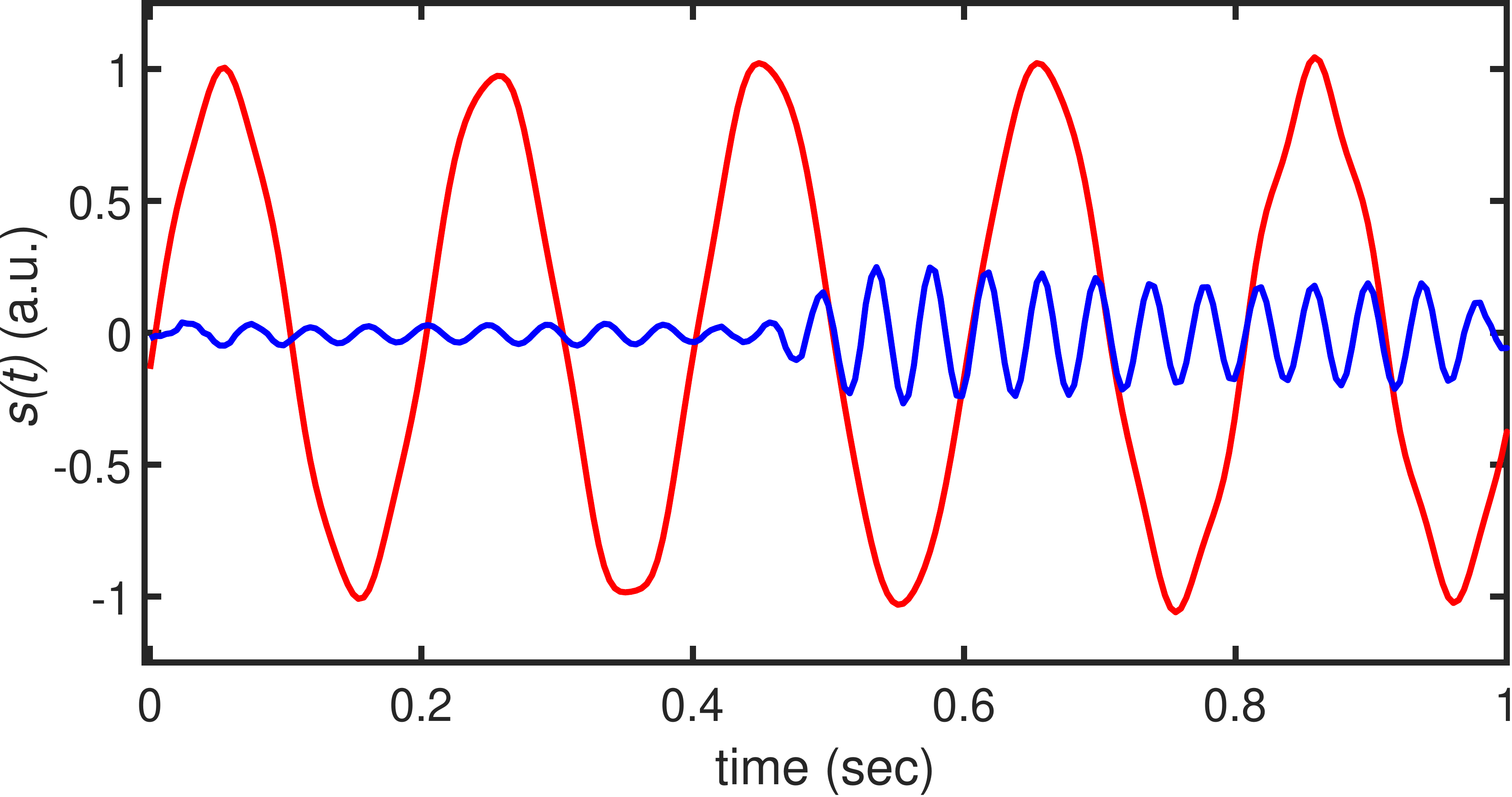}}
\caption{An example of a non-stationary function $s(t)$ comprised of two sinusoidal components, in arbitrary units (a.u.), for $t\in[0,1]$. The two original components are shown in \textbf{(a)}, and the SSD components are found using classical SSD \textbf{(b)}, gate-based QSSD \textbf{(c)} and pulse-based QSSD \textbf{(d)}. A sampling frequency $F_S=256$ Hz was chosen. The non-stationary character of the function is picked up upon by the algorithm in all variations of SSD.}
\label{fig:nonstationary}}
\end{figure}

As a first proof of principle, we applied SSD to the following non-stationary toy signal, composed of two harmonic sinusoidal functions:
\begin{equation}
    s(t) = \sin(10\pi t) + \frac{1}{4}\sin(50\pi t)\cdot\theta\left(t-\frac{1}{2}\right),
\end{equation}
where $\theta(\cdot)$ is the Heaviside function, for $t\in[0,1]$. The classical SSD and QSSD results are presented and compared, in Fig. \ref{fig:nonstationary}. The figure shows a clear separation of harmonic functions, while also pinpointing the moment when the non-stationary harmonic function begins. In Fig. \ref{fig:scatter}, the error $\epsilon$ is graphed for both components, for all SSD variations.

\subsection{Local Field Potential signal}

Next we applied SSD to simulated local field potentials (LFPs), which represent the relatively slow varying temporal components of the neural signal, picked up from within a few hundreds of microns of a recording electrode. Brain waves can be separated into separate characteristic frequency bands called alpha [8-12 Hz], beta [13-30 Hz], gamma [30-90 Hz] and theta [4-7 Hz] bands, corresponding to the frequency range in which their dominant frequency lies \cite{ssd}. We simulated these LFPs by superimposing four components, each in every aforementioned band \cite{macaque3}. Because the alpha and theta components have overlapping frequency contents, we grouped them together as

\onecolumngrid

\twocolumngrid

\begin{figure}[H]
\captionsetup{justification=justified}
\captionsetup[subfigure]{labelformat=empty}
\centering{
\subfloat[\textbf{(a)}]{\includegraphics[scale=0.25]{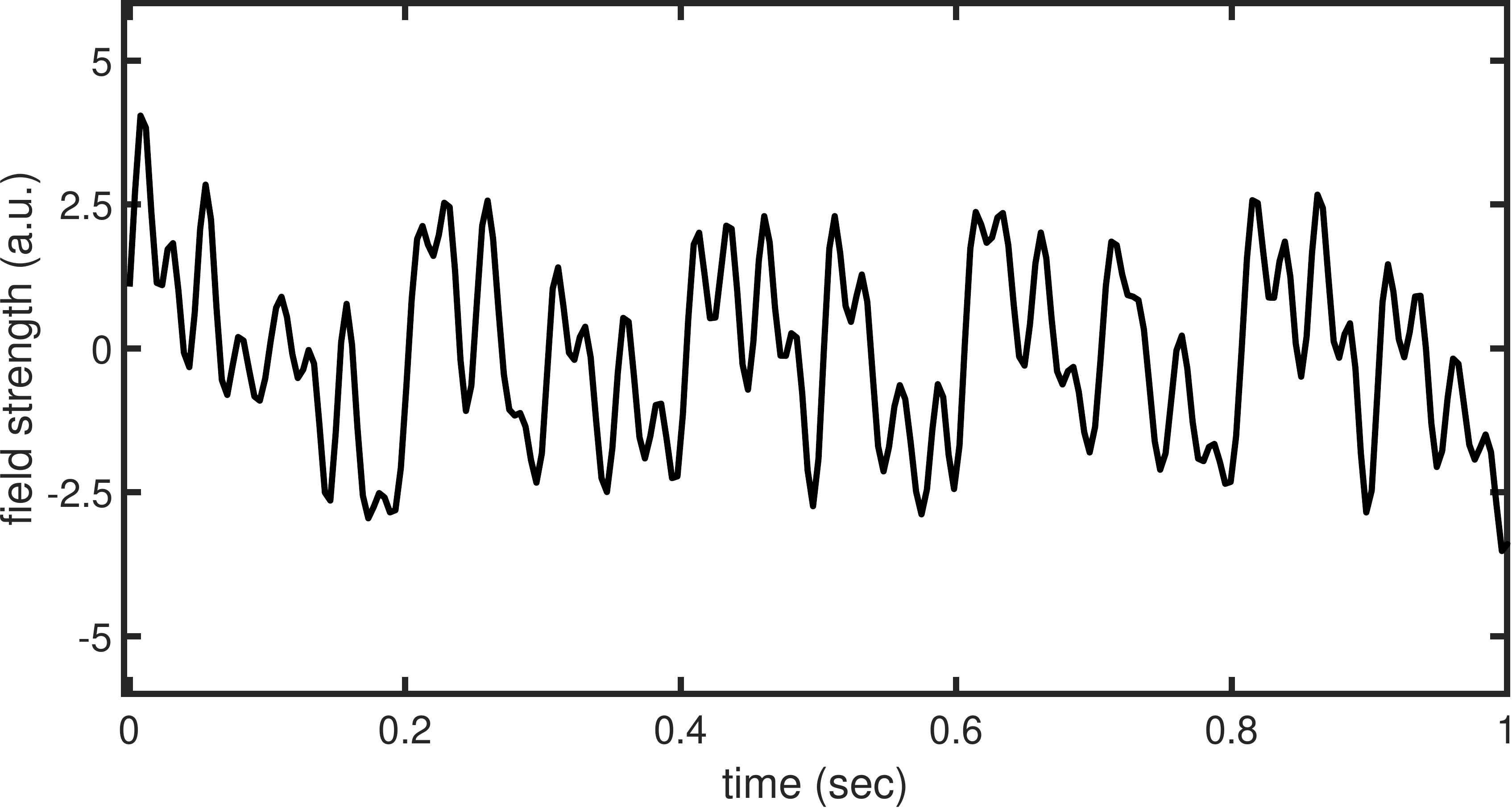}}

\vspace{0.5cm}

\textbf{(b)}
\subfloat[]{\includegraphics[scale=0.16]{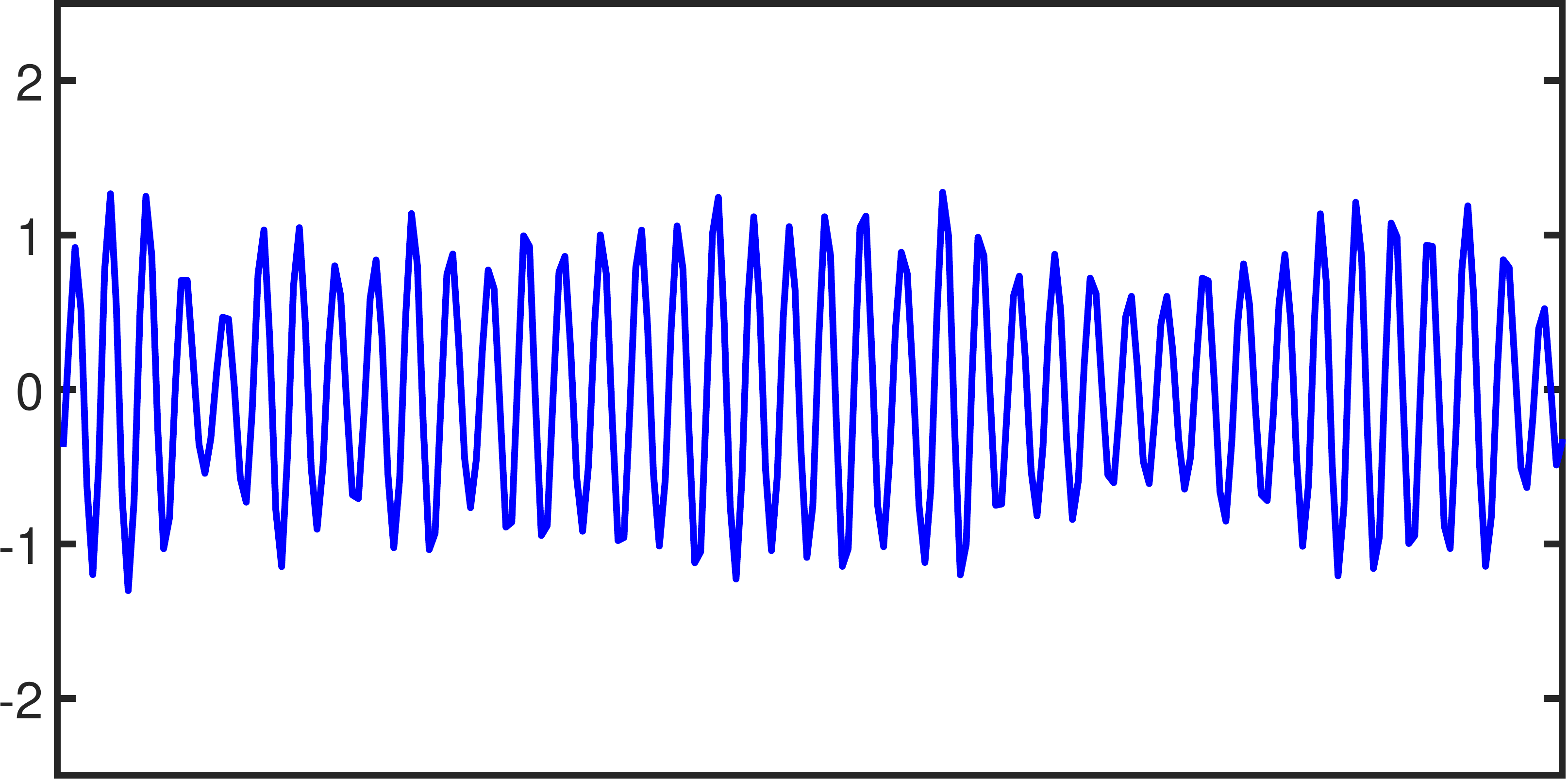}}\;
\subfloat[]{\includegraphics[scale=0.16]{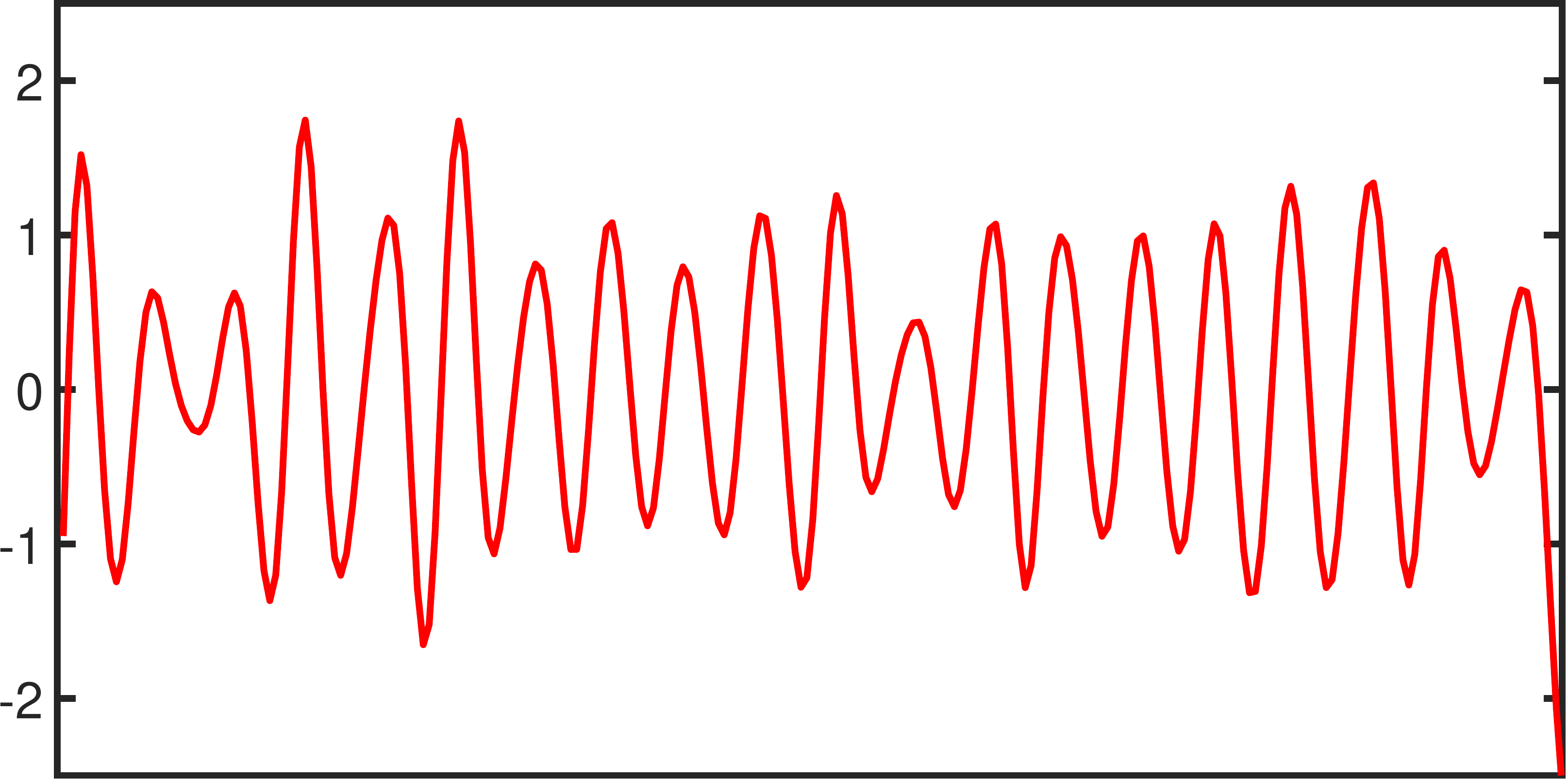}}\;
\subfloat[]{\includegraphics[scale=0.16]{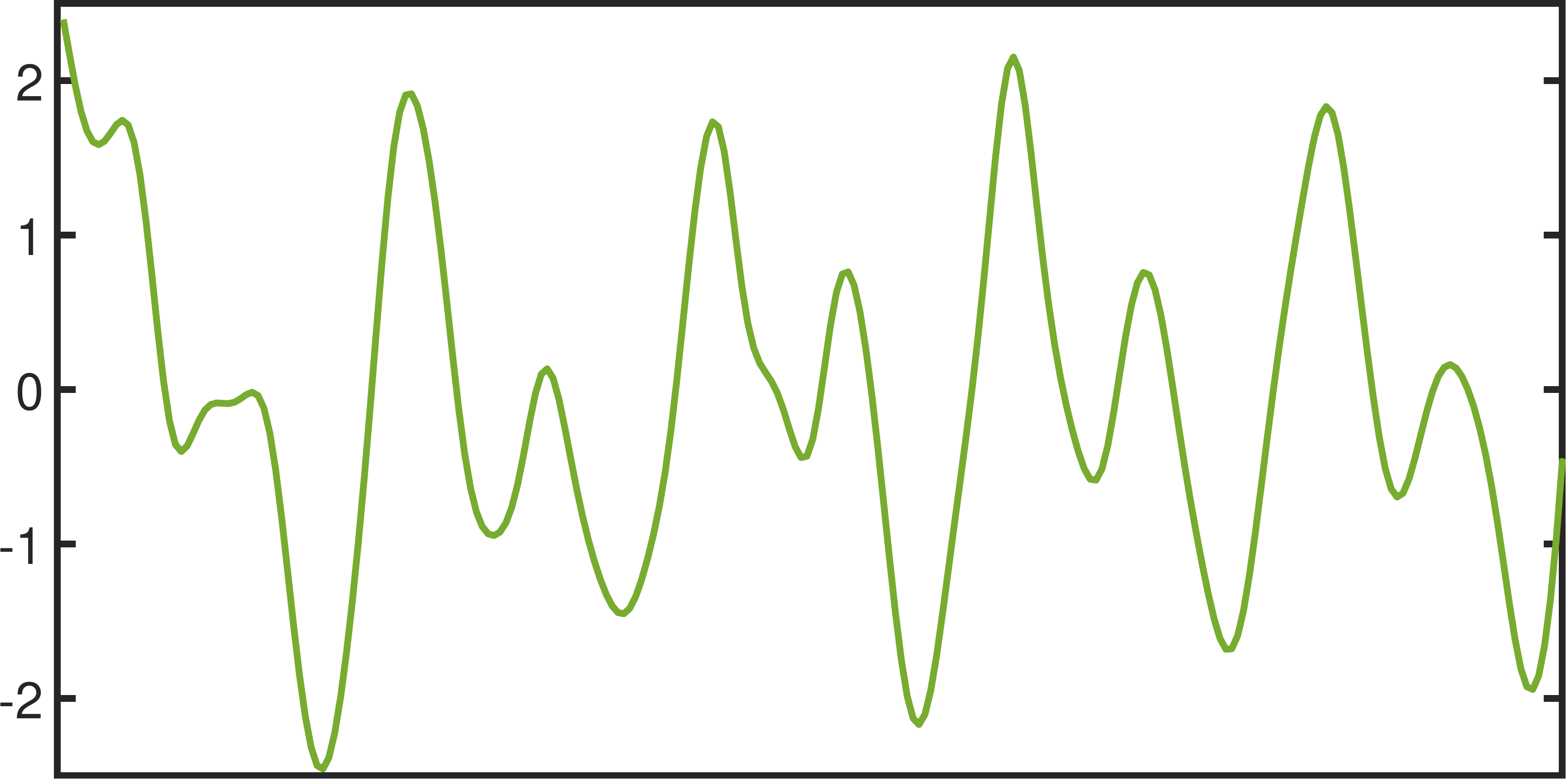}}

\textbf{(c)}
\subfloat[]{\includegraphics[scale=0.16]{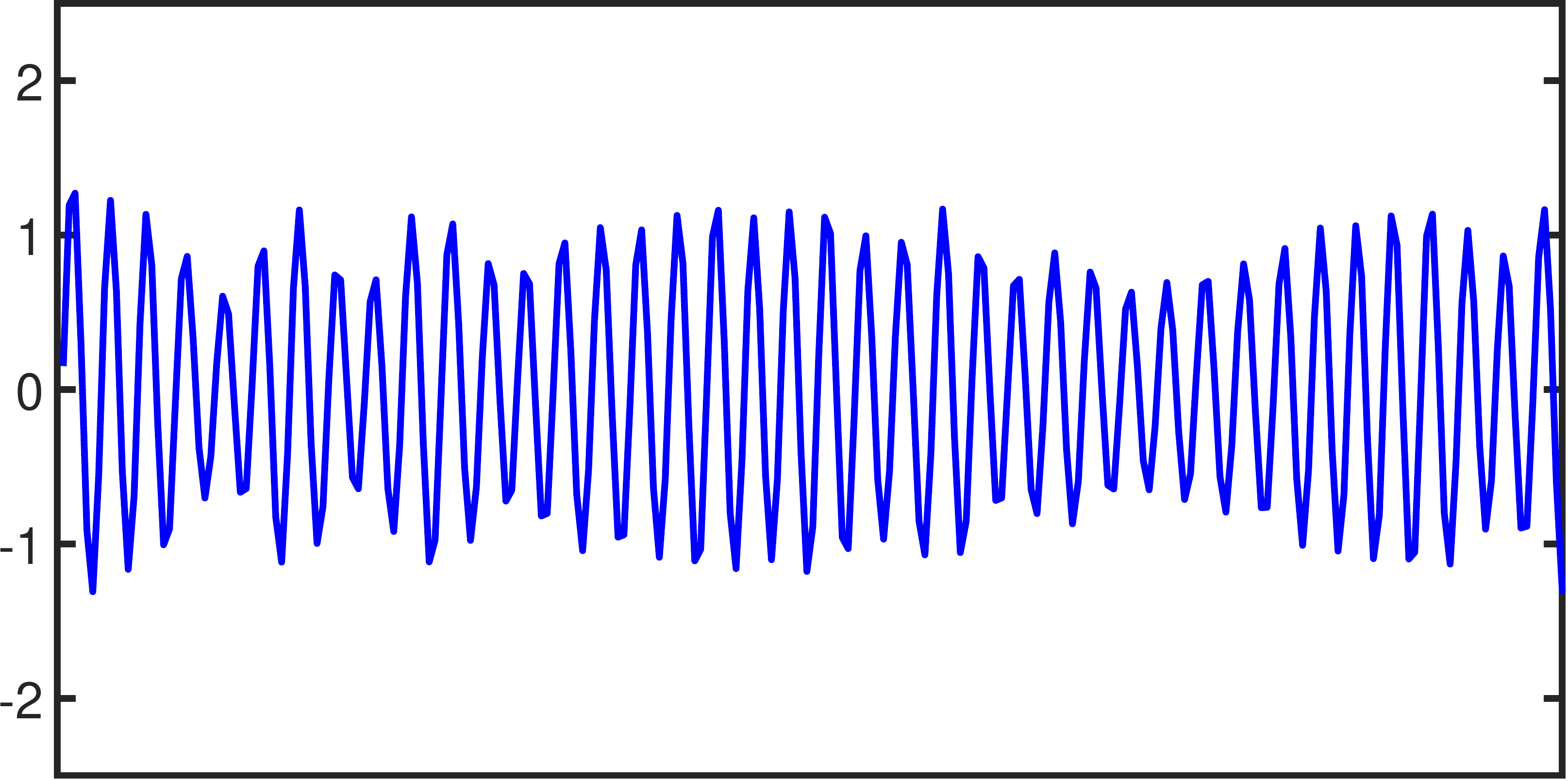}}\;
\subfloat[]{\includegraphics[scale=0.16]{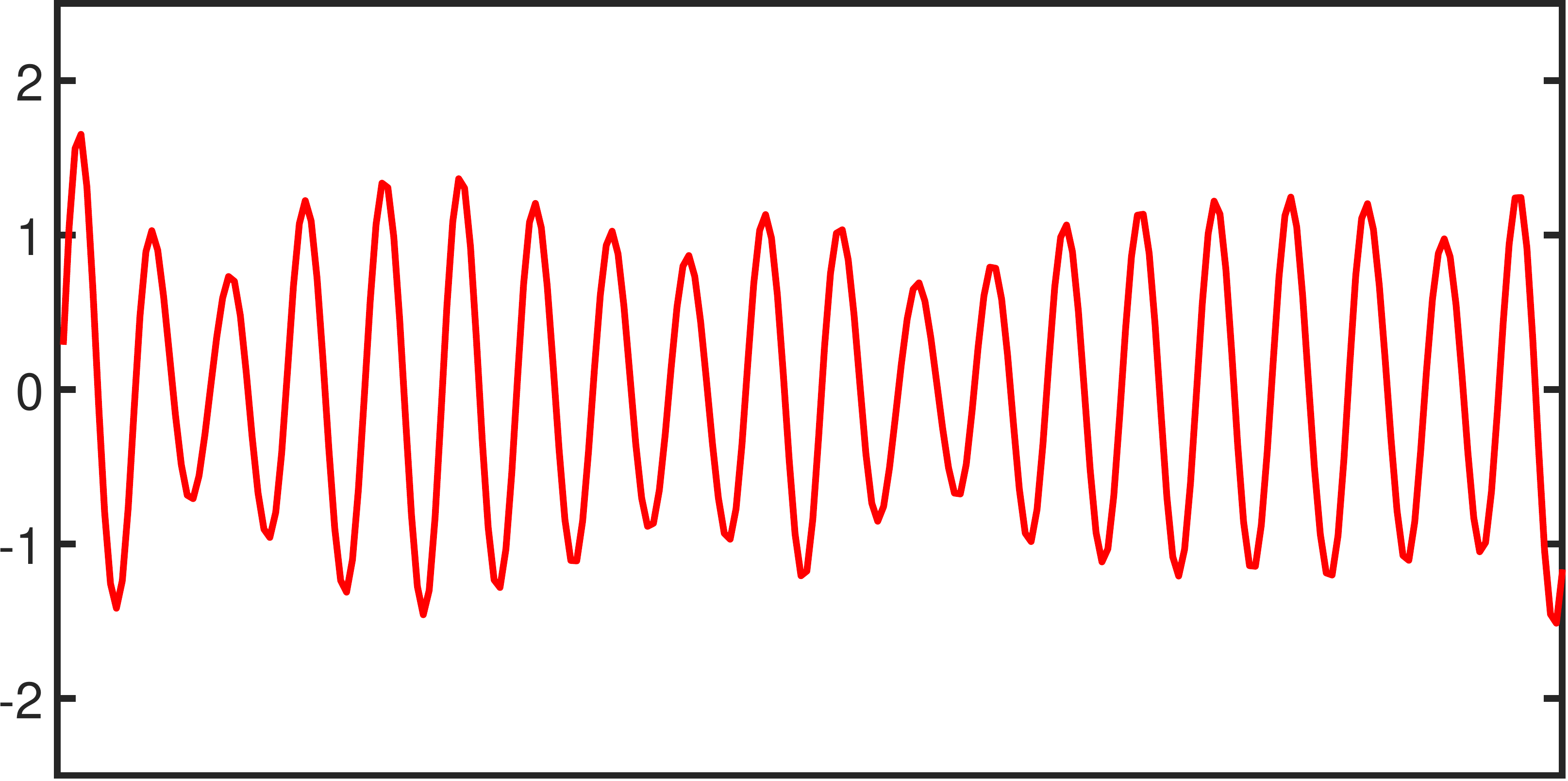}}\;
\subfloat[]{\includegraphics[scale=0.16]{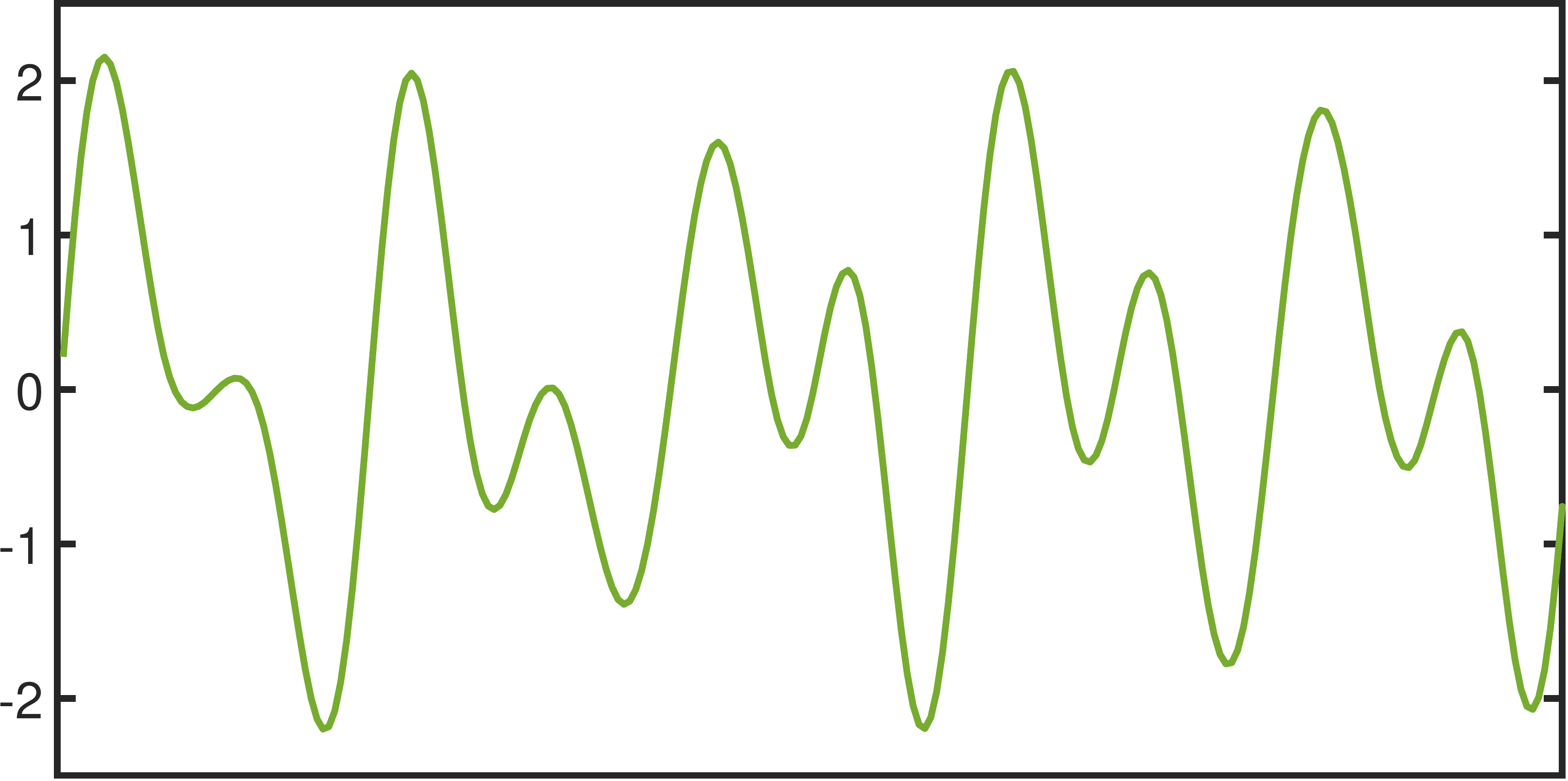}}

\textbf{(d)}
\subfloat[]{\includegraphics[scale=0.16]{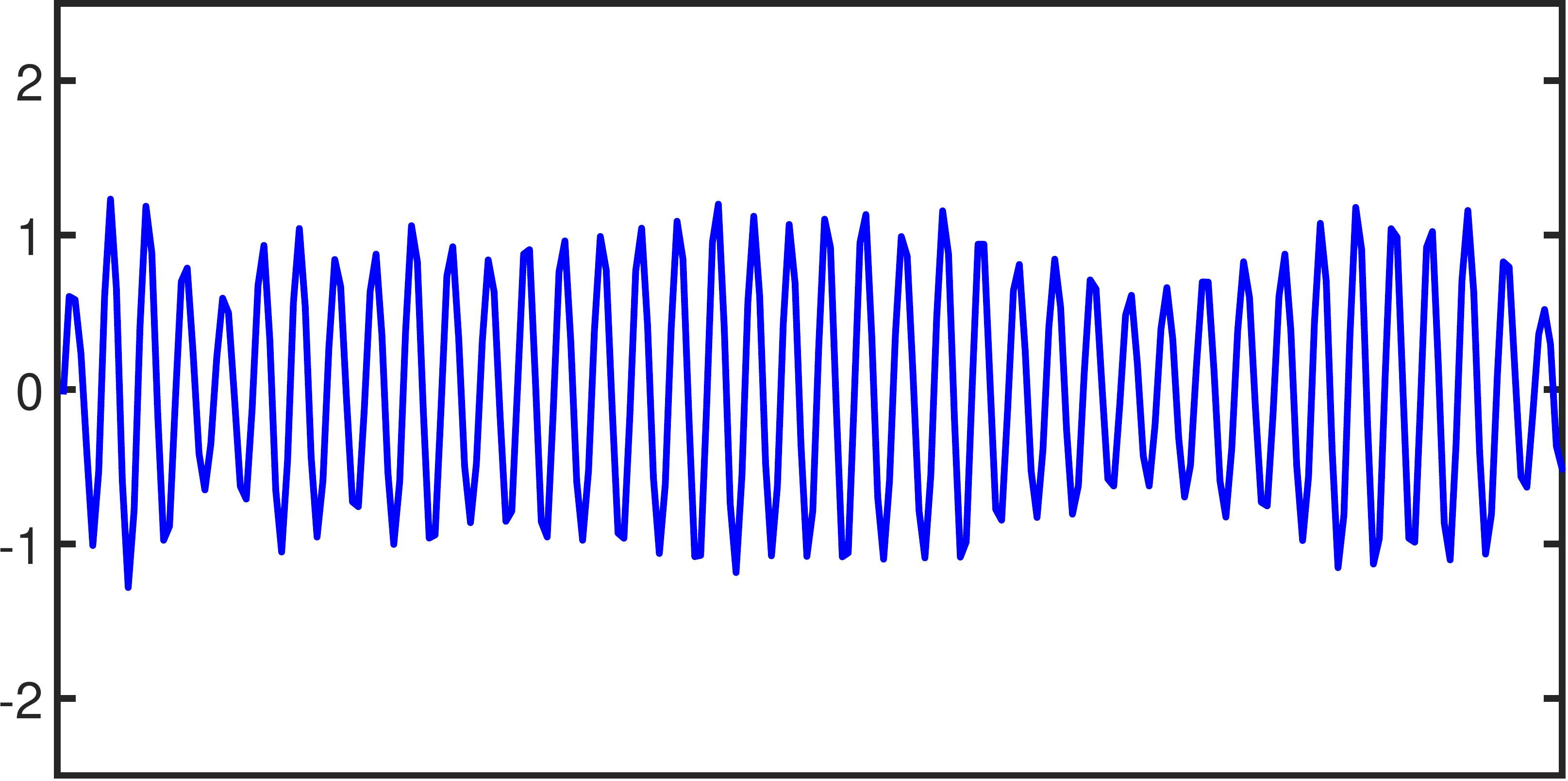}}\;
\subfloat[]{\includegraphics[scale=0.16]{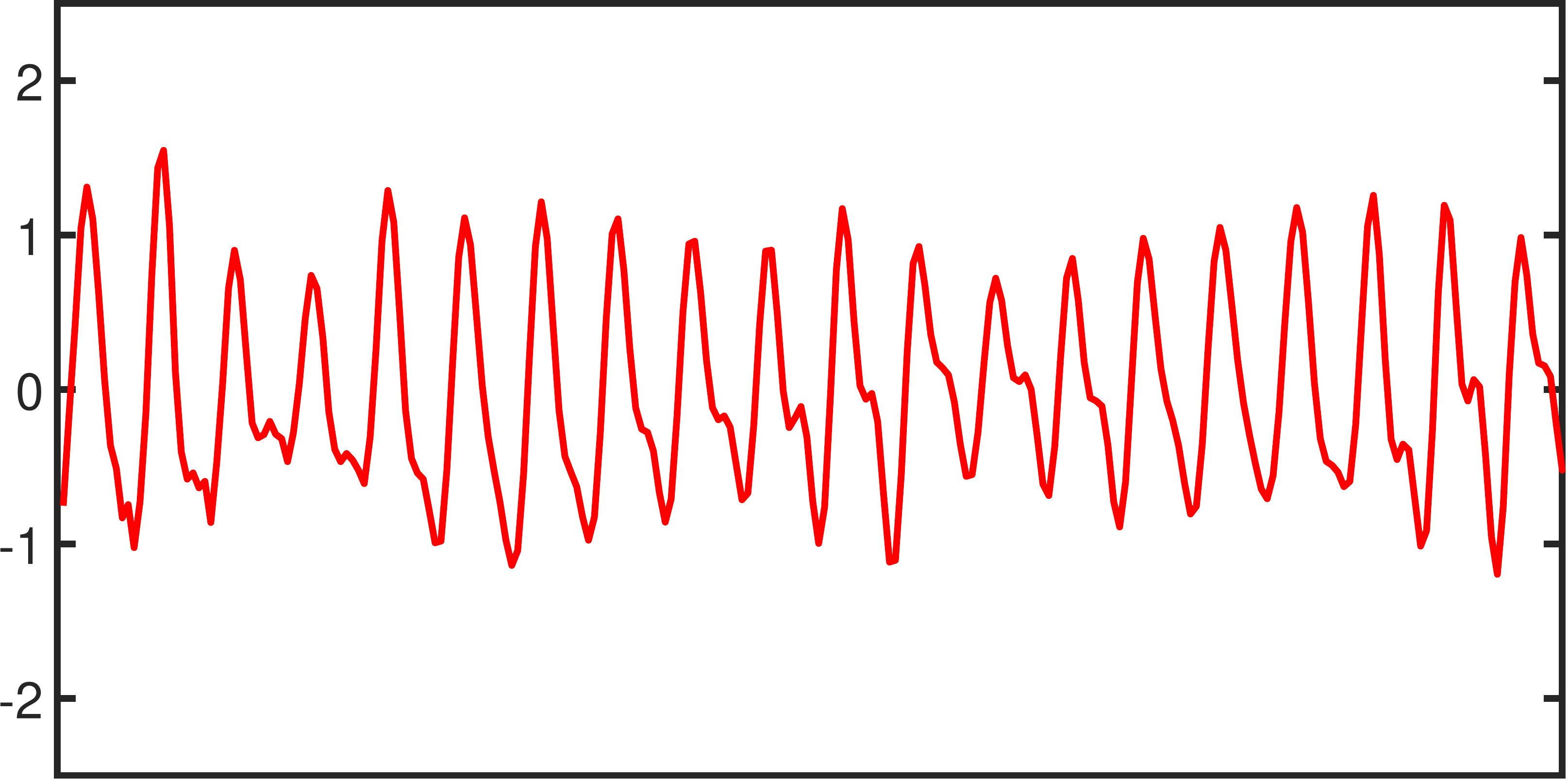}}\;
\subfloat[]{\includegraphics[scale=0.16]{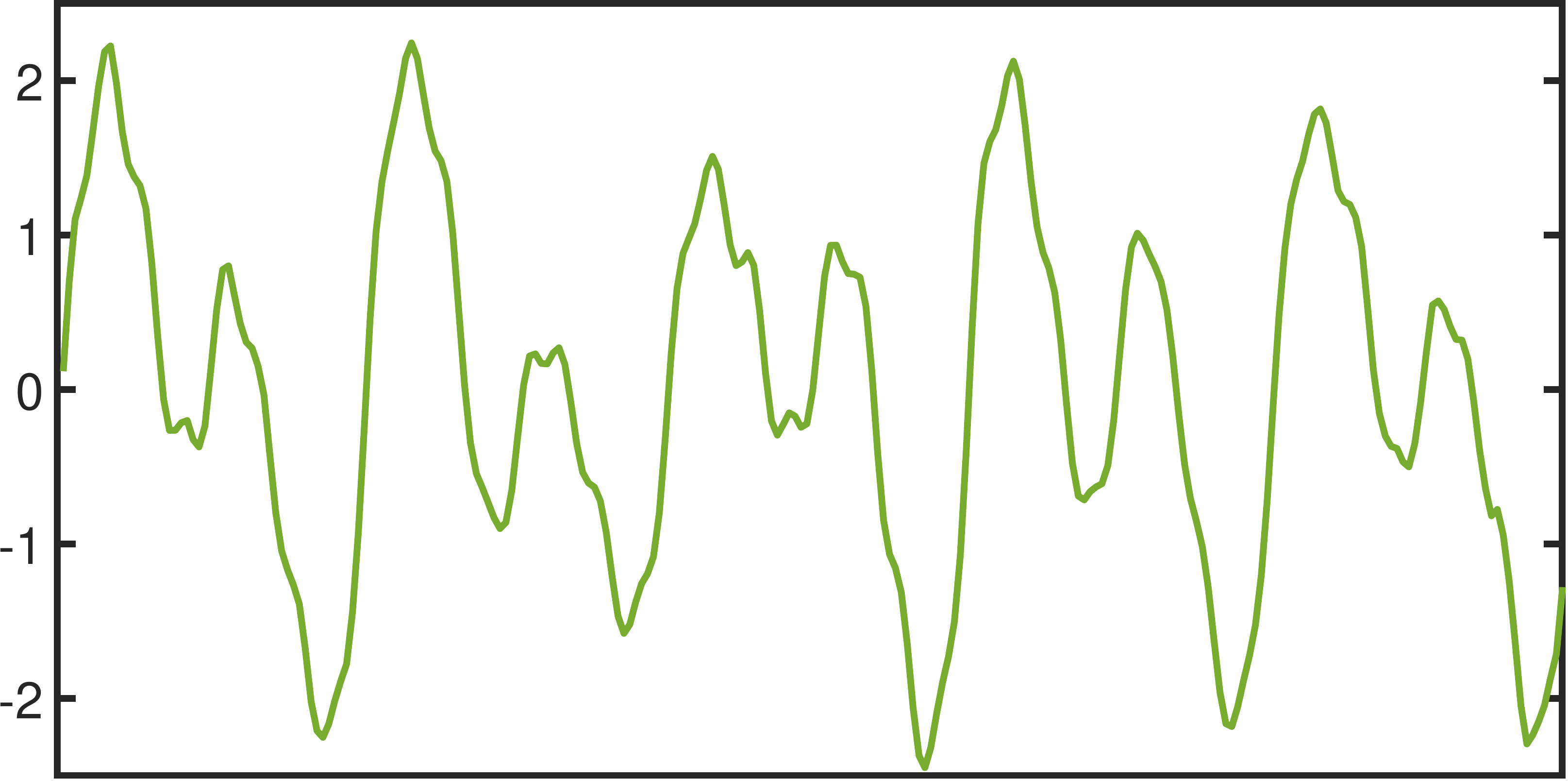}}

\textbf{(e)}
\subfloat[]{\includegraphics[scale=0.16]{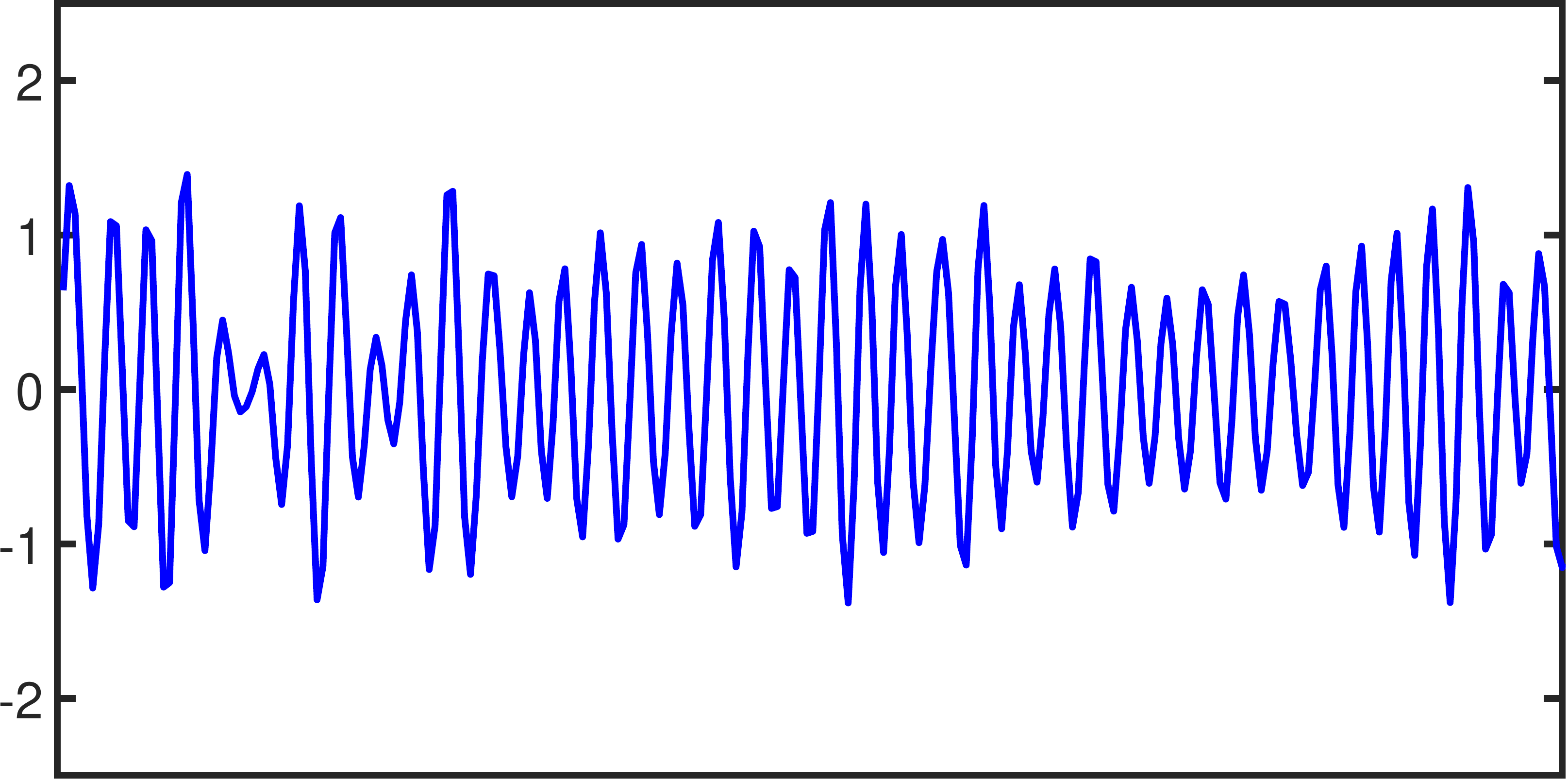}}\;
\subfloat[]{\includegraphics[scale=0.16]{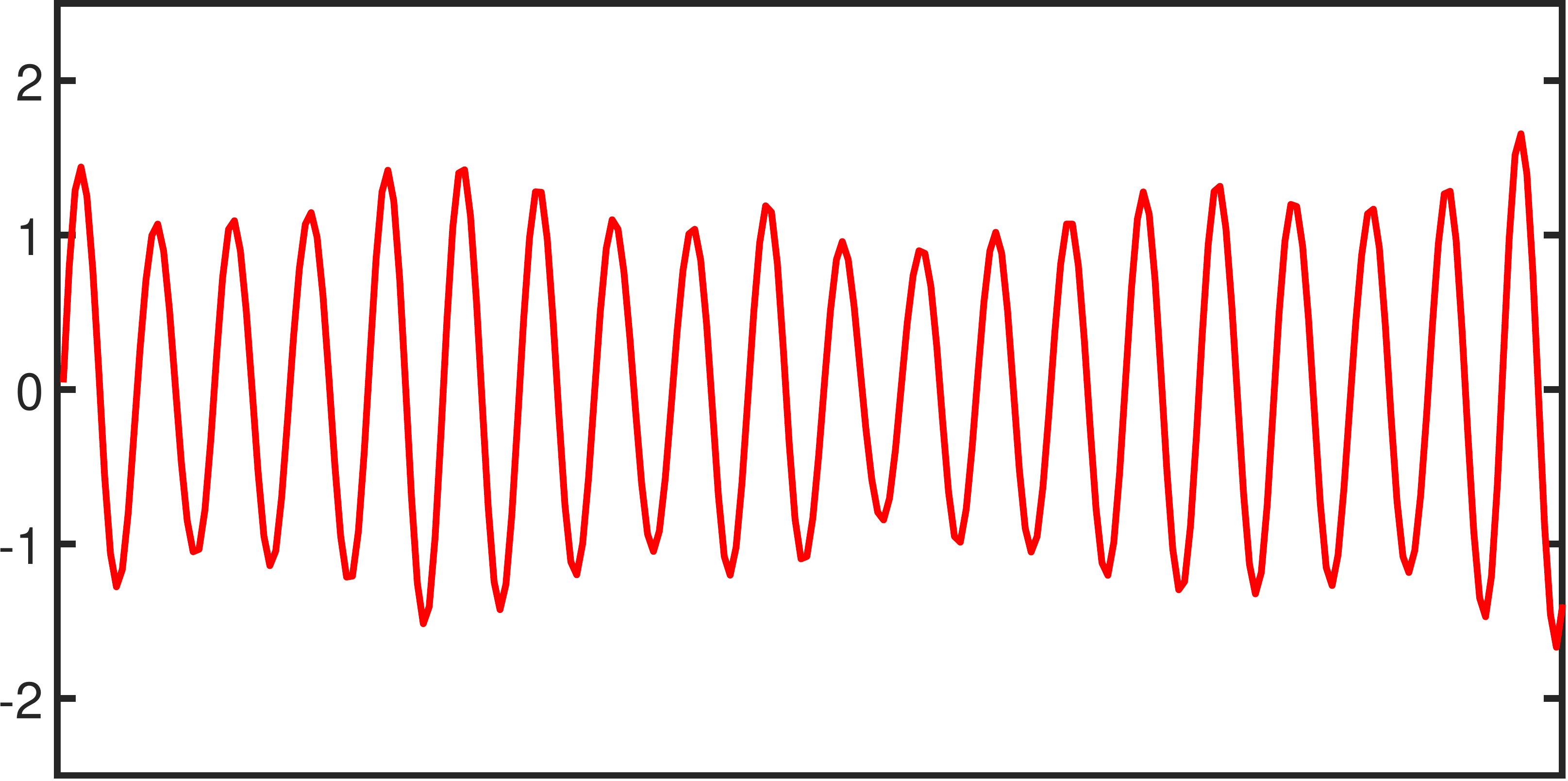}}\;
\subfloat[]{\includegraphics[scale=0.16]{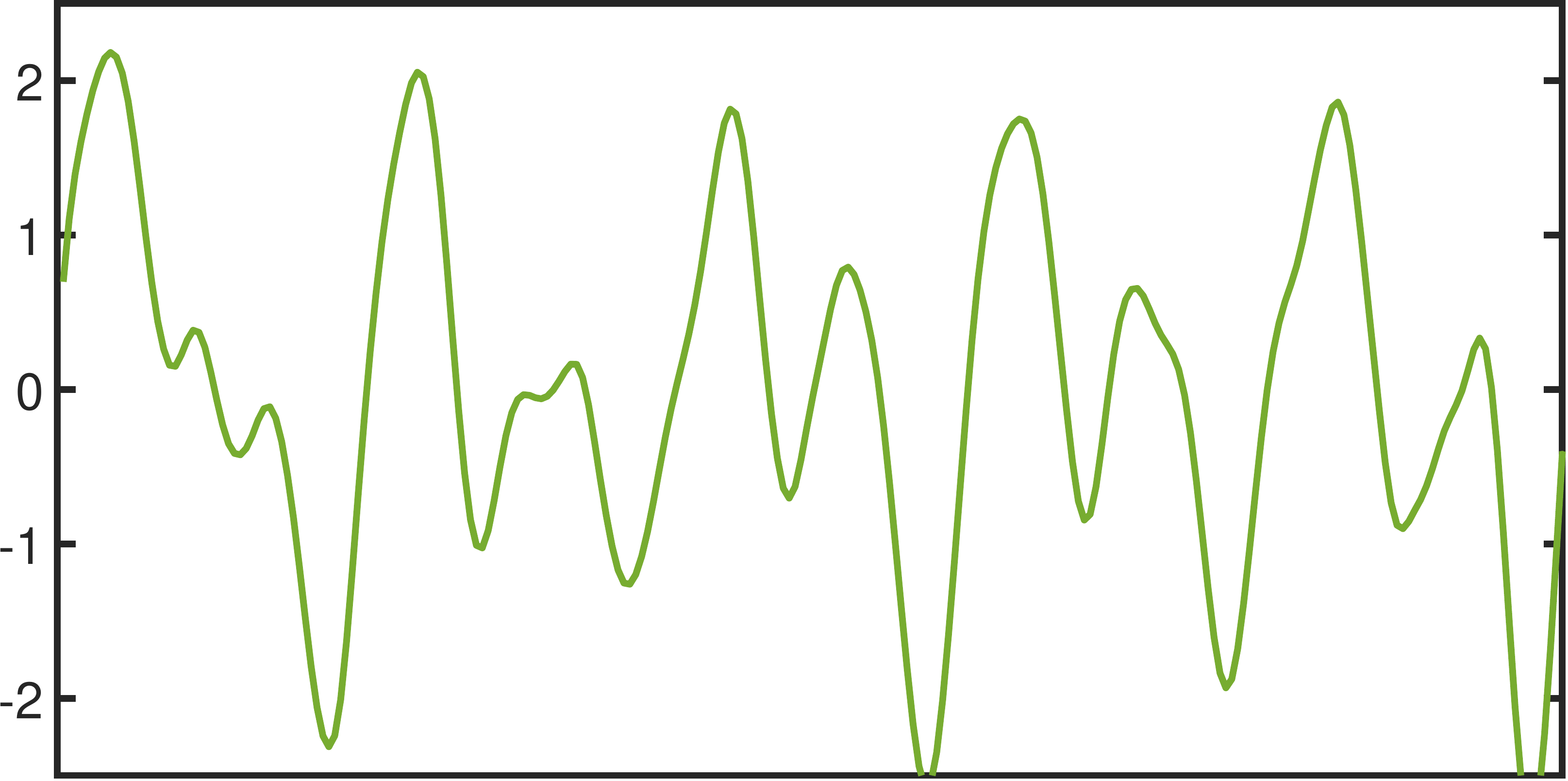}}
}
\caption{\textbf{(a)} The simulated LFP data, where the potential field strength (in arbitrary units) is plotted as a function of time for $t\in[0,1]$, at a sampling frequency of $F_S = 256$ Hz. The signal consists of 4 components, where the alpha and theta components are grouped together because their frequency ranges show strong overlap. The beta, gamma and alpha-theta components are shown in \textbf{(b)}. Every column is colour coded to represent a different frequency component: beta (\textbf{\textcolor{blue}{blue}}), gamma (\textbf{\textcolor{red}{red}}) and alpha-theta (\textbf{\textcolor{plotgreen}{green}}). The SSD recovered components are shown for classical SSD \textbf{(c)}, gate-based QSSD \textbf{(d)} and pulse-based QSSD \textbf{(e)}. The field strength units are arbitrary in \textbf{(b-e)}, and time runs from $t=0$ to $t=1$ sec, omitted for clarity.}
\label{fig:biomedical}
\end{figure}

\noindent SSD cannot distinguish them. In Fig. \ref{fig:biomedical}, we show that SSD is capable of unentangling the superimposed signal into its characteristic components within reasonable accuracy bounds. The corresponding errors are graphed in Fig. \ref{fig:scatter} for all SSD variations.

\subsection{Gravitational wave event GW150914}

Black holes and gravitational waves are characteristic predictions of general relativity, and are therefore favourable probes of the theory. The waveforms of such gravitational waves are constrained by the underlying mechanisms such as their generation and propagation. The analysis of such waves is crucial to the understanding of relativistic gravitational physics, and may provide useful insights in black hole physics, the structure of neutron stars, supernovae and the cosmic gravitational wave background. Gravitational wave analysis will likely become classically intractable in the age of Big Data and third generation detectors such as the Einstein Telescope \cite{thirdgen}, and can greatly benefit from quantum data analysis algorithms. We applied SSD to gravitational wave analysis since it serves as an interesting addition to the classical gravitational wave data analysis pipeline. Not only is it capable of providing waveforms with a more crisp quality, it is also able to filter out glitches, lowering the risk of a wrong signal-to-noise ratio (SNR) threshold exceedance.\\

\begin{figure}[H]
\captionsetup{justification=justified}
\captionsetup[subfigure]{labelformat=empty}
\centering{

\textbf{(a)}
\subfloat[]{\includegraphics[width=0.4\textwidth]{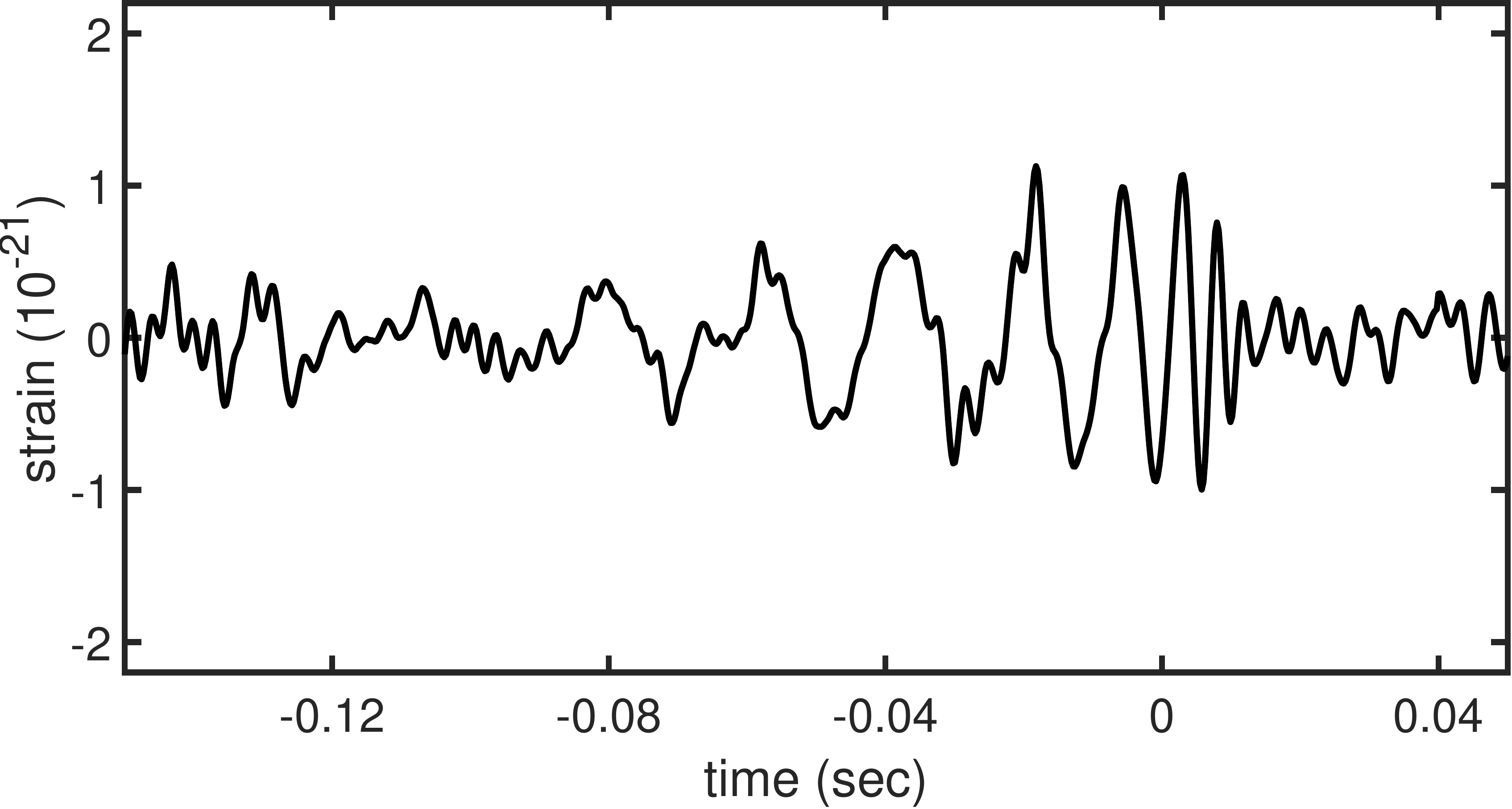}}\;
\subfloat[]{\includegraphics[width=0.418\textwidth]{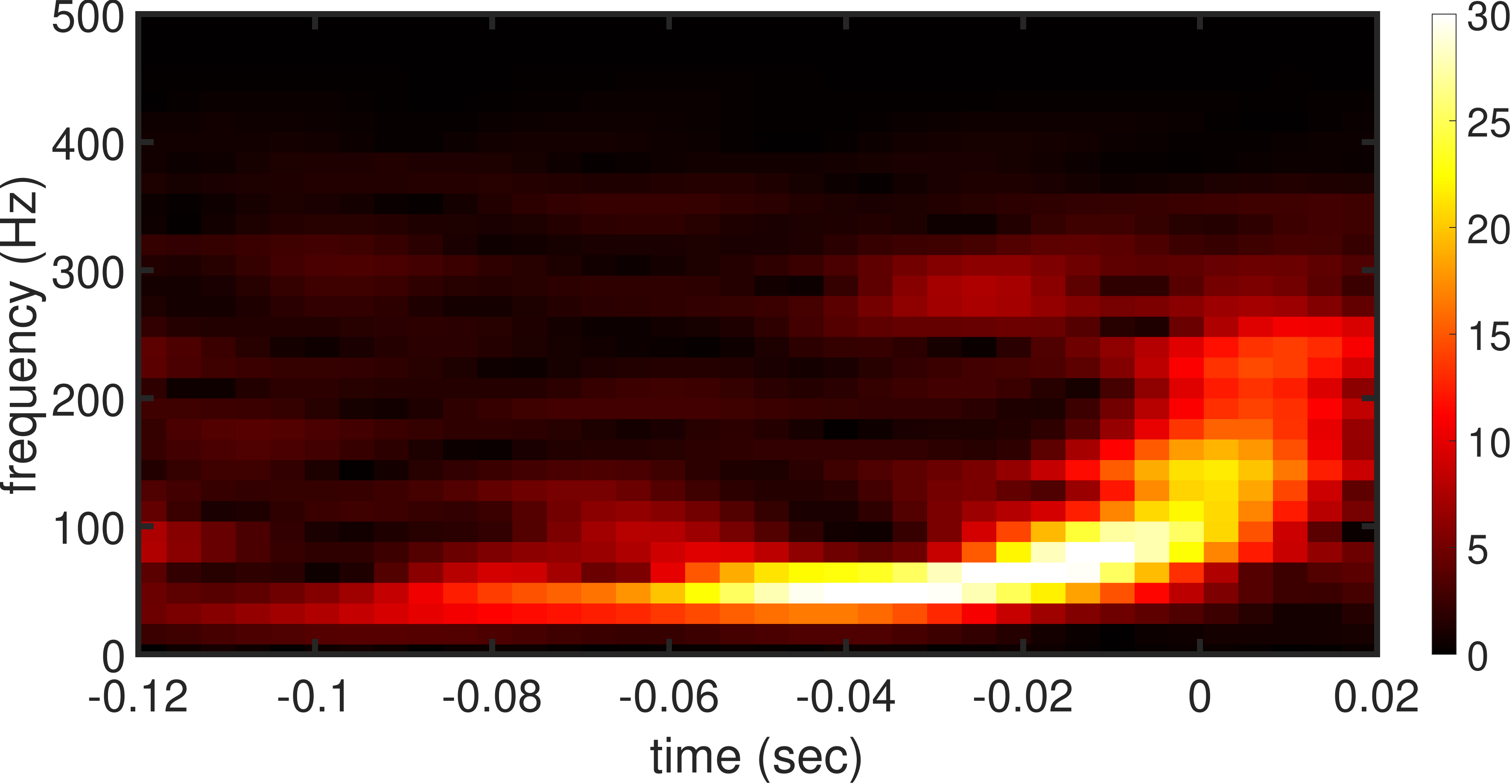}}\;

\textbf{(b)}
\subfloat[]{\includegraphics[width=0.4\textwidth]{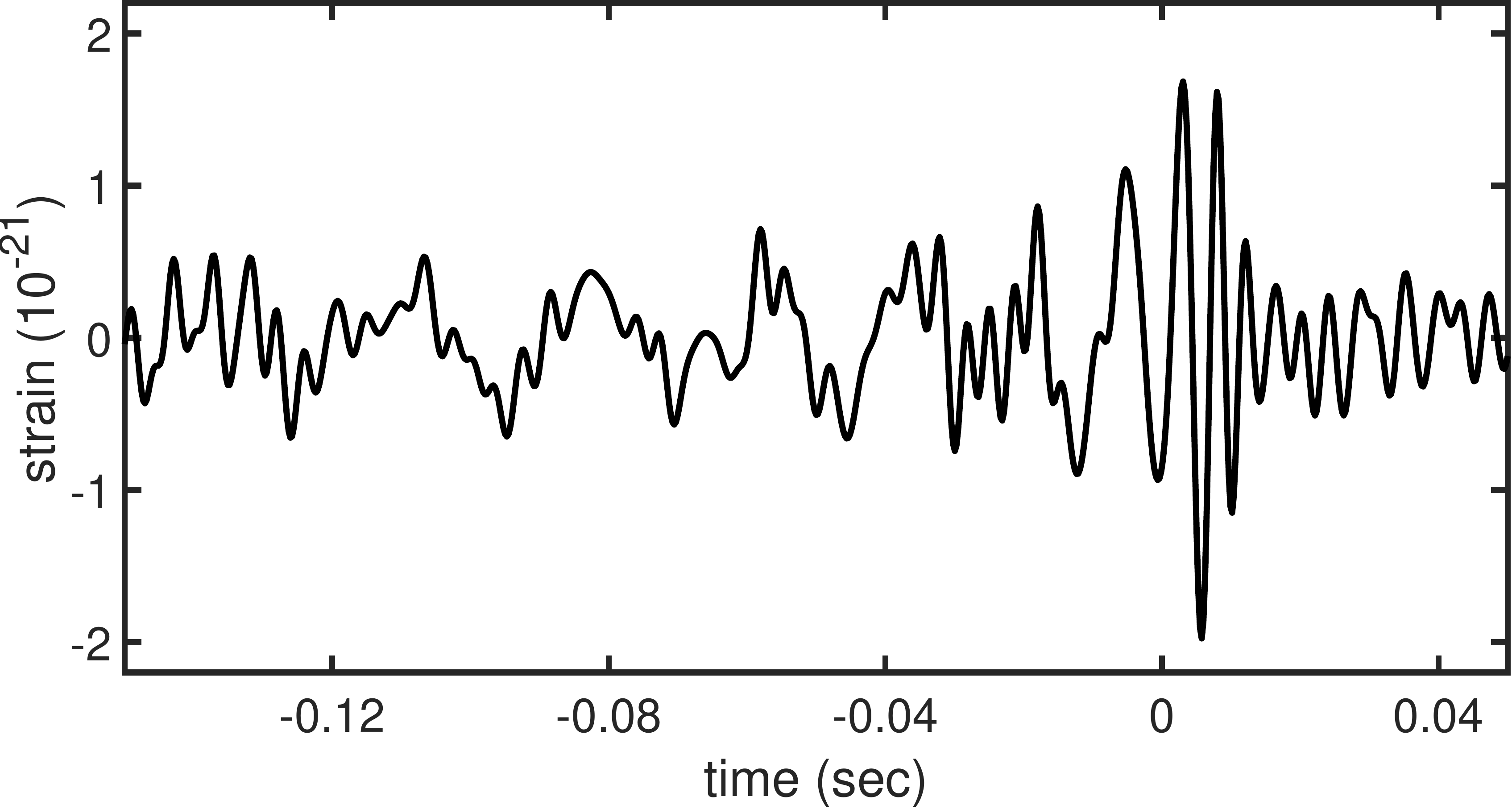}}\;
\subfloat[]{\includegraphics[width=0.418\textwidth]{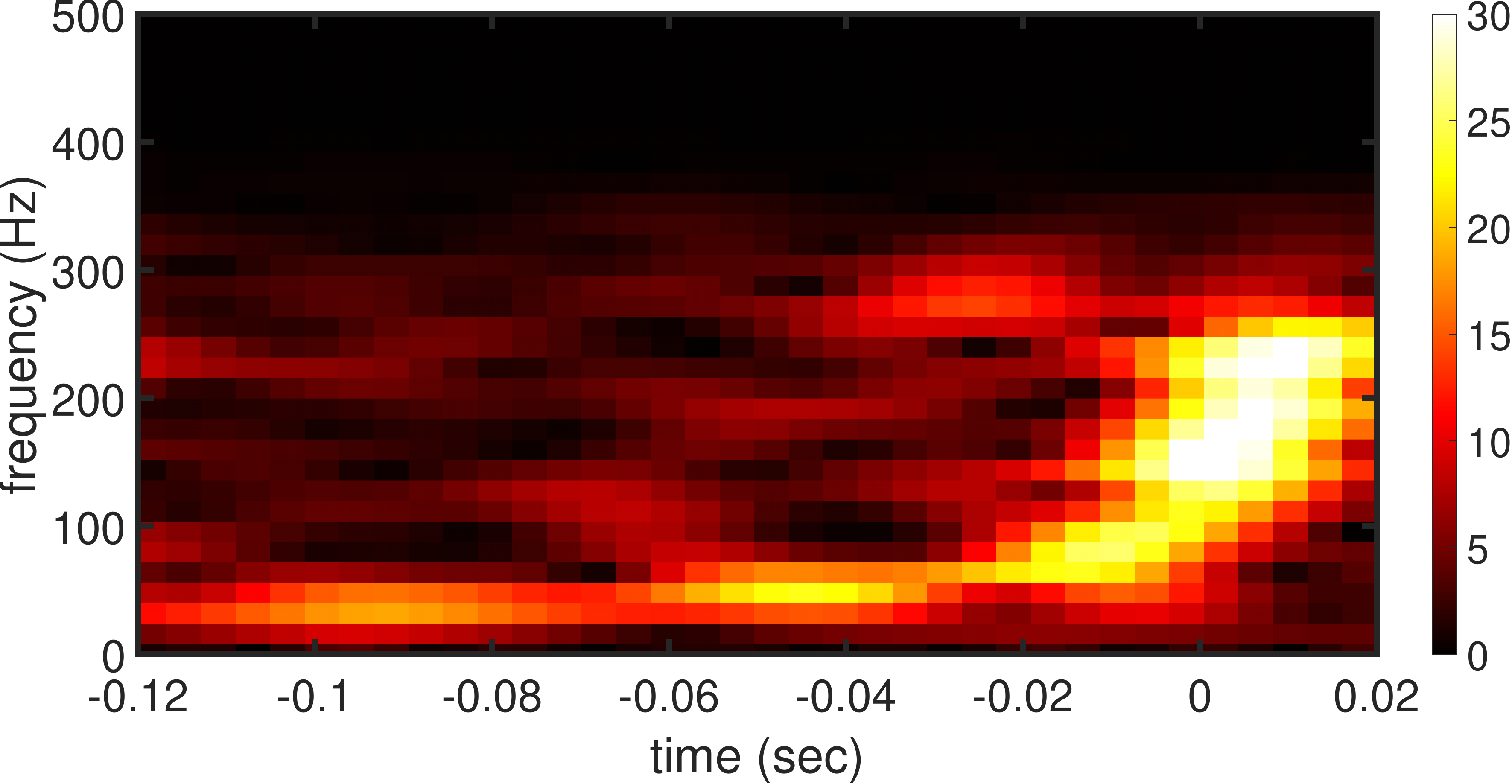}}\;

\textbf{(c)}
\subfloat[]{\includegraphics[width=0.4\textwidth]{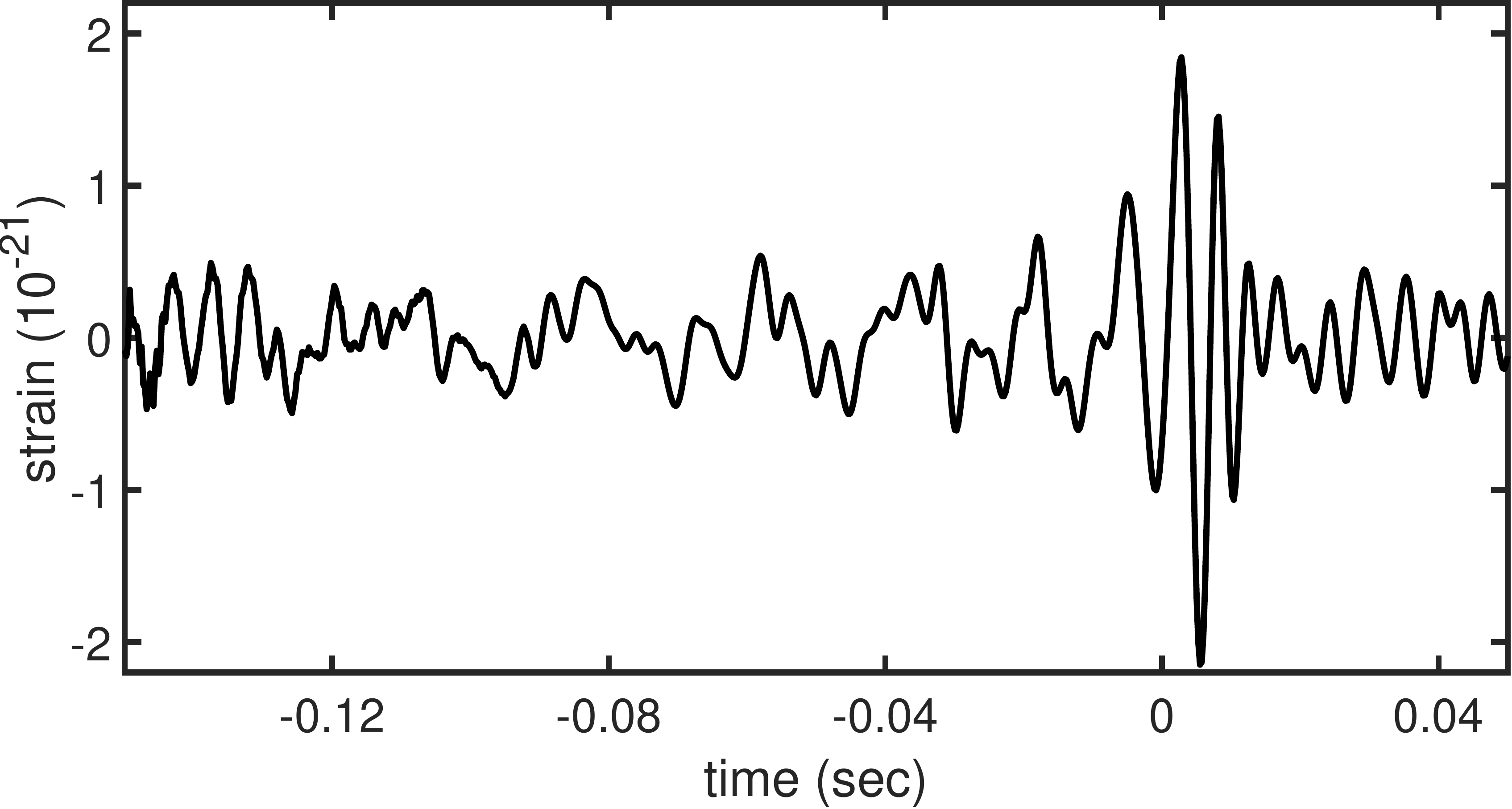}}\;
\subfloat[]{\includegraphics[width=0.418\textwidth]{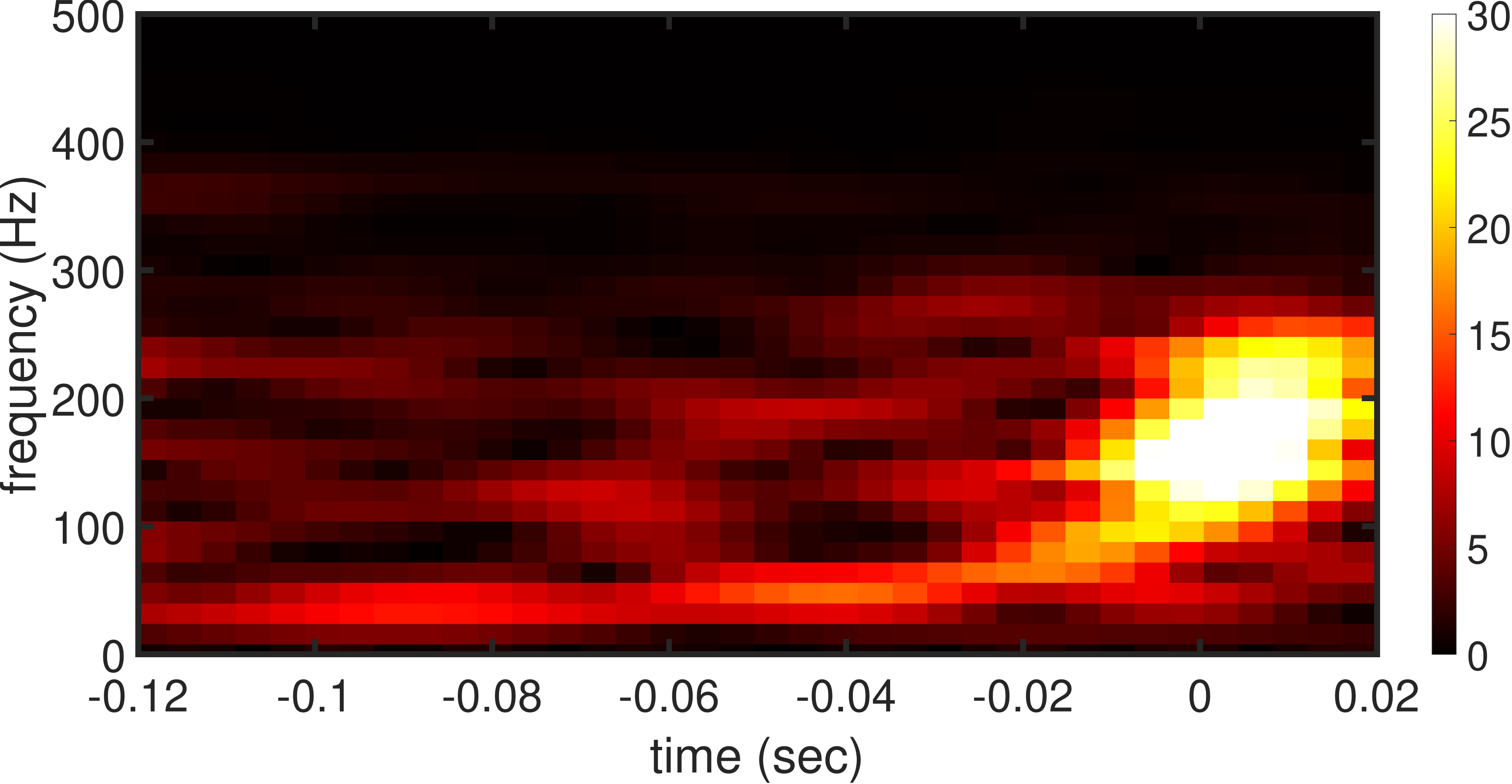}}\;

\textbf{(d)}
\subfloat[]{\includegraphics[width=0.4\textwidth]{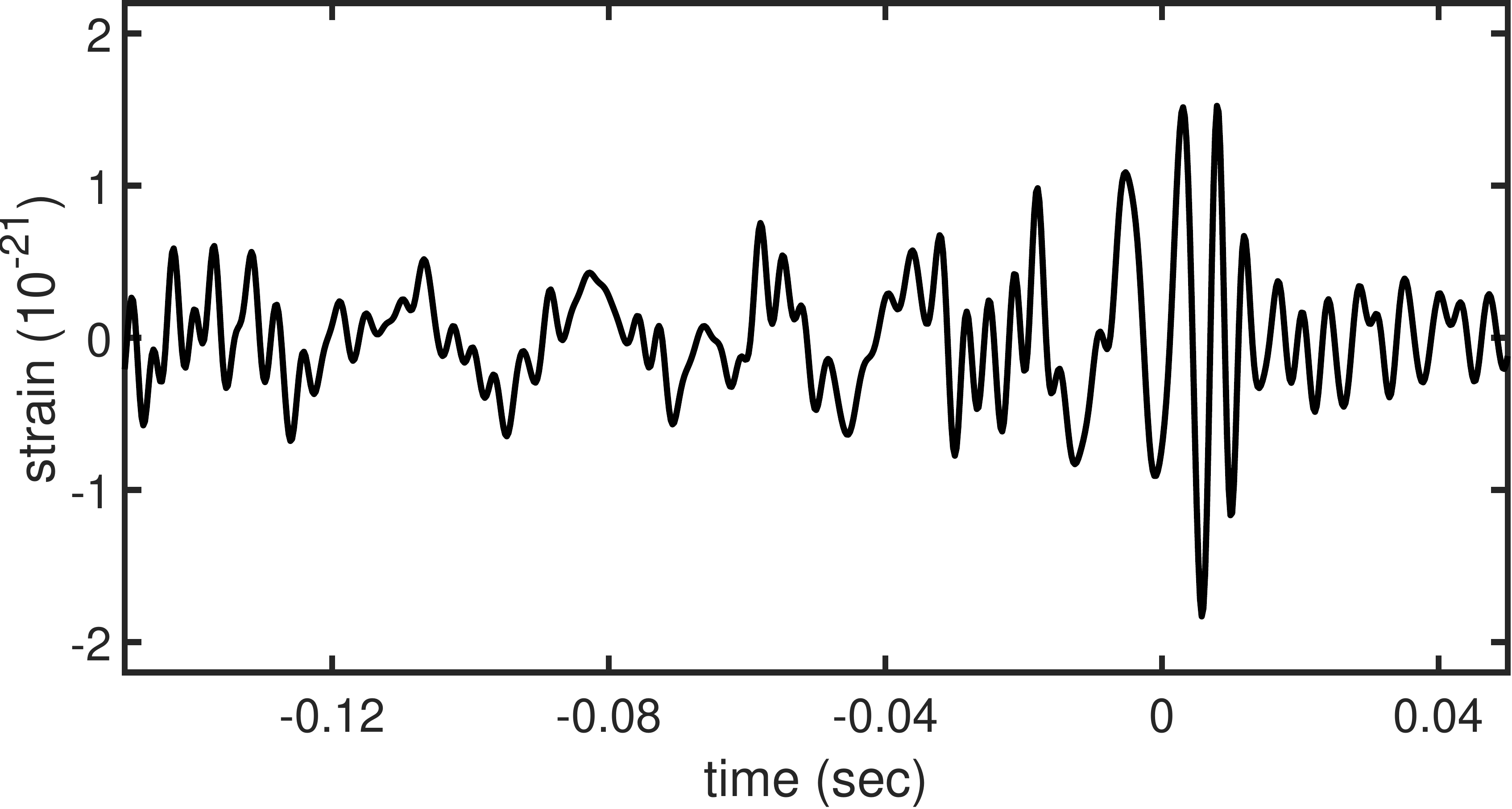}}\;
\subfloat[]{\includegraphics[width=0.418\textwidth]{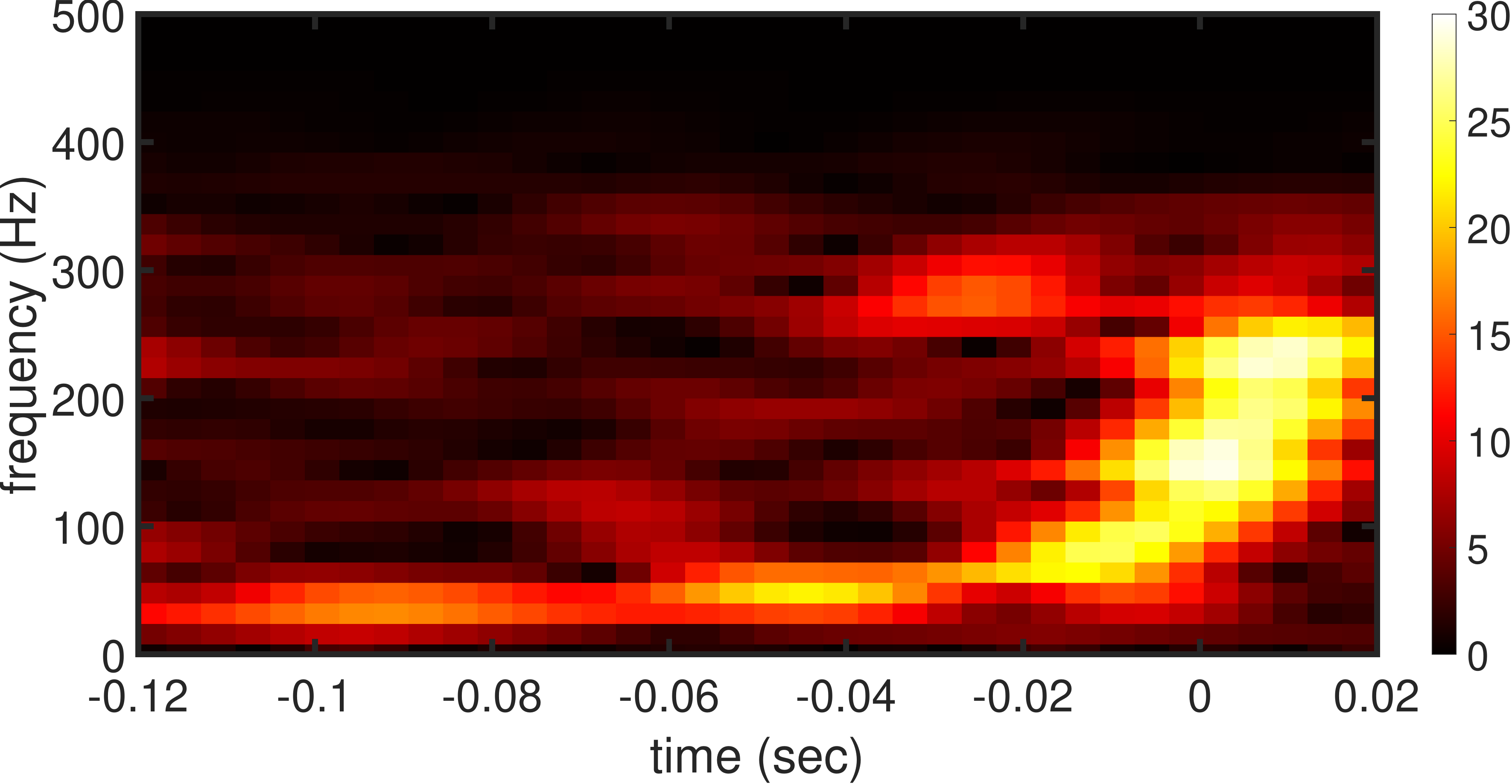}}\;
}

\caption{\textbf{(a)} The GW150914 event gravitational wave strain amplitude, measured at the LIGO Hanford Observatory, after noise whitening, band-passing and notching (\textit{left}) \cite{ligodata}, presented with its corresponding time-frequency spectrogram (\textit{right)}. Time is relative to the event time, on September 14, 2015 at 09:50:45 UTC. \textbf{(b-d)} The extracted gravitational wave components (\textit{left column}) with their respective spectrograms (\textit{right column}), obtained through classical SSD \textbf{(b)}, gate-based QSSD \textbf{(c)} and pulse-based QSSD \textbf{(d)}. For SSD calculations, the detector sampling frequency $F_S = 4096$ Hz was chosen. The spectrograms were obtained by shifting Blackman windows of size 256 with a mutual overlap of 240 points. Results show that through SSD, we can obtain a signal with a higher quality chirp. The spectrogram amplitude is normalised with respect to the strongest amplitude present.}
\label{fig:gravitational}
\end{figure}

\begin{figure}[H]
\centering{
\captionsetup{justification=justified}
\includegraphics[scale=0.35]{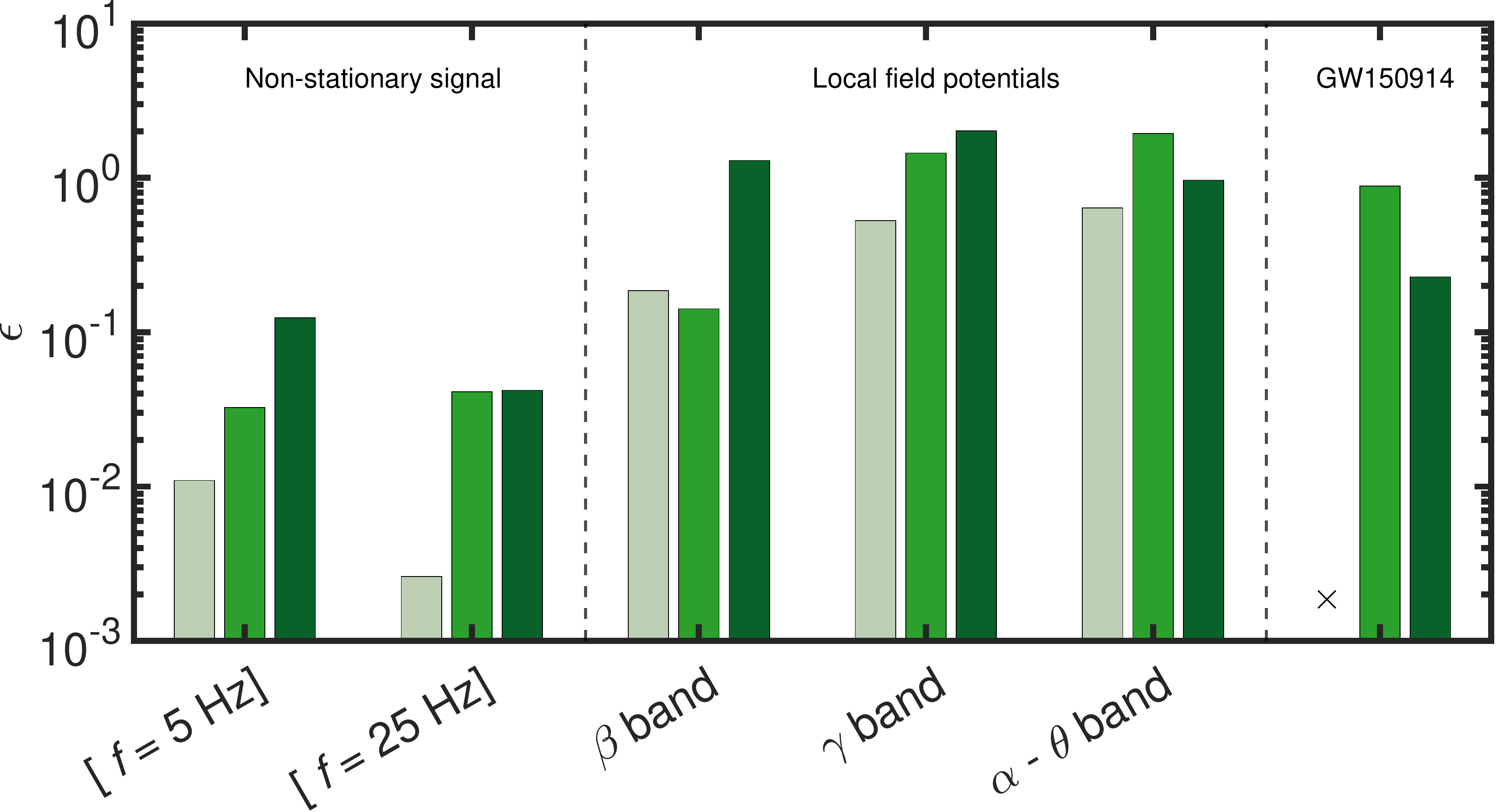}}
\caption{The error $\epsilon$, given by Eq. (\ref{eq:error}) of the retrieved SSD components for all application examples. The error of the components of the non-stationary signal and the LFPs are with respect to the simulated signal. For GW150914, the error is taken with respect to the classical SSD components. Results are shown for classical SSD (\textbf{\textcolor{lightgreen}{light green}}), gate-based QSSD (\textbf{\textcolor{normalgreen}{green}}) and pulse-based QSSD (\textbf{\textcolor{darkgreen}{dark green}}). For general problem-agnostic quantum circuit resources, gate-based QSSD seems to provide more accurate signals than pulse-based QSSD, when the component closely resembles a simple harmonic function.}
\label{fig:scatter}
\end{figure}

To showcase the applicability of SSD, we applied it to the the first measured gravitational wave event GW150914, eminent from a black hole binary merger, consistent of a pair of black holes with inferred source masses $M_1 = 29.1_{-4.4}^{+3.7} M_\odot$ and $M_2 = 36.2 _{-3.8}^{+5.2} M_\odot$ \cite{firstGW, hanford}. We have taken the raw data as measured by the LIGO Hanford detector around the merger time, after which we applied noise whitening, band-passing and notching to obtain an approximate gravitational waveform \cite{ligodata}. Through SSD we hope to separate the relevant information about the binary merger from any other unwanted component or noise, thus obtaining a subset of SSD components that can provide a more crisp waveform. The results of SSD filtering are presented in Fig. \ref{fig:gravitational}, as well as their respective time-frequency spectrograms. Since we are analysing a real signal rather than a simulated one, we cannot fairly compare the quality of a components to any original benchmark, because we do not know a priori what the \textit{true} gravitational wave signal should look like, given the right physical parameters. For this reason, we can only provide a figure of merit of the QSSD components with respect to the classical SSD components. The errors are given in Fig. \ref{fig:scatter}, where the comparison with classical SSD is missing for GW150914 because of the aforementioned argument.\\

\onecolumngrid

\twocolumngrid

\onecolumngrid

\twocolumngrid

\null\newpage
\null\newpage

\onecolumngrid

\twocolumngrid

\section{\label{sec:sectionVI}Conclusion and discussion}
In this work we have established a unified framework that combines the promising method of singular spectrum decomposition for classical time series analysis with the variational quantum singular value decomposition algorithm into quantum singular spectrum decomposition (QSSD). The singular value decomposition subroutine is designed to be performed on quantum hardware, while the retrieval of the SSD components is done classically. Using analytical gradient optimisers, we circumvent finite difference errors, and we guarantee optimisation at every iteration. QSSD could find interesting applications to near-term gravitational wave analysis because of the variational formulation. Here, we demonstrated this algorithm for providing a more crisp waveform for matched filtering.\\

QSSD translates a classical algorithm into a hybrid quantum-classical framework. In contrast to the classical algorithm, the quantum formulation does not have efficient access to all matrix elements of $U$ and $V$. Unpacking the dense trajectory matrix into its unitary basis elements and reading out all the columns of $U$ and $V$ are two processes that required us to perform an exponential number of quantum circuit runs, in the number of qubits $q_m$. An efficient quantum algorithm would rely on knowing only few parameters that are readily measured on a quantum device, while employing few resources. To mitigate the effects of the exponential scaling laws, we have adopted randomised SVD. While rendering QSSD more economic, the same trick can be applied to the classical algorithm, providing no inherent \textit{quantum} speedup. An efficient rendition of SSD would circumvent measuring $\ket{u_i}$ and $\ket{v_i}$, and instead use a different pipeline to read out SSD components. For further research, we pose the question whether this is possible within this variational framework.\\

The quality of SSD is related to the quality of the singular vectors, not the singular values, as dictated by Eq. (\ref{eq:rescale}), since a rescaling is performed after every iteration to ensure optimal energy extraction. It is therefore crucial for the convergence of the quantum routine to select a circuit ansatz that can approximate the solution space well. For the gate-based approach we have adopted the\textit{ hardware efficient ansatz} (HEA), and for the pulse-based approach we have assumed full control. Despite the HEA being often invoked as a standard method for quantum state preparation, there is a lack of understanding on its capabilities. Since data analysis does not know an effective design or model, we pose the question if an effective model can be established for specific time series applications, and if certain ansätze, either gate-based and pulse-based, can improve convergence for finding the orthonormal matrices.\\

In conclusion, it remains an open problem whether a true quantum speedup can be achieved through the method of QSSD. The variational formulation of this quantum algorithm does not allow for polynomial scaling laws in the number of qubits, and requires measuring exponentially many quantum amplitudes. At the same time, as mentioned above, several aspects of the theory can be improved upon, which are not necessarily restricted to QSSD applications only. For quantum state preparation purposes, it is interesting to investigate if one could apply \textit{quantum natural gradient} (QNG) theory to the optimiser routine to improve convergence \cite{qng}, or quantify the geometry of the search space to find better initial states \cite{geometry2q}. Overall, results from this preliminary attempt to translate SSD to a quantum computing framework show that this should be likely achievable with proper improvements in the theory and methods employed. In turn, this may pave the way to a true quantum speedup of QSSD over classical SSD. We leave the adaptations of all these ideas to future studies of QSSD and alternative approaches to time series analysis on a quantum computer.\\

\onecolumngrid

\section*{Acknowledgements}
We thank Robert de Keijzer, Madhav Mohan and Menica Dibenedetto for discussions. This research is financially supported by the Dutch Ministry of Economic Affairs and Climate Policy (EZK), as part of the Quantum Delta NL programme, and by the Netherlands Organisation for Scientific Research (NWO) under Grant No. 680.92.18.05.

\section*{Data Availability}
The data that support the findings of this study are available from the corresponding author upon request. The GW150914 gravitational wave event data is publicly available \cite{hanford}. Simulations were performed using QuTiP \cite{qutip}.

\appendix
\section{Preliminaries - Singular Spectrum Analysis}
\label{app:appendixPrel}
Suppose a time series is sampled $N$ times at a steady rate $F_S$ to obtain the string $x(n)=\{x_1,x_2,\cdots,x_N\}$. Then, the SSA algorithm works as follows:
\begin{enumerate}
    \item The time series is turned into an $M\times K$ Hankel matrix, where $M$ is the embedding dimension and $K=N+1-M$. This gives the trajectory matrix
    \begin{equation}
        X=\begin{bmatrix}
    x_1 & x_2 & \cdots & x_K\\
    x_2 & x_3 & \cdots & x_{K+1}\\
    \vdots & \vdots & \ddots & \smash{\vdots}\\
    x_M & x_{M+1} & \cdots & x_N
    \end{bmatrix}.
    \end{equation}
    Such a matrix encapsulates correlations between the data points. By tuning the embedding dimension just right, SSA can faithfully extract sub-components. If $M$ is too low, not enough correlation can be captured through the trajectory matrix. If $M$ is too large, however, ghost oscillations called spurious components will be captured.
    \item An SVD is performed on the trajectory matrix, yielding $X=U\Sigma V^\top$, with all relevant notation defined in eq. (\ref{eq:SVD}). One can further decompose a rectangular matrix $X$ of rank $n$ into a sum of rank-1 matrices according to
    \begin{equation}
        X=\sum_{i=1}^r X_i\myeq\sum_{i=1}^r \sigma_i \boldsymbol{u}_i \boldsymbol{v}^\top_i. 
    \end{equation}
    The set $\{\sigma_i,\boldsymbol{u}_i,\boldsymbol{v}_i\}$ is referred to as the $i$-th eigentriple for fixed $i$.
    \item The index set $\{1,\cdots,r\}$ is partitioned into $\mathfrak{r}\leq r$ disjoint subsets $\{I_1,\cdots,I_\mathfrak{r}\}$. Each element $I_i$ contains a set of principal components that capture a specific component in a time series (like trend, oscillations, noise, etc.) The original trajectory matrix is then decomposed into
    \begin{equation}
        X=\sum_{i=1}^{\mathfrak{r}} X_{I_i},
    \end{equation}
    where each matrix on the right hand side represents a specific component series $g_c(n)$.
    \item Sub-components are reconstructed by Hankelisation of each $X_{I_i}$, where the matrix components of each cross-diagonal are averaged over. Reverse engineering those Hankel matrices gives the embedded sub-signals $g_c(n)$.
\end{enumerate}

\section{Pseudo-unitary continuation}\label{app:pseudounitary}
In this Appendix, we provide a protocol for the most economic extension of a pseudo-unitary matrix to a square unitary matrix that can be readily prepared for a Hadamard test. Let
\begin{equation}
   e_j = (O|O|\cdots | O | P | O | \cdots |O|O)
\end{equation}
be a $2^{q_m}\times 2^{q_n}$ pseudo-unitary, with $q_n>q_m$. The location of the $P$ operator inside the string, starting from 0 to $2^{q_n-q_m}-1$ can be turned into a binary string $i_0i_1\cdots i_{q_n-q_m-2}i_{q_n-q_m-1}$. Using the definition of the Pauli operators (\ref{eq:pauli}), we can find the continuation of the pseudo-unitaries as
\begin{equation}
    (O|O|\cdots |O | P | O| \cdots |O|O) \mapsto \left(\bigotimes_{j=0}^{q_n-q_m-1} \sigma^{i_j}\right) \otimes P \myeq e_j^{\text{cont}}
\end{equation}
for $\sigma^0=\mathbb{I}, \sigma^1=\sigma^x$. Such $q_n$-qubit operators are readily prepared on a quantum computer, while preserving the structure of the original pseudo-unitary. Such protocol works equally well for transpose structures. Then, the evaluated singular values yield:
\begin{equation}
    x_i = \sum_j a_j \bra{\psi_i}U^\dagger(\boldsymbol{\alpha})e_j V(\boldsymbol{\beta}) \ket{\psi_i} \mapsto
    \sum_j a_j \bra{\psi_i}(\mathbb{I}\otimes U)^\dagger(\boldsymbol{\alpha})e_j^{\text{cont}} V(\boldsymbol{\beta}) \ket{\psi_i}.
\end{equation}

\section{Analytic gradients}\label{app:appgradient}

Because the constituent gates of the quantum circuits are generated by Hermitian generators, analytical gradients are readily calculated. The loss function is given by
\begin{equation}
    \mathcal{L}(\boldsymbol{\alpha},\boldsymbol{\beta})=\sum_{i=1}^T w_i  \cdot \bra{\psi_i} U^\dagger(\boldsymbol{\alpha})XV(\boldsymbol{\beta})\ket{\psi_i},
\end{equation}
and both circuits are parameterised by the hardware efficient ansatz
\begin{equation}
    \mathcal{U}(\boldsymbol{\theta})=\prod_{d=0}^{D-1} \left(\mathcal{U}_{\text{ent}}\times\bigotimes_{n=1}^N R_Y(\theta_{n+Nd})\right).
\end{equation}
Loss function gradients can therefore directly be related to derivatives of the circuit representation with respect to the variational parameters according to
\begin{equation}
    \frac{\partial\mathcal{L}}{\partial\alpha_\mu}=\sum_{i=1}^T w_i \cdot \bra{\psi_i^{(U)}} \frac{\partial U^\dagger(\boldsymbol{\alpha})}{\partial \alpha_\mu}XV(\boldsymbol{\beta})\ket{\psi_i^{(V)}}.
\end{equation}
Because the entanglement operations are unparameterised, we can relate this derivative to a shift in the $\alpha_\mu$ parameter:
\begin{multline}
    {\frac{\partial U^\dagger(\boldsymbol{\alpha})}{\partial \alpha_\mu}=\prod_{d=0}^{D-1}\left(\bigotimes_{n=1}^{q_m} \frac{\partial R_Y^\dagger(\alpha_{n+q_md})}{\partial\alpha_\mu} \times \mathcal{U}_{\text{ent}}\right)}=\\\prod_{d=0}^{D-1}\left(\bigotimes_{n=1}^{q_m} \frac{i}{2}Y\delta_{n+q_md,\mu} R_Y^\dagger(\alpha_{n+q_md}) \times \mathcal{U}_{\text{ent}}\right)=\\\frac{1}{2}\prod_{d=0}^{D-1}\left(\bigotimes_{n=1}^{q_m}  R_Y^\dagger(\alpha_{n+q_md}+\pi \delta_{n+q_md,\mu}) \times \mathcal{U}_{\text{ent}}\right)=\frac{1}{2}U^\dagger(\boldsymbol{\alpha}+\pi \boldsymbol{e}_\mu).
\end{multline}
In a similar spirit, we find
\begin{multline}
    {\frac{\partial V(\boldsymbol{\beta})}{\partial \beta_\nu}=\prod_{d=0}^{D-1}\left(\bigotimes_{n=1}^{q_n} \frac{\partial R_Y(\beta_{n+q_nd})}{\partial\beta_\mu} \times \mathcal{U}_{\text{ent}}\right)}=\\\prod_{d=0}^{D-1}\left(\bigotimes_{n=1}^{q_n} -\frac{i}{2}Y\delta_{n+q_nd,\nu} R_Y(\beta_{n+q_nd}) \times \mathcal{U}_{\text{ent}}\right)=\\\frac{1}{2}\prod_{d=0}^{D-1}\left(\bigotimes_{n=1}^{q_n}  R_Y(\beta_{n+q_nd}-\pi \delta_{n+q_nd,\nu}) \times \mathcal{U}_{\text{ent}}\right)=\frac{1}{2}V(\boldsymbol{\beta}-\pi \boldsymbol{e}_\nu).
\end{multline}
This completes the proof.

\section{Quantum Optimal Control for SVD}\label{app:qoc}

In this Appendix, we prove that QOC finds a suitable application in performing SVD on a quantum computer, by showing that the quantities we must measure for optimisation can be prepared on a quantum computer. Additionally, we formally lay out some of the QOC theory laid out in Ref. \cite{robert} and tweak it to describe an SVD application. The Lagrangian $\mathcal{L}_{\text{QOC}}$ for SVD applications is given by
\begin{equation}
    \mathcal{L}_{\text{QOC}} = J_1(U,V)+J_2(Z_U,Z_V)+J_3(U,Z_U,\eta_U)+J_4(V,Z_V,\eta_V),
\end{equation}
where $U$ and $V$ are the unitary representations of the pulses on circuit $U$ and $V$ respectively, $Z_U$ and $Z_V$ represent the controls over the circuits, and $\eta_U$ and $\eta_V$ are the adjoints. The first term represents the loss function we want to minimise regardless of optimal control:
\begin{equation}
    J_1(U,V) = \sum_{i=1}^s w_i \cdot \mathfrak{Re}\bigg\{\text{Tr}\left[U^\dagger(T)XV(T)\ket{\psi_i^{(V)}}\bra{\psi_i^{(U)}}\right]\bigg\}=\mathfrak{Re}\bigg\{\text{Tr}\left[U^\dagger(T)XV(T)\rho\right]\bigg\},
\end{equation}
where we defined the artificial density matrix
\begin{equation}
    \rho = \sum_{i=1}^s w_i \cdot \ket{\psi_i^{(V)}}\bra{\psi_i^{(U)}}.
\end{equation}
Furthermore, we include the regularisation of the pulse norms
\begin{equation}
    J_2(Z_U,Z_V) = \frac{\lambda}{2} \int_0^T \sum_{l=1}^L \left(|Z_{U,l}(t)|^2+|Z_{V,l}(t)|^2\right) dt
\end{equation}
for some regularisation parameter $\lambda$, which we will fix at $\lambda = 10^{-4}.$ We also add two Lagrange multiplier terms that makes sure that the unitaries that are generated by the pulses are constrained to satisfy the Schrödinger equation. These are given by
\begin{equation}
    J_3(U,Z_U,\eta_U) = \mathfrak{Re}\bigg\{\int_0^T \langle\eta_U,\partial_t U(t) + iH_U(t)U(t)\rangle_{L^2([0,T];\mathbb{C}^{2^{q_m}\times 2^{q_m}})} dt\bigg\},
\end{equation}
\begin{equation}
    J_4(V,Z_V,\eta_V) = \mathfrak{Re}\bigg\{\int_0^T \langle\eta_V,\partial_t V(t) + iH_V(t)V(t)\rangle_{L^2([0,T];\mathbb{C}^{2^{q_n}\times 2^{q_n}})} dt\bigg\},
\end{equation}
where $H_{U/V}(t) = H_d + \sum_{l=1}^L \left(Z_{l,U/V}(t)Q_l + \overline{Z_{l,U/V}(t)}Q_l^\dagger\right)$ denotes the Hamiltonian that drives the circuit $U$ and $V$ respectively, and where the inner product $\langle\cdot\rangle_{L^2([0,T];\mathbb{C}^{2^m\times 2^m})}$ is given by
\begin{equation}
    \langle A,B\rangle_{L^2([0,T];\mathbb{C}^{2^m\times 2^m})} = \mathfrak{Re}\bigg\{\int_0^T \text{Tr}\left[A^\dagger B\right]dt\bigg\}.
\end{equation}
First we show how adjoints are calculated. This follows from the Fréchet derivates $\frac{\partial}{\partial S}$ of the Lagrangian with respect to the quantity $S$, be it the circuit unitary representation, the pulses or the adjoints. If we vary a Lagrangian $J$ with respect to a certain quantity $S$ with $\epsilon\delta S$ with $|\epsilon|\ll 1$, and if we obtain
\begin{equation}
    J(S+\epsilon\delta S)-J(S) = \epsilon\langle f(S),\delta S\rangle_{L^2([0,T];\mathbb{C}^{2^m\times 2^m})},
\end{equation}
then $f(S)$ is said to be the Fréchet derivative. In our next calculation, we omit $\epsilon$ by making it implicit in the variation, and we sloppily write $\frac{\partial J}{\partial S}$ as $\langle f(S),\delta S\rangle_{L^2([0,T];\mathbb{C}^{2^m\times 2^m})}$. We find
\begin{equation}
    \frac{\partial J_1}{\partial U} = J_1(U+\delta U,V)-J_1(U,V)=\mathfrak{Re}\bigg\{\text{Tr}\left[\delta U^\dagger(T)XV(T)\rho\right]\bigg\}=\mathfrak{Re}\bigg\{\text{Tr}\left[XV(T)\rho\delta U^\dagger(T)\right]\bigg\},
\end{equation}
\begin{equation}
    \frac{\partial J_1}{\partial V} = J_1(U,V+\delta V)-J_1(U,V)=\mathfrak{Re}\bigg\{\text{Tr}\left[U^\dagger(T)X\delta V(T)\rho\right]\bigg\}=\mathfrak{Re}\bigg\{\text{Tr}\left[\rho U^\dagger(T) X\delta V(T)\right]\bigg\},
\end{equation}
\begin{multline}
    \frac{\partial J_3}{\partial U} = J_3(U+\delta U,Z_U,\eta_U)-J_3(U,Z_U,\eta_U)=\mathfrak{Re}\bigg\{\int_0^T\text{Tr}\left[\eta_U(t)\left(\partial_t \delta U(t) + iH_U(t)\delta U(t)\right)\right]dt\bigg\} \\ = \mathfrak{Re}\bigg\{\text{Tr}\left[\eta_U(T)\delta U(T)\right]+\int_0^T\text{Tr}\left[(-\partial_t\eta_U^\dagger(t)+i\eta_U^\dagger(t)H_U(t))\delta U(t)\right]\bigg\},
\end{multline}
\begin{multline}
    \frac{\partial J_4}{\partial V} = J_4(V+\delta V,Z_V,\eta_V)-J_4(V,Z_V,\eta_V)=\mathfrak{Re}\bigg\{\int_0^T\text{Tr}\left[\eta_V(t)\left(\partial_t \delta V(t) + iH_V(t)\delta V(t)\right)\right]dt\bigg\} \\ = \mathfrak{Re}\bigg\{\text{Tr}\left[\eta_V(T)\delta V(T)\right]+\int_0^T\text{Tr}\left[(-\partial_t\eta_V^\dagger(t)+i\eta_V^\dagger(t)H_V(t))\delta V(t)\right]\bigg\}.
\end{multline}
Stationary of the action under the Lagrangian $\mathcal{L}_{\text{QOC}}$ requires all its Fréchet derivatives to vanish. This gives us
\begin{equation}
    \frac{\partial \mathcal{L}_{\text{QOC}}}{\partial U} = \frac{\partial J_1}{\partial U} + \frac{\partial J_3}{\partial U} = \mathfrak{Re}\bigg\{\text{Tr}[\left(XV(T)\rho+\eta_U(T)\right)\delta U^\dagger (T)]+\int_0^T \text{Tr}\left[-\partial_t\eta_U^\dagger(t)+i\eta_U^\dagger(t)H_U(t)\right]dt\bigg\}=0.
\end{equation}
where we have used the property $\mathfrak{Re}\bigg\{\text{Tr}\left[AB^\dagger\right]\bigg\}=\mathfrak{Re}\bigg\{\text{Tr}\left[BA^\dagger\right]\bigg\}$. This gives us the propagator equation for the adjoint
\begin{equation}\label{eq:adj1}
    i\partial_t \eta_U(t) - H_U(t)^\dagger \eta_U(t) = 0 \quad \text{with boundary condition} \quad  \eta_U(T) = -iXV(T)\rho.
\end{equation}
In a similar fashion we find
\begin{equation}\label{eq:adj2}
    i\partial_t \eta_V(t) - H_V(t)^\dagger \eta_V(t) = 0 \quad \text{with boundary condition} \quad  \eta_V(T) = -i\rho U^\dagger(T) X^\dagger.
\end{equation}
The derivatives with respect to the pulse controls yield
\begin{multline}
    \frac{\partial J_3}{\partial Z_U} = J_3(U,Z_U+\delta Z_U,\eta_U) - J_3(U,Z_U,\eta_U) =  \mathfrak{Re}\bigg\{\int_0^T \text{Tr}\left[i\eta_U^\dagger (t) \sum_{l=1}^L\left(Q_l \delta Z_{U,l}(t)+Q_l^\dagger \overline{\delta Z_{U,l}(t)}\right)U(t)\right]dt\bigg\} \\ =  \mathfrak{Re}\bigg\{\int_0^T \text{Tr}\left[iU(t)\eta_U^\dagger(t)Q_l^\dagger - i\eta_U(t)U^\dagger(t)Q_l^\dagger\right] \overline{\delta Z_{U,l}(t)}dt\bigg\},
\end{multline}
\begin{multline}
    \frac{\partial J_4}{\partial Z_V} = J_4(V,Z_V+\delta Z_V,\eta_V) - J_4(V,Z_V,\eta_V) = \mathfrak{Re}\bigg\{\int_0^T \text{Tr}\left[i\eta_V^\dagger (t) \sum_{l=1}^L\left(Q_l \delta Z_{V,l}(t)+Q_l^\dagger \overline{\delta Z_{V,l}(t)}\right)V(t)\right]dt\bigg\} \\ =  \mathfrak{Re}\bigg\{\int_0^T \text{Tr}\left[iV(t)\eta_V^\dagger(t)Q_l^\dagger - i\eta_V(t)V^\dagger(t)Q_l^\dagger\right] \overline{\delta Z_{V,l}(t)}dt\bigg\}.
\end{multline}
If QOC is to implemented for a quantum SVD application, we need to be able to measure quantities $\mathcal{Q}$ defined by
\begin{equation}
    \mathcal{Q} = \text{Tr}\left[i V(t) \eta_V^\dagger(t)Q_l^\dagger\right] = \text{Tr}\left[-iV(t)\rho U^\dagger(T)XV(T)V^\dagger(t)Q_l^\dagger\right]=-i\text{Tr}\left[\rho U^\dagger(T)XV(T)V^\dagger(t)Q_l^\dagger V(t)\right]
\end{equation}
on a quantum computer. Defining $\Upsilon_{V,l}=V^\dagger(t)Q_l^\dagger V(t)$, we find that we need to measure quantities of the form
\begin{equation}
    \mathcal{Q} = -i\sum_{i=1}^s w_i \cdot \bra{\psi_i}U^\dagger(T)XV(T)\Upsilon_{V,l}\ket{\psi_i},
\end{equation}
which we can sample through Hadamard tests. A similar calculation for the other circuit yields a similar conclusion. Therefore, a QOC implementation of quantum SVD is possible. With this implementation, we can analytically calculate gradients too, to find the update for the pulses at iteration $k$:
\begin{equation}
    Z_{l}^{k+1}(t) = Z_{l}^{k}(t) + \eta \frac{\partial}{\partial Z} \mathcal{L}_{\text{QOC}}.
\end{equation}
We find that
\begin{equation}
    \frac{\partial\mathcal{L}_{\text{QOC}}}{\partial Z_U} = \frac{\partial J_2}{\partial Z_U}+\frac{\partial J_3}{\partial Z_U} = \lambda Z_l - \text{Tr}\left[Q_l^\dagger(\eta_U(t)U^\dagger(t)+U(t)\eta_U^\dagger(t))\right],
\end{equation}
\begin{equation}
    \frac{\partial\mathcal{L}_{\text{QOC}}}{\partial Z_V} = \frac{\partial J_2}{\partial Z_V}+\frac{\partial J_4}{\partial Z_V} = \lambda Z_l - \text{Tr}\left[Q_l^\dagger(\eta_V(t)V^\dagger(t)+V(t)\eta_V^\dagger(t))\right].
\end{equation}
Because of the dependence of $U$-adjoints on $V(t)$ and vice versa, this allows for a coupled gradient calculation on a quantum computer for SVD applications.

\twocolumngrid
\bibliographystyle{apsrev4-1}
\bibliography{Bibliography.bib}

\begin{thebibliography}{29}%
\makeatletter
\providecommand \@ifxundefined [1]{%
 \@ifx{#1\undefined}
}%
\providecommand \@ifnum [1]{%
 \ifnum #1\expandafter \@firstoftwo
 \else \expandafter \@secondoftwo
 \fi
}%
\providecommand \@ifx [1]{%
 \ifx #1\expandafter \@firstoftwo
 \else \expandafter \@secondoftwo
 \fi
}%
\providecommand \natexlab [1]{#1}%
\providecommand \enquote  [1]{``#1''}%
\providecommand \bibnamefont  [1]{#1}%
\providecommand \bibfnamefont [1]{#1}%
\providecommand \citenamefont [1]{#1}%
\providecommand \href@noop [0]{\@secondoftwo}%
\providecommand \href [0]{\begingroup \@sanitize@url \@href}%
\providecommand \@href[1]{\@@startlink{#1}\@@href}%
\providecommand \@@href[1]{\endgroup#1\@@endlink}%
\providecommand \@sanitize@url [0]{\catcode `\\12\catcode `\$12\catcode
  `\&12\catcode `\#12\catcode `\^12\catcode `\_12\catcode `\%12\relax}%
\providecommand \@@startlink[1]{}%
\providecommand \@@endlink[0]{}%
\providecommand \url  [0]{\begingroup\@sanitize@url \@url }%
\providecommand \@url [1]{\endgroup\@href {#1}{\urlprefix }}%
\providecommand \urlprefix  [0]{URL }%
\providecommand \Eprint [0]{\href }%
\providecommand \doibase [0]{http://dx.doi.org/}%
\providecommand \selectlanguage [0]{\@gobble}%
\providecommand \bibinfo  [0]{\@secondoftwo}%
\providecommand \bibfield  [0]{\@secondoftwo}%
\providecommand \translation [1]{[#1]}%
\providecommand \BibitemOpen [0]{}%
\providecommand \bibitemStop [0]{}%
\providecommand \bibitemNoStop [0]{.\EOS\space}%
\providecommand \EOS [0]{\spacefactor3000\relax}%
\providecommand \BibitemShut  [1]{\csname bibitem#1\endcsname}%
\let\auto@bib@innerbib\@empty
\bibitem [{\citenamefont {Bonizzi}\ \emph {et~al.}(2014)\citenamefont
  {Bonizzi}, \citenamefont {Karel}, \citenamefont {Meste},\ and\ \citenamefont
  {Peeters}}]{ssd}%
  \BibitemOpen
  \bibfield  {author} {\bibinfo {author} {\bibfnamefont {P.}~\bibnamefont
  {Bonizzi}}, \bibinfo {author} {\bibfnamefont {J.~M.}\ \bibnamefont {Karel}},
  \bibinfo {author} {\bibfnamefont {O.}~\bibnamefont {Meste}}, \ and\ \bibinfo
  {author} {\bibfnamefont {R.~L.}\ \bibnamefont {Peeters}},\ }\href@noop {}
  {\bibfield  {journal} {\bibinfo  {journal} {Advances in Adaptive Data
  Analysis}\ }\textbf {\bibinfo {volume} {6}},\ \bibinfo {pages} {1450011}
  (\bibinfo {year} {2014})}\BibitemShut {NoStop}%
\bibitem [{\citenamefont {Bonizzi}\ \emph {et~al.}(2015)\citenamefont
  {Bonizzi}, \citenamefont {Karel}, \citenamefont {Zeemering},\ and\
  \citenamefont {Peeters}}]{sleep}%
  \BibitemOpen
  \bibfield  {author} {\bibinfo {author} {\bibfnamefont {P.}~\bibnamefont
  {Bonizzi}}, \bibinfo {author} {\bibfnamefont {J.}~\bibnamefont {Karel}},
  \bibinfo {author} {\bibfnamefont {S.}~\bibnamefont {Zeemering}}, \ and\
  \bibinfo {author} {\bibfnamefont {R.}~\bibnamefont {Peeters}},\ }in\ \href
  {\doibase 10.1109/CIC.2015.7408648} {\emph {\bibinfo {booktitle} {2015
  Computing in Cardiology Conference (CinC)}}}\ (\bibinfo {year} {2015})\ pp.\
  \bibinfo {pages} {309--312}\BibitemShut {NoStop}%
\bibitem [{\citenamefont {Lowet}\ \emph {et~al.}(2017)\citenamefont {Lowet},
  \citenamefont {Roberts}, \citenamefont {Peter}, \citenamefont {Gips},\ and\
  \citenamefont {De~Weerd}}]{macaque}%
  \BibitemOpen
  \bibfield  {author} {\bibinfo {author} {\bibfnamefont {E.}~\bibnamefont
  {Lowet}}, \bibinfo {author} {\bibfnamefont {M.~J.}\ \bibnamefont {Roberts}},
  \bibinfo {author} {\bibfnamefont {A.}~\bibnamefont {Peter}}, \bibinfo
  {author} {\bibfnamefont {B.}~\bibnamefont {Gips}}, \ and\ \bibinfo {author}
  {\bibfnamefont {P.}~\bibnamefont {De~Weerd}},\ }\href@noop {} {\bibfield
  {journal} {\bibinfo  {journal} {Elife}\ }\textbf {\bibinfo {volume} {6}},\
  \bibinfo {pages} {e26642} (\bibinfo {year} {2017})}\BibitemShut {NoStop}%
\bibitem [{\citenamefont {Lowet}\ \emph {et~al.}(2016)\citenamefont {Lowet},
  \citenamefont {Roberts}, \citenamefont {Bonizzi}, \citenamefont {Karel},\
  and\ \citenamefont {De~Weerd}}]{macaque2}%
  \BibitemOpen
  \bibfield  {author} {\bibinfo {author} {\bibfnamefont {E.}~\bibnamefont
  {Lowet}}, \bibinfo {author} {\bibfnamefont {M.~J.}\ \bibnamefont {Roberts}},
  \bibinfo {author} {\bibfnamefont {P.}~\bibnamefont {Bonizzi}}, \bibinfo
  {author} {\bibfnamefont {J.}~\bibnamefont {Karel}}, \ and\ \bibinfo {author}
  {\bibfnamefont {P.}~\bibnamefont {De~Weerd}},\ }\href@noop {} {\bibfield
  {journal} {\bibinfo  {journal} {PloS one}\ }\textbf {\bibinfo {volume}
  {11}},\ \bibinfo {pages} {e0146443} (\bibinfo {year} {2016})}\BibitemShut
  {NoStop}%
\bibitem [{\citenamefont {Cline}\ and\ \citenamefont
  {Dhillon}(2006)}]{svdcomplexity}%
  \BibitemOpen
  \bibfield  {author} {\bibinfo {author} {\bibfnamefont {A.~K.}\ \bibnamefont
  {Cline}}\ and\ \bibinfo {author} {\bibfnamefont {I.~S.}\ \bibnamefont
  {Dhillon}},\ }\href@noop {} {\emph {\bibinfo {title} {Handbook of Linear
  Algebra}}}\ (\bibinfo  {publisher} {CRC Press},\ \bibinfo {year}
  {2006})\BibitemShut {NoStop}%
\bibitem [{\citenamefont {Preskill}(2018)}]{nisq}%
  \BibitemOpen
  \bibfield  {author} {\bibinfo {author} {\bibfnamefont {J.}~\bibnamefont
  {Preskill}},\ }\href {\doibase 10.22331/q-2018-08-06-79} {\bibfield
  {journal} {\bibinfo  {journal} {Quantum}\ }\textbf {\bibinfo {volume} {2}},\
  \bibinfo {pages} {79} (\bibinfo {year} {2018})}\BibitemShut {NoStop}%
\bibitem [{\citenamefont {Kandala}\ \emph {et~al.}(2018)\citenamefont
  {Kandala}, \citenamefont {Temme}, \citenamefont {Corcoles}, \citenamefont
  {Mezzacapo}, \citenamefont {Chow},\ and\ \citenamefont
  {Gambetta}}]{kandalarda}%
  \BibitemOpen
  \bibfield  {author} {\bibinfo {author} {\bibfnamefont {A.}~\bibnamefont
  {Kandala}}, \bibinfo {author} {\bibfnamefont {K.}~\bibnamefont {Temme}},
  \bibinfo {author} {\bibfnamefont {A.~D.}\ \bibnamefont {Corcoles}}, \bibinfo
  {author} {\bibfnamefont {A.}~\bibnamefont {Mezzacapo}}, \bibinfo {author}
  {\bibfnamefont {J.~M.}\ \bibnamefont {Chow}}, \ and\ \bibinfo {author}
  {\bibfnamefont {J.~M.}\ \bibnamefont {Gambetta}},\ }\href@noop {} {\bibfield
  {journal} {\bibinfo  {journal} {arXiv preprint arXiv:1805.04492}\ } (\bibinfo
  {year} {2018})}\BibitemShut {NoStop}%
\bibitem [{\citenamefont {O'Malley}\ \emph {et~al.}(2016)\citenamefont
  {O'Malley}, \citenamefont {Babbush}, \citenamefont {Kivlichan}, \citenamefont
  {Romero}, \citenamefont {McClean}, \citenamefont {Barends}, \citenamefont
  {Kelly}, \citenamefont {Roushan}, \citenamefont {Tranter}, \citenamefont
  {Ding}, \citenamefont {Campbell}, \citenamefont {Chen}, \citenamefont {Chen},
  \citenamefont {Chiaro}, \citenamefont {Dunsworth}, \citenamefont {Fowler},
  \citenamefont {Jeffrey}, \citenamefont {Lucero}, \citenamefont {Megrant},
  \citenamefont {Mutus}, \citenamefont {Neeley}, \citenamefont {Neill},
  \citenamefont {Quintana}, \citenamefont {Sank}, \citenamefont {Vainsencher},
  \citenamefont {Wenner}, \citenamefont {White}, \citenamefont {Coveney},
  \citenamefont {Love}, \citenamefont {Neven}, \citenamefont {Aspuru-Guzik},\
  and\ \citenamefont {Martinis}}]{firstvqe?}%
  \BibitemOpen
  \bibfield  {author} {\bibinfo {author} {\bibfnamefont {P.~J.~J.}\
  \bibnamefont {O'Malley}}, \bibinfo {author} {\bibfnamefont {R.}~\bibnamefont
  {Babbush}}, \bibinfo {author} {\bibfnamefont {I.~D.}\ \bibnamefont
  {Kivlichan}}, \bibinfo {author} {\bibfnamefont {J.}~\bibnamefont {Romero}},
  \bibinfo {author} {\bibfnamefont {J.~R.}\ \bibnamefont {McClean}}, \bibinfo
  {author} {\bibfnamefont {R.}~\bibnamefont {Barends}}, \bibinfo {author}
  {\bibfnamefont {J.}~\bibnamefont {Kelly}}, \bibinfo {author} {\bibfnamefont
  {P.}~\bibnamefont {Roushan}}, \bibinfo {author} {\bibfnamefont
  {A.}~\bibnamefont {Tranter}}, \bibinfo {author} {\bibfnamefont
  {N.}~\bibnamefont {Ding}}, \bibinfo {author} {\bibfnamefont {B.}~\bibnamefont
  {Campbell}}, \bibinfo {author} {\bibfnamefont {Y.}~\bibnamefont {Chen}},
  \bibinfo {author} {\bibfnamefont {Z.}~\bibnamefont {Chen}}, \bibinfo {author}
  {\bibfnamefont {B.}~\bibnamefont {Chiaro}}, \bibinfo {author} {\bibfnamefont
  {A.}~\bibnamefont {Dunsworth}}, \bibinfo {author} {\bibfnamefont {A.~G.}\
  \bibnamefont {Fowler}}, \bibinfo {author} {\bibfnamefont {E.}~\bibnamefont
  {Jeffrey}}, \bibinfo {author} {\bibfnamefont {E.}~\bibnamefont {Lucero}},
  \bibinfo {author} {\bibfnamefont {A.}~\bibnamefont {Megrant}}, \bibinfo
  {author} {\bibfnamefont {J.~Y.}\ \bibnamefont {Mutus}}, \bibinfo {author}
  {\bibfnamefont {M.}~\bibnamefont {Neeley}}, \bibinfo {author} {\bibfnamefont
  {C.}~\bibnamefont {Neill}}, \bibinfo {author} {\bibfnamefont
  {C.}~\bibnamefont {Quintana}}, \bibinfo {author} {\bibfnamefont
  {D.}~\bibnamefont {Sank}}, \bibinfo {author} {\bibfnamefont {A.}~\bibnamefont
  {Vainsencher}}, \bibinfo {author} {\bibfnamefont {J.}~\bibnamefont {Wenner}},
  \bibinfo {author} {\bibfnamefont {T.~C.}\ \bibnamefont {White}}, \bibinfo
  {author} {\bibfnamefont {P.~V.}\ \bibnamefont {Coveney}}, \bibinfo {author}
  {\bibfnamefont {P.~J.}\ \bibnamefont {Love}}, \bibinfo {author}
  {\bibfnamefont {H.}~\bibnamefont {Neven}}, \bibinfo {author} {\bibfnamefont
  {A.}~\bibnamefont {Aspuru-Guzik}}, \ and\ \bibinfo {author} {\bibfnamefont
  {J.~M.}\ \bibnamefont {Martinis}},\ }\href {\doibase
  10.1103/PhysRevX.6.031007} {\bibfield  {journal} {\bibinfo  {journal} {Phys.
  Rev. X}\ }\textbf {\bibinfo {volume} {6}},\ \bibinfo {pages} {031007}
  (\bibinfo {year} {2016})}\BibitemShut {NoStop}%
\bibitem [{\citenamefont {Peruzzo}\ \emph {et~al.}(2014)\citenamefont
  {Peruzzo}, \citenamefont {McClean}, \citenamefont {Shadbolt}, \citenamefont
  {Yung}, \citenamefont {Zhou}, \citenamefont {Love}, \citenamefont
  {Aspuru-Guzik},\ and\ \citenamefont {O’brien}}]{vqe}%
  \BibitemOpen
  \bibfield  {author} {\bibinfo {author} {\bibfnamefont {A.}~\bibnamefont
  {Peruzzo}}, \bibinfo {author} {\bibfnamefont {J.}~\bibnamefont {McClean}},
  \bibinfo {author} {\bibfnamefont {P.}~\bibnamefont {Shadbolt}}, \bibinfo
  {author} {\bibfnamefont {M.-H.}\ \bibnamefont {Yung}}, \bibinfo {author}
  {\bibfnamefont {X.-Q.}\ \bibnamefont {Zhou}}, \bibinfo {author}
  {\bibfnamefont {P.~J.}\ \bibnamefont {Love}}, \bibinfo {author}
  {\bibfnamefont {A.}~\bibnamefont {Aspuru-Guzik}}, \ and\ \bibinfo {author}
  {\bibfnamefont {J.~L.}\ \bibnamefont {O’brien}},\ }\href@noop {} {\bibfield
   {journal} {\bibinfo  {journal} {Nature communications}\ }\textbf {\bibinfo
  {volume} {5}},\ \bibinfo {pages} {1} (\bibinfo {year} {2014})}\BibitemShut
  {NoStop}%
\bibitem [{\citenamefont {Colless}\ \emph {et~al.}(2018)\citenamefont
  {Colless}, \citenamefont {Ramasesh}, \citenamefont {Dahlen}, \citenamefont
  {Blok}, \citenamefont {Kimchi-Schwartz}, \citenamefont {McClean},
  \citenamefont {Carter}, \citenamefont {de~Jong},\ and\ \citenamefont
  {Siddiqi}}]{molecularspectra}%
  \BibitemOpen
  \bibfield  {author} {\bibinfo {author} {\bibfnamefont {J.~I.}\ \bibnamefont
  {Colless}}, \bibinfo {author} {\bibfnamefont {V.~V.}\ \bibnamefont
  {Ramasesh}}, \bibinfo {author} {\bibfnamefont {D.}~\bibnamefont {Dahlen}},
  \bibinfo {author} {\bibfnamefont {M.~S.}\ \bibnamefont {Blok}}, \bibinfo
  {author} {\bibfnamefont {M.~E.}\ \bibnamefont {Kimchi-Schwartz}}, \bibinfo
  {author} {\bibfnamefont {J.~R.}\ \bibnamefont {McClean}}, \bibinfo {author}
  {\bibfnamefont {J.}~\bibnamefont {Carter}}, \bibinfo {author} {\bibfnamefont
  {W.~A.}\ \bibnamefont {de~Jong}}, \ and\ \bibinfo {author} {\bibfnamefont
  {I.}~\bibnamefont {Siddiqi}},\ }\href@noop {} {\bibfield  {journal} {\bibinfo
   {journal} {Physical Review X}\ }\textbf {\bibinfo {volume} {8}},\ \bibinfo
  {pages} {011021} (\bibinfo {year} {2018})}\BibitemShut {NoStop}%
\bibitem [{\citenamefont {Wang}\ \emph {et~al.}(2021)\citenamefont {Wang},
  \citenamefont {Song},\ and\ \citenamefont {Wang}}]{vqsvd}%
  \BibitemOpen
  \bibfield  {author} {\bibinfo {author} {\bibfnamefont {X.}~\bibnamefont
  {Wang}}, \bibinfo {author} {\bibfnamefont {Z.}~\bibnamefont {Song}}, \ and\
  \bibinfo {author} {\bibfnamefont {Y.}~\bibnamefont {Wang}},\ }\href {\doibase
  10.22331/q-2021-06-29-483} {\bibfield  {journal} {\bibinfo  {journal}
  {Quantum}\ }\textbf {\bibinfo {volume} {5}},\ \bibinfo {pages} {483}
  (\bibinfo {year} {2021})}\BibitemShut {NoStop}%
\bibitem [{\citenamefont {de~Keijzer}\ \emph {et~al.}(2022)\citenamefont
  {de~Keijzer}, \citenamefont {Tse},\ and\ \citenamefont
  {Kokkelmans}}]{robert}%
  \BibitemOpen
  \bibfield  {author} {\bibinfo {author} {\bibfnamefont {R.}~\bibnamefont
  {de~Keijzer}}, \bibinfo {author} {\bibfnamefont {O.}~\bibnamefont {Tse}}, \
  and\ \bibinfo {author} {\bibfnamefont {S.}~\bibnamefont {Kokkelmans}},\
  }\href@noop {} {\bibfield  {journal} {\bibinfo  {journal} {arXiv preprint
  arXiv:2202.08908}\ } (\bibinfo {year} {2022})}\BibitemShut {NoStop}%
\bibitem [{\citenamefont {Halko}\ \emph {et~al.}(2011)\citenamefont {Halko},
  \citenamefont {Martinsson},\ and\ \citenamefont {Tropp}}]{halko}%
  \BibitemOpen
  \bibfield  {author} {\bibinfo {author} {\bibfnamefont {N.}~\bibnamefont
  {Halko}}, \bibinfo {author} {\bibfnamefont {P.-G.}\ \bibnamefont
  {Martinsson}}, \ and\ \bibinfo {author} {\bibfnamefont {J.~A.}\ \bibnamefont
  {Tropp}},\ }\href@noop {} {\bibfield  {journal} {\bibinfo  {journal} {SIAM
  review}\ }\textbf {\bibinfo {volume} {53}},\ \bibinfo {pages} {217} (\bibinfo
  {year} {2011})}\BibitemShut {NoStop}%
\bibitem [{\citenamefont {Vautard}\ and\ \citenamefont {Ghil}(1989)}]{ssa}%
  \BibitemOpen
  \bibfield  {author} {\bibinfo {author} {\bibfnamefont {R.}~\bibnamefont
  {Vautard}}\ and\ \bibinfo {author} {\bibfnamefont {M.}~\bibnamefont {Ghil}},\
  }\href@noop {} {\bibfield  {journal} {\bibinfo  {journal} {Physica D:
  Nonlinear Phenomena}\ }\textbf {\bibinfo {volume} {35}},\ \bibinfo {pages}
  {395} (\bibinfo {year} {1989})}\BibitemShut {NoStop}%
\bibitem [{\citenamefont {Vautard}\ \emph {et~al.}(1992)\citenamefont
  {Vautard}, \citenamefont {Yiou},\ and\ \citenamefont {Ghil}}]{m=n3}%
  \BibitemOpen
  \bibfield  {author} {\bibinfo {author} {\bibfnamefont {R.}~\bibnamefont
  {Vautard}}, \bibinfo {author} {\bibfnamefont {P.}~\bibnamefont {Yiou}}, \
  and\ \bibinfo {author} {\bibfnamefont {M.}~\bibnamefont {Ghil}},\ }\href@noop
  {} {\bibfield  {journal} {\bibinfo  {journal} {Physica D: Nonlinear
  Phenomena}\ }\textbf {\bibinfo {volume} {58}},\ \bibinfo {pages} {95}
  (\bibinfo {year} {1992})}\BibitemShut {NoStop}%
\bibitem [{\citenamefont {Eckart}\ and\ \citenamefont
  {Young}(1936)}]{eckartyoung}%
  \BibitemOpen
  \bibfield  {author} {\bibinfo {author} {\bibfnamefont {C.}~\bibnamefont
  {Eckart}}\ and\ \bibinfo {author} {\bibfnamefont {G.}~\bibnamefont {Young}},\
  }\href@noop {} {\bibfield  {journal} {\bibinfo  {journal} {Psychometrika}\
  }\textbf {\bibinfo {volume} {1}},\ \bibinfo {pages} {211} (\bibinfo {year}
  {1936})}\BibitemShut {NoStop}%
\bibitem [{\citenamefont {Erichson}\ \emph {et~al.}(2016)\citenamefont
  {Erichson}, \citenamefont {Voronin}, \citenamefont {Brunton},\ and\
  \citenamefont {Kutz}}]{rsvd}%
  \BibitemOpen
  \bibfield  {author} {\bibinfo {author} {\bibfnamefont {N.~B.}\ \bibnamefont
  {Erichson}}, \bibinfo {author} {\bibfnamefont {S.}~\bibnamefont {Voronin}},
  \bibinfo {author} {\bibfnamefont {S.~L.}\ \bibnamefont {Brunton}}, \ and\
  \bibinfo {author} {\bibfnamefont {J.~N.}\ \bibnamefont {Kutz}},\ }\href@noop
  {} {\bibfield  {journal} {\bibinfo  {journal} {arXiv preprint
  arXiv:1608.02148}\ } (\bibinfo {year} {2016})}\BibitemShut {NoStop}%
\bibitem [{\citenamefont {Fan}(1951)}]{KyFan}%
  \BibitemOpen
  \bibfield  {author} {\bibinfo {author} {\bibfnamefont {K.}~\bibnamefont
  {Fan}},\ }\href@noop {} {\bibfield  {journal} {\bibinfo  {journal}
  {Proceedings of the National Academy of Sciences of the United States of
  America}\ }\textbf {\bibinfo {volume} {37}},\ \bibinfo {pages} {760}
  (\bibinfo {year} {1951})}\BibitemShut {NoStop}%
\bibitem [{\citenamefont {Schuld}\ \emph {et~al.}(2019)\citenamefont {Schuld},
  \citenamefont {Bergholm}, \citenamefont {Gogolin}, \citenamefont {Izaac},\
  and\ \citenamefont {Killoran}}]{schuld}%
  \BibitemOpen
  \bibfield  {author} {\bibinfo {author} {\bibfnamefont {M.}~\bibnamefont
  {Schuld}}, \bibinfo {author} {\bibfnamefont {V.}~\bibnamefont {Bergholm}},
  \bibinfo {author} {\bibfnamefont {C.}~\bibnamefont {Gogolin}}, \bibinfo
  {author} {\bibfnamefont {J.}~\bibnamefont {Izaac}}, \ and\ \bibinfo {author}
  {\bibfnamefont {N.}~\bibnamefont {Killoran}},\ }\href {\doibase
  10.1103/PhysRevA.99.032331} {\bibfield  {journal} {\bibinfo  {journal} {Phys.
  Rev. A}\ }\textbf {\bibinfo {volume} {99}},\ \bibinfo {pages} {032331}
  (\bibinfo {year} {2019})}\BibitemShut {NoStop}%
\bibitem [{\citenamefont {Koch}\ \emph {et~al.}(2022)\citenamefont {Koch},
  \citenamefont {Boscain}, \citenamefont {Calarco}, \citenamefont {Dirr},
  \citenamefont {Filipp}, \citenamefont {Glaser}, \citenamefont {Kosloff},
  \citenamefont {Montangero}, \citenamefont {Schulte-Herbr{\"u}ggen},
  \citenamefont {Sugny} \emph {et~al.}}]{koch}%
  \BibitemOpen
  \bibfield  {author} {\bibinfo {author} {\bibfnamefont {C.~P.}\ \bibnamefont
  {Koch}}, \bibinfo {author} {\bibfnamefont {U.}~\bibnamefont {Boscain}},
  \bibinfo {author} {\bibfnamefont {T.}~\bibnamefont {Calarco}}, \bibinfo
  {author} {\bibfnamefont {G.}~\bibnamefont {Dirr}}, \bibinfo {author}
  {\bibfnamefont {S.}~\bibnamefont {Filipp}}, \bibinfo {author} {\bibfnamefont
  {S.~J.}\ \bibnamefont {Glaser}}, \bibinfo {author} {\bibfnamefont
  {R.}~\bibnamefont {Kosloff}}, \bibinfo {author} {\bibfnamefont
  {S.}~\bibnamefont {Montangero}}, \bibinfo {author} {\bibfnamefont
  {T.}~\bibnamefont {Schulte-Herbr{\"u}ggen}}, \bibinfo {author} {\bibfnamefont
  {D.}~\bibnamefont {Sugny}},  \emph {et~al.},\ }\href@noop {} {\bibfield
  {journal} {\bibinfo  {journal} {arXiv preprint arXiv:2205.12110}\ } (\bibinfo
  {year} {2022})}\BibitemShut {NoStop}%
\bibitem [{\citenamefont {Hubregtsen}\ \emph {et~al.}(2021)\citenamefont
  {Hubregtsen}, \citenamefont {Pichlmeier}, \citenamefont {Stecher},\ and\
  \citenamefont {Bertels}}]{expressibilityhaar}%
  \BibitemOpen
  \bibfield  {author} {\bibinfo {author} {\bibfnamefont {T.}~\bibnamefont
  {Hubregtsen}}, \bibinfo {author} {\bibfnamefont {J.}~\bibnamefont
  {Pichlmeier}}, \bibinfo {author} {\bibfnamefont {P.}~\bibnamefont {Stecher}},
  \ and\ \bibinfo {author} {\bibfnamefont {K.}~\bibnamefont {Bertels}},\
  }\href@noop {} {\bibfield  {journal} {\bibinfo  {journal} {Quantum Machine
  Intelligence}\ }\textbf {\bibinfo {volume} {3}},\ \bibinfo {pages} {1}
  (\bibinfo {year} {2021})}\BibitemShut {NoStop}%
\bibitem [{\citenamefont {Liang}\ \emph {et~al.}(2005)\citenamefont {Liang},
  \citenamefont {Bressler}, \citenamefont {Buffalo}, \citenamefont {Desimone},\
  and\ \citenamefont {Fries}}]{macaque3}%
  \BibitemOpen
  \bibfield  {author} {\bibinfo {author} {\bibfnamefont {H.}~\bibnamefont
  {Liang}}, \bibinfo {author} {\bibfnamefont {S.~L.}\ \bibnamefont {Bressler}},
  \bibinfo {author} {\bibfnamefont {E.~A.}\ \bibnamefont {Buffalo}}, \bibinfo
  {author} {\bibfnamefont {R.}~\bibnamefont {Desimone}}, \ and\ \bibinfo
  {author} {\bibfnamefont {P.}~\bibnamefont {Fries}},\ }\href@noop {}
  {\bibfield  {journal} {\bibinfo  {journal} {Biological cybernetics}\ }\textbf
  {\bibinfo {volume} {92}},\ \bibinfo {pages} {380} (\bibinfo {year}
  {2005})}\BibitemShut {NoStop}%
\bibitem [{\citenamefont {Punturo}\ \emph {et~al.}(2010)\citenamefont
  {Punturo}, \citenamefont {Abernathy}, \citenamefont {Acernese}, \citenamefont
  {Allen}, \citenamefont {Andersson}, \citenamefont {Arun}, \citenamefont
  {Barone}, \citenamefont {Barr}, \citenamefont {Barsuglia}, \citenamefont
  {Beker} \emph {et~al.}}]{thirdgen}%
  \BibitemOpen
  \bibfield  {author} {\bibinfo {author} {\bibfnamefont {M.}~\bibnamefont
  {Punturo}}, \bibinfo {author} {\bibfnamefont {M.}~\bibnamefont {Abernathy}},
  \bibinfo {author} {\bibfnamefont {F.}~\bibnamefont {Acernese}}, \bibinfo
  {author} {\bibfnamefont {B.}~\bibnamefont {Allen}}, \bibinfo {author}
  {\bibfnamefont {N.}~\bibnamefont {Andersson}}, \bibinfo {author}
  {\bibfnamefont {K.}~\bibnamefont {Arun}}, \bibinfo {author} {\bibfnamefont
  {F.}~\bibnamefont {Barone}}, \bibinfo {author} {\bibfnamefont
  {B.}~\bibnamefont {Barr}}, \bibinfo {author} {\bibfnamefont {M.}~\bibnamefont
  {Barsuglia}}, \bibinfo {author} {\bibfnamefont {M.}~\bibnamefont {Beker}},
  \emph {et~al.},\ }\href@noop {} {\bibfield  {journal} {\bibinfo  {journal}
  {Classical and Quantum Gravity}\ }\textbf {\bibinfo {volume} {27}},\ \bibinfo
  {pages} {194002} (\bibinfo {year} {2010})}\BibitemShut {NoStop}%
\bibitem [{\citenamefont {Abbott}(2016)}]{firstGW}%
  \BibitemOpen
  \bibfield  {author} {\bibinfo {author} {\bibfnamefont {B.~P.}\ \bibnamefont
  {Abbott}} (\bibinfo {collaboration} {LIGO Scientific Collaboration and Virgo
  Collaboration}),\ }\href {\doibase 10.1103/PhysRevLett.116.061102} {\bibfield
   {journal} {\bibinfo  {journal} {Phys. Rev. Lett.}\ }\textbf {\bibinfo
  {volume} {116}},\ \bibinfo {pages} {061102} (\bibinfo {year}
  {2016})}\BibitemShut {NoStop}%
\bibitem [{\citenamefont {{LIGO-Virgo collaboration}}(2019)}]{hanford}%
  \BibitemOpen
  \bibfield  {author} {\bibinfo {author} {\bibnamefont {{LIGO-Virgo
  collaboration}}},\ }\href@noop {} {\enquote {\bibinfo {title} {{Data release
  for event GW150914}},}\ }\bibinfo {howpublished}
  {\url{https://www.gw-openscience.org/events/GW150914/}} (\bibinfo {year}
  {2019})\BibitemShut {NoStop}%
\bibitem [{\citenamefont {Center}(2017)}]{ligodata}%
  \BibitemOpen
  \bibfield  {author} {\bibinfo {author} {\bibfnamefont {G.~W. O.~S.}\
  \bibnamefont {Center}},\ }\href@noop {} {\enquote {\bibinfo {title} {{Signal
  processing with GW150914 open data}},}\ }\bibinfo {howpublished}
  {\url{https://www.gw-openscience.org/s/events/GW150914/GW150914_tutorial.html}}
  (\bibinfo {year} {2017})\BibitemShut {NoStop}%
\bibitem [{\citenamefont {Stokes}\ \emph {et~al.}(2020)\citenamefont {Stokes},
  \citenamefont {Izaac}, \citenamefont {Killoran},\ and\ \citenamefont
  {Carleo}}]{qng}%
  \BibitemOpen
  \bibfield  {author} {\bibinfo {author} {\bibfnamefont {J.}~\bibnamefont
  {Stokes}}, \bibinfo {author} {\bibfnamefont {J.}~\bibnamefont {Izaac}},
  \bibinfo {author} {\bibfnamefont {N.}~\bibnamefont {Killoran}}, \ and\
  \bibinfo {author} {\bibfnamefont {G.}~\bibnamefont {Carleo}},\ }\href@noop {}
  {\bibfield  {journal} {\bibinfo  {journal} {Quantum}\ }\textbf {\bibinfo
  {volume} {4}},\ \bibinfo {pages} {269} (\bibinfo {year} {2020})}\BibitemShut
  {NoStop}%
\bibitem [{\citenamefont {Katabarwa}\ \emph {et~al.}(2022)\citenamefont
  {Katabarwa}, \citenamefont {Sim}, \citenamefont {Koh},\ and\ \citenamefont
  {Dallaire-Demers}}]{geometry2q}%
  \BibitemOpen
  \bibfield  {author} {\bibinfo {author} {\bibfnamefont {A.}~\bibnamefont
  {Katabarwa}}, \bibinfo {author} {\bibfnamefont {S.}~\bibnamefont {Sim}},
  \bibinfo {author} {\bibfnamefont {D.~E.}\ \bibnamefont {Koh}}, \ and\
  \bibinfo {author} {\bibfnamefont {P.-L.}\ \bibnamefont {Dallaire-Demers}},\
  }\href@noop {} {\bibfield  {journal} {\bibinfo  {journal} {Quantum}\ }\textbf
  {\bibinfo {volume} {6}},\ \bibinfo {pages} {782} (\bibinfo {year}
  {2022})}\BibitemShut {NoStop}%
\bibitem [{\citenamefont {Johansson}\ \emph {et~al.}(2012)\citenamefont
  {Johansson}, \citenamefont {Nation},\ and\ \citenamefont {Nori}}]{qutip}%
  \BibitemOpen
  \bibfield  {author} {\bibinfo {author} {\bibfnamefont {J.~R.}\ \bibnamefont
  {Johansson}}, \bibinfo {author} {\bibfnamefont {P.~D.}\ \bibnamefont
  {Nation}}, \ and\ \bibinfo {author} {\bibfnamefont {F.}~\bibnamefont
  {Nori}},\ }\href@noop {} {\bibfield  {journal} {\bibinfo  {journal} {Computer
  Physics Communications}\ }\textbf {\bibinfo {volume} {183}},\ \bibinfo
  {pages} {1760} (\bibinfo {year} {2012})}\BibitemShut {NoStop}%
\end{thebibliography}%
\onecolumngrid

\end{document}